\documentclass[aps,prd,onecolumn,groupedaddress,showpacs,nofootinbib,amssymb]{revtex4-2}
\usepackage[dvips]{graphicx}
\usepackage{amssymb}
\usepackage{amsmath}
\usepackage{graphicx,,color}
\usepackage{amsfonts}
\usepackage{bm}
\usepackage{cancel}
\usepackage{comment}
\usepackage{floatflt}
\usepackage{slashed}
\usepackage{appendix}
\usepackage{array}
\usepackage{tabularx}

\newcommand\be{\begin{equation}}
\newcommand\ee{\end{equation}}

\allowdisplaybreaks[4]

\begin{document}

\title{Electroweak Phase Transition in Singlet Extensions of The Standard Model with Dimension-Six Operators}
\author{V.K. Oikonomou$^{1,2}$, Apostolos Giovanakis$^{1}$}
\affiliation{$^{1)}$Department of Physics, Aristotle University of
Thessaloniki, Thessaloniki 54124, Greece} \affiliation{$^{2)}$L.N.
Gumilyov Eurasian National University - Astana, 010008,
Kazakhstan}


\tolerance=5000

\begin{abstract}
The significance of the electroweak phase transition is undeniable,
and although initially it was believed that it was second-order, it is now believed that it is a first-order
transition. However, it is not a strong first-order phase
transition in the context of the Standard Model and the remedy to this issue is to use the Higgs portal
and directly couple the Higgs to a hidden scalar sector. This can
result in a strong electroweak phase transition,
while the couplings to a hidden scalar are constrained by several phenomenological constraints, such as the sphaleron rate criterion and the branching ratio of the Higgs to invisible channels. In this work, we consider the standard singlet extensions of the Standard Model, including dimension-six
non-renormalizable operators that couple a real singlet scalar field with the Higgs doublet. As a result, we examine the effects of those Higgs-singlet couplings on the
electroweak phase transition. The effective theory, where the
non-renormalizable couplings originate from, is considered to be
active beyond 15$\,$TeV. As
we show, the Universe experiences a two-step electroweak phase transition, a primary
phase transition in the singlet sector at a high temperature, and
then a subsequent first-order phase transition from the singlet vacuum to the electroweak vacuum. The singlet's
phase transition can either be second-order or first-order,
depending on the singlet mass and its couplings to the Higgs. In particular, we show that the dimension-six operator assists in generating a strong electroweak
phase transition in regions of the parameter space that were
excluded in the previous singlet extensions of the Standard Model. This is further apparent for low singlet masses \(m_S < m_H/2\) which are rarely taken into account in the literature due to the invisible branching ratio of the Higgs boson. In some limited cases in the parameter space, the
electroweak phase transition is weakened by the presence of the
higher order operator.
\end{abstract}

\pacs{04.50.Kd, 95.36.+x, 98.80.-k, 98.80.Cq,11.25.-w}

\maketitle

\section{Introduction}

The most mysterious eras of our Universe are the primordial eras,
and specifically the inflationary era and the reheating era. These
two eras are strongly related and to date no direct evidence of
the occurrence of these two exists. Much light on these two
mysterious eras is expected to be shed by the upcoming stage 4
Cosmic Microwave Background (CMB) experiments
\cite{CMB-S4:2016ple,SimonsObservatory:2019qwx} and of course by
the future interferometric gravitational wave experiments, like
the LISA, BBO, DECIGO, and the Einstein telescope
\cite{Hild:2010id,Baker:2019nia,Smith:2019wny,Crowder:2005nr,Smith:2016jqs,Seto:2001qf,Kawamura:2020pcg,Bull:2018lat,LISACosmologyWorkingGroup:2022jok}.
The stage 4 CMB experiments will directly probe the B-modes of the
CMB polarization, if they exist, and the interferometers will
probe the stochastic primordial gravitational waves. The latter is
already confirmed to exist at nanohertz frequencies, in 2023 by
the NANOGrav collaboration \cite{nanograv} and by the EPTA
\cite{Antoniadis:2023ott}, the PPTA \cite{Reardon:2023gzh} and the
CPTA \cite{Xu:2023wog} collaborations. The future seems to be
fruitful in this research line.

Moreover, due to the complete absence of particle detections at
the LHC since the detection of the Higgs particle, and recall that
the LHC functions at a center-of-mass energy $13.6\,$TeV, modern
high energy physics relies on gravitational wave experiments and
astrophysical observations in order to address fundamental
problems related to baryogenesis, electroweak phase transition,
and dark matter. However, we must mention that the particles
colliding at the LHC are basically hadrons and they have some non-negligible form factors, so the actual theories and energies
probed at the LHC might actually correspond to lower energies than
$13.6\,$TeV center-of-mass. To date, the electroweak phase
transition is believed to be a first-order phase transition
\cite{Carrington:1991hz}, although earlier studies indicated that
it was a second-order phase transition \cite{Kajantie:1995kf}. The
order of the electroweak phase transition has triggered a great
deal of heated debate in the last decades. In the past years, it
was shown in numerous studies using perturbative calculations that
the electroweak phase transition is a first-order phase transition
\cite{Carrington:1991hz}. On the other hand, it is understood that
this topic is still controversial. In fact, recent simulations and
studies indicate that with the Higgs mass being $\sim 125\,$GeV,
the phase transition is even a crossover, and not a first-order
phase transition, in the customary sense, see for example Refs.
\cite{Wagner:2023vqw,Laine:2012jy} and for an older lattice
simulation, see Ref. \cite{Kajantie:1995kf} which points out to a
second-order phase transition. A first-order phase transition
would provide non-equilibrium conditions and thus one of the
Sakharov criteria \cite{Sakharov:1967dj} would hold true and the
baryogenesis could occur during the electroweak phase transition.
First-order phase transitions also produce a stochastic
gravitational wave background, and are thus observationally
important
\cite{Apreda:2001us,Schabinger:2005ei,Kusenko:2006rh,McDonald:1993ex,Chala:2018ari,Davoudiasl:2004be,Baldes:2016rqn,Noble:2007kk,Zhou:2020ojf,
Weir:2017wfa,Hindmarsh:2020hop,Han:2020ekm,Child:2012qg,Fairbairn:2013uta,LISACosmologyWorkingGroup:2022jok,Caprini:2015zlo,Huber:2015znp,
Delaunay:2007wb,Chung:2012vg,Barenboim:2012nh,Senaha:2020mop,Grojean:2006bp,Katz:2014bha,Alves:2018jsw,Athron:2023xlk}.
The problem with the electroweak phase transition in the Standard
Model (SM) is that it is not strong enough, and thus singlet
extensions of the SM active through the Higgs portal have been
proposed to solve it
\cite{Profumo:2007wc,Damgaard:2013kva,Ashoorioon:2009nf,OConnell:2006rsp,Gonderinger:2012rd,Profumo:2010kp,Gonderinger:2009jp,Barger:2008jx,
Cheung:2012nb,Barger:2007im,Cline:2013gha,Burgess:2000yq,Kakizaki:2015wua,Enqvist:2014zqa,Chala:2018ari,Noble:2007kk,Katz:2014bha,Espinosa:1993bs,Alanne:2014bra,Cline:2012hg,Beniwal:2017eik,Curtin:2014jma,Chiang:2018gsn,Dev:2019njv,Ghorbani:2018yfr,Ghorbani:2020xqv,Espinosa:2011ax,Espinosa:2007qk,
Kurup:2017dzf,Alves:2018jsw,Athron:2023xlk}. In this work, we
consider the standard real singlet extension, including higher
order non-renormalizable interactions between the Higgs field and
a real singlet scalar field and we concretely examine the effects
of these Higgs-singlet couplings on the electroweak phase
transition. The dimension-six non-renormalizable operators are
considered to originate from a weakly coupled effective theory
that remains active well beyond 15$\,$TeV. As we demonstrate, the
Universe experiences two phase transitions, a primary phase
transition in the singlet scalar sector at high temperatures, and
then the ordinary electroweak phase transition. The singlet scalar
phase transition can either be a second-order or a first-order
phase transition, which depends solely on the singlet scalar mass
and its couplings to the Higgs boson. More importantly, we
demonstrate that due to the presence of the dimension-six
non-renormalizable operator, the electroweak phase transition is
stronger in some regions of the parameter space, which were
excluded in the standard singlet extensions of the SM without the
higher order operators. However, we show that, in some limited
cases for some parameter values, the electroweak phase transition
is actually weakened by the presence of the higher order operator.

This paper is organized as follows: In section II, we present the effective potential for the SM enriched with the singlet extension,
including the higher order non-renormalizable operators. In
section III, we discuss the constraints from the electroweak
baryogenesis, which must be taken into account in our analysis, in
sections IV and V, we showcase
the allowed parameter space of our model and we also discuss various phenomenological constraints and
constraints from invisible Higgs decay. Finally, in section VI, we
study and present all the different possibilities for the
electroweak phase transition in our singlet-extended SM, and the
conclusions follow at the end of the article.

\section{SM Effective Potential with Singlet Extensions and Higher Order Operators}

\begin{figure}
\centering
\includegraphics[width=20pc]{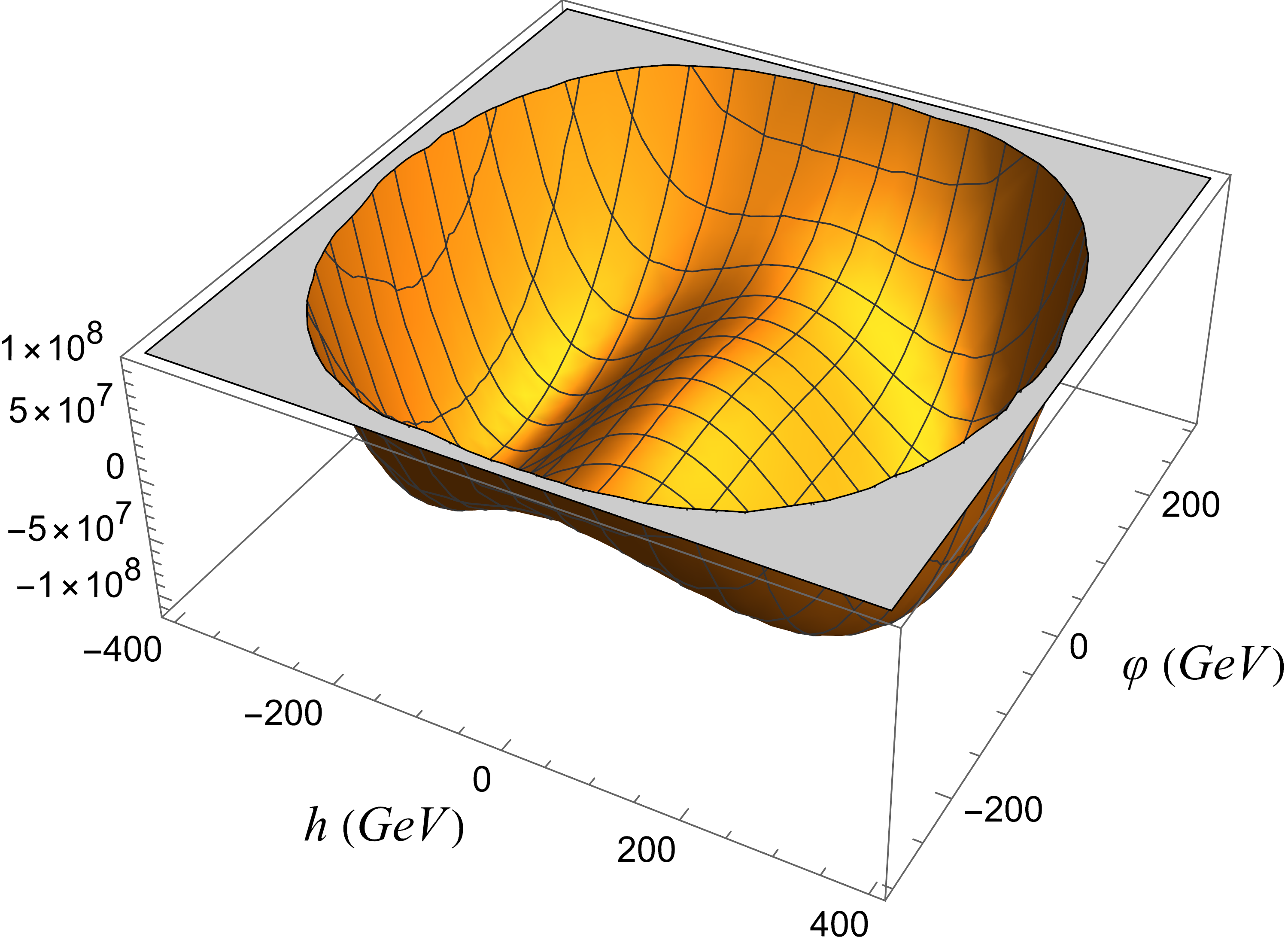}
\includegraphics[width=20pc]{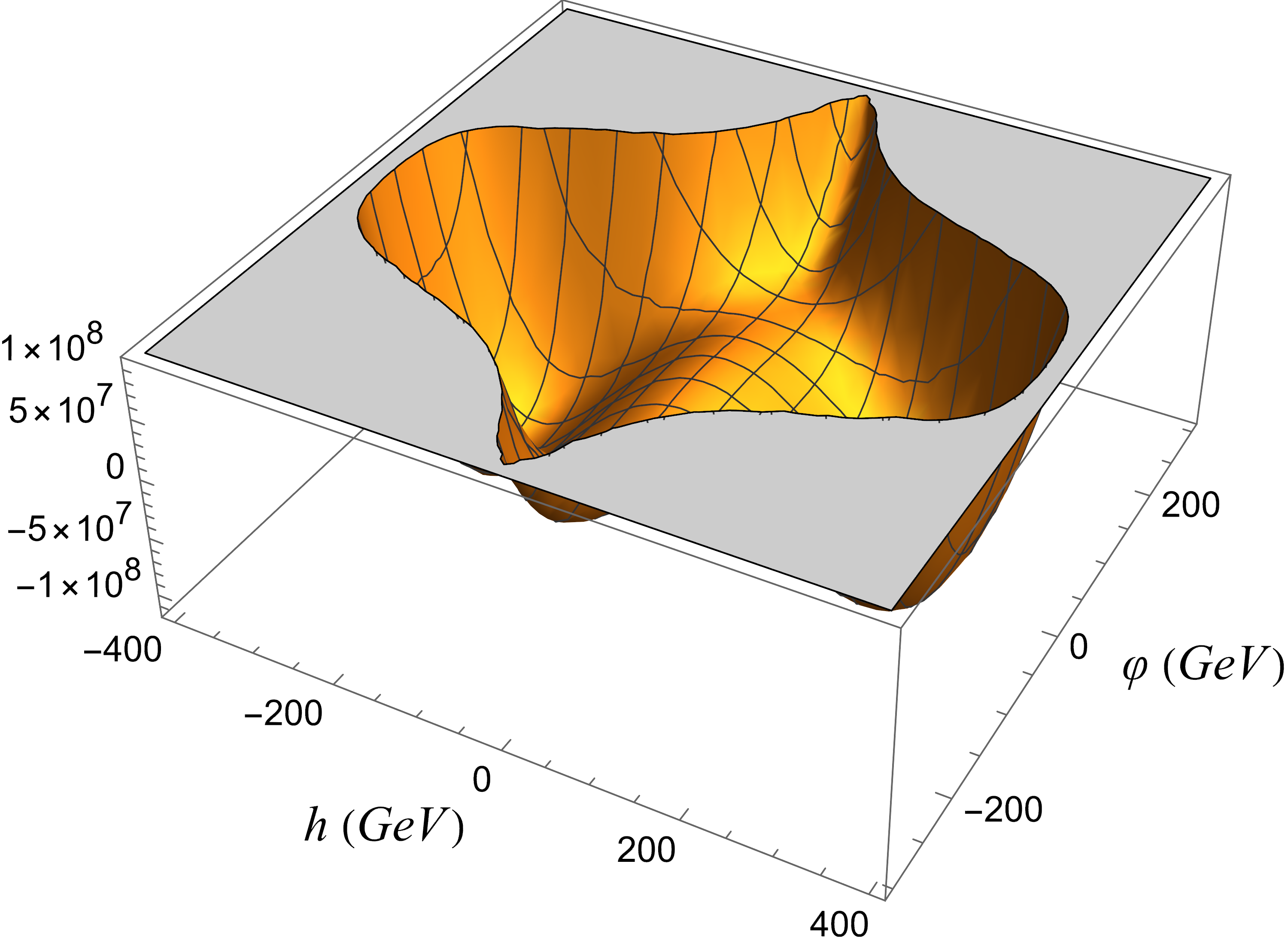}
\caption{The tree-level potential in the field space for \(\lambda
= 0\) (left) and \(\lambda/M^2 = 2 \times 10^{-5}\) GeV\(^{-2}\)
(right), where the global minimum is the electroweak vacuum. We
assumed that \(m_S = 62.5\) GeV, \(\lambda_{HS} = 0.1\), and \(a =
0.1\).}\label{treelevelpotential}
\end{figure}

We extend the SM by introducing a real singlet scalar field
\(\phi\)
\cite{Profumo:2007wc,Damgaard:2013kva,Ashoorioon:2009nf,OConnell:2006rsp,Gonderinger:2012rd,Profumo:2010kp,Gonderinger:2009jp,Barger:2008jx,
Cheung:2012nb,Barger:2007im,Cline:2013gha,Burgess:2000yq,Kakizaki:2015wua,Enqvist:2014zqa,Chala:2018ari,Noble:2007kk,Katz:2014bha,Espinosa:1993bs,Alanne:2014bra,Cline:2012hg,Beniwal:2017eik,Curtin:2014jma,Chiang:2018gsn,Dev:2019njv,Ghorbani:2018yfr,Ghorbani:2020xqv,Espinosa:2011ax,Espinosa:2007qk,
Kurup:2017dzf,Alves:2018jsw,Athron:2023xlk} equipped with a
\(\mathbb{Z}_2\) discrete symmetry, under which \(\phi \to -
\phi\) and all other SM fields remain unaffected. In addition, we
shall assume that a higher dimensional operator of the singlet
scalar field is weakly coupled to the Higgs sector. This higher order non-renormalizable operator originates from an
effective theory active at a scale \(M\), which will be assumed to
be way higher than the electroweak scale, of the order \(M = 15 -
100\) TeV, a fact that is further motivated by the lack of new
particle observations in the LHC, beyond the electroweak symmetry
breaking scale \cite{Chala:2018ari}. The dimensionless coupling
\(\lambda\) is the Wilson coefficient of the higher order
effective theory. In general, the coefficient of the higher order term is
considered to be small such that \(\lambda/M^2 < 10^{-4}\) GeV\(^{-2}\)
and the contribution of the higher dimensional operator to the
invisible Higgs decays is omitted. The tree-level potential for
the Higgs doublet and the singlet scalar field is given by,
\begin{equation}\label{eq:1.1}
    V_0 (H,\phi) = - \mu_H^{2} |H|^2 + \lambda_H |H|^4 - \frac{\mu^2_S}{2}  \phi^2 + \frac{\lambda_S}{4} \phi^4 + \lambda_{HS} |H|^2 \phi^2 + \frac{\lambda}{M^2}|H|^2 \phi^4,
\end{equation}
where \(m_H = \sqrt{2}\mu_H \) is the Higgs boson mass,
\(\lambda_H > 0 \) is the Higgs self-coupling, and \(\lambda_{HS}\)
is the Higgs-singlet interaction coupling, which is assumed to be
positive. In this work, we consider one of the simplest
higher dimensional operators in the real singlet extensions to the SM. The \(D > 6\) operators are suppressed and it can
be easily shown that their effect on the electroweak phase
transition is not so strong for valid values of the Wilson
coefficients. In principle, dimension-eight and higher order
operators can be added, due to some non-perturbative physics
motivated Lagrangian terms of the form $\sim |H|^2\cos
(\frac{\phi}{M})$, but the dominant term is always the dimension-six
operator. The tree-level potential of the singlet scalar field is
determined by the four parameters \(\mu_S\), \(\lambda_S\),
\(\lambda_{HS}\) and \(\lambda\) and the energy scale \(M\). The
SM Higgs doublet is parameterized as follows,
\begin{equation}\label{eq:1.2}
    H = \frac{1}{\sqrt{2}}\begin{pmatrix}
        \chi_1 + i \chi_2 \\
        h + i \chi_3
        \end{pmatrix},
\end{equation}
where \(h\) is the Higgs boson, \(\chi_1, \chi_2, \chi_3\) are
three Goldstone bosons, and \(\upsilon = \mu_H /\sqrt{\lambda_H}
\) is the electroweak symmetry breaking minimum of the Higgs
scalar field \(h\) at zero temperature, without the singlet
sector. The tree-level potential (\ref{eq:1.1}) in the unitary
gauge takes the following form,
\begin{equation}\label{eq:1.3}
    V_0 (h,\phi) = - \frac{\mu^{2}_H}{2} h^2 + \frac{\lambda_H}{4} h^4 - \frac{\mu^2_S}{2}  \phi^2 + \frac{\lambda_S}{4} \phi^4 + \frac{\lambda_{HS}}{2} h^2 \phi^2 +  \frac{\lambda}{2M^2}h^2 \phi^4,
\end{equation}
which is shown in Fig. \ref{treelevelpotential}. In the context of
the SM, we consider only the dominant contributions to the
one-loop finite temperature effective potential coming from the
gauge bosons, the top quark (the heaviest fermion), the Higgs and
Goldstone bosons. The effective masses of the Higgs boson, the
singlet scalar field and those fields which are coupled to the
background fields \(h\) and \(\phi\) are,
\begin{equation}\label{effectivemasshiggs}
    m^2_h (h,\phi) = - \mu^2_H + 3\lambda_H h^2 + \lambda_{HS} \phi^2 + \frac{\lambda}{M^2}\phi^4,
\end{equation}
\begin{equation}\label{effectivemasschi}
    m^2_{\chi} (h,\phi) = - \mu^2_H + \lambda_H h^2 + \lambda_{HS} \phi^2 +\frac{\lambda}{M^2}\phi^4,
\end{equation}
\begin{equation}\label{effectivemasssinglet}
    m^2_S (h,\phi) = - \mu^2_S + 3\lambda_S \phi^2 + \lambda_{HS} h^2 + \frac{6\lambda}{M^2}h^2 \phi^2,
\end{equation}
\begin{equation}\label{effectivemassW}
    m^2_W (h) = \frac{g^2}{4} h^2,
\end{equation}
\begin{equation}\label{effectivemassZ}
    m^2_Z (h) = \frac{g^2 + g^{\prime 2}}{4} h^2,
\end{equation}
\begin{equation}\label{effectivemasstop}
    m^2_t (h) = \frac{y^2_t}{2} h^2,
\end{equation}
where \(g\),\(g^{\prime}\) and \(y_t\) are the \(SU(2)_L\),
\(U(1)_Y\) and top quark Yukawa couplings, respectively, and \(m_H
= 125\) GeV, \(m_W = 80.4\) GeV, \(m_Z = 91.2\) GeV and \(m_t =
173\) GeV at the current vacuum state of the Universe (\(h =
\upsilon\)).

\par The full effective potential is defined as the sum of the
tree-level potential and the one-loop finite-temperature effective
potential which splits into a zero-temperature and a
temperature-dependent part, which will be presented explicitly in
the next two sections.

\subsection{Zero-Temperature Corrections}

The zero-temperature one-loop contribution to the effective
potential is called the Coleman-Weinberg (CW) potential
\cite{Coleman:1973jx} and is computed as the sum of all one-particle
irreducible Feynman diagrams with zero
external momenta and a single loop. In the \(\overline{\rm MS}\) renormalization
scheme, it is written as,
\begin{equation}\label{eq:10}
    V^i_1 (h,\phi) = (-1)^{F_i} n_i \frac{m^4_{i}(h,\phi)}{64 \pi^2} \left[ \ln{\left( \frac{m^2_{i}(h,\phi)}{\mu^2_R}\right)} - C_i \right],
\end{equation}
where \(i = \{ h, \chi, \phi, W, Z, t \}\) counts the particles
that mainly contribute to the CW potential and couple with the Higgs
doublet and the singlet scalar field, \(F_i = 1\) \((0)\) for
fermions (bosons), \(n_i\) is the number of degrees of freedom of
each particle \(i\), \(\mu_R\) denotes the renormalization scale,
and \(C_i = 3/2\) \((5/6)\) for scalars and fermions (gauge
bosons). The degrees of freedom of each particle \(i\) are,
\begin{equation}\label{eq:11}
n_h = 1, \quad n_{\chi} = 3, \quad n_{\phi} = 1, \quad n_{W} = 6, \quad n_{Z} = 3, \quad n_{t} = 12.
\end{equation}
The choice of the renormalization scale leads to an uncertainty in
the critical temperature \(T_c\) and to other related quantities
\cite{Chiang:2018gsn,Athron:2022jyi, Croon:2020cgk,
Gould:2021oba}. Different schemes can be followed to compute the
CW potential and handle this uncertainty, such as the on-shell
(OS) and on-shell-like schemes (OS-like). In the literature, the
OS-like scheme is commonly adopted, since it has a fixed
prescription for the renormalization scales for each particle and
the tree-level relations among the parameters are valid at higher
loop corrections, like in the OS scheme. For instance, in this
scheme, the CW potential takes the form,
\begin{equation}\label{OS}
     V_1^{(OS)} (h,\phi) = \sum_{i} \frac{(-1)^{F_i} n_i}{64 \pi^2} \left[  m^4_{i}(h,\phi)\left( \ln  \frac{m^2_{i}(h,\phi)}{m^2_{i}(\upsilon,0)} - \frac{3}{2} \right) + 2 m^2_{i}(h,\phi) m^2_{i}(\upsilon,0) \right].
\end{equation}
However, both OS schemes suffer from an infrared divergence which
originates from the Goldstone bosons as they acquire a zero mass
at zero temperature. This subtlety can be rectified, by the
proposed methods in Refs. \cite{Martin:2014bca,
Elias-Miro:2014pca}, but in our study, we extensively work in the
\(\overline{\rm MS}\) scheme.

In Ref. \cite{Chiang:2018gsn}, it was concluded that varying
\(\mu_R\) from \(m_t/2\) to \(2 m_t\) the OS-like and
\(\overline{\rm MS}\) schemes agreed with each other within the
scale uncertainties which were \((3.8 - 6.2) \% \) in the critical
temperature \(T_c\) and \((10 - 23 )\% \) in the ratio
\(\upsilon_c/T_c\). In the case of \(\mu_R = m_t/2\) the two
results were approximately identical with the highest accuracy. In
our study we have chosen \(\mu_R = m_t/2\). A similar analysis was
also developed in Ref. \cite{Athron:2022jyi}. Lastly, the
renormalization scale-dependence could be eliminated by the
Renormalization Group Equations (RGE) improvement for the CW
potential \cite{Andreassen:2014eha,
Andreassen:2014gha,Bando:1992np,Tamarit:2014dua,Quiros:1999jp},
but this approach is left to be implemented in future studies.

\subsection{Finite-Temperature Corrections}

The temperature dependence of the finite-temperature one-loop
potential is interpreted in terms of the free energy of an ideal
gas. The effective potential at finite temperature contains the
effective potential at zero temperature, which was presented in the
previous section. Thus, we focus on its temperature-dependent
component\footnote{This component vanishes at zero temperature.}
for each particle \(i\) and it is equal to,
\begin{equation}\label{T-potential}
V_T^i(h,\phi,T) = (-1)^{F_i} \frac{n_iT^4}{2 \pi^2} \int_{0}^{\infty} dx \, x^2 \ln \left[ 1 - (-1)^{F_i} \exp \left( - \sqrt{x^2 + \frac{m^2_{i}  (h,\phi)}{T^2} }\right) \right],
\end{equation}
where the thermal functions are integrals defined as follows,
\begin{equation}\label{thermalfunction}
J_{B/F} \left(y^2\right) = \int_{0}^{\infty} dx \, x^2 \ln \left[ 1 \mp \exp \left( - \sqrt{x^2 + y^2 }\right) \right],
\end{equation}
where the subscript \(B\) \((F)\) stands for bosons (fermions).
The thermal functions can be computed numerically, but they also
admit a high-temperature expansion (for \(|y^2|\ll1\))
\cite{Quiros:1999jp}, which will be applied in this work, and is
given by,
\begin{equation}\label{bosonthermalfunction}
    J_B(y^2) =- \frac{\pi^4}{45} + \frac{\pi^2}{12} y^2 - \frac{\pi}{6} y^3 - \frac{1}{32} y^4 \log \left(\frac{y^2}{a_b} \right) + O (y^6),
\end{equation}
\begin{equation}\label{fermionthermalfunction}
    J_F(y^2) = \frac{7\pi^4}{360} - \frac{\pi^2}{24} y^2 - \frac{1}{32} y^4 \log \left(\frac{y^2}{a_f} \right) + O (y^6),
\end{equation}
where \(a_b = \pi^2 \exp{\left(3/2 - 2 \gamma_E \right)}\), \(a_f
= 16 \pi^2 \exp{\left(3/2 - 2 \gamma_E\right)}\), \(\zeta\)
denotes the Riemann \(\zeta\)-function, and \(\gamma_E\) is the
Euler-Mascheroni constant. Numerical analysis shows that the
high-temperature expansion up to the logarithmic term is accurate
to better than \(5 \% \) for \(y \leq 1.6\) for fermions and \(y
\leq 2.2\) for bosons \cite{Laine:2016hma, Anderson:1991zb}.
Moreover, this approximation for both the potential and its
derivatives also agrees with the exact form to better than
approximately \( 10 \% \) for the values \(y \lesssim (1 - 3)\),
depending on the function \cite{Curtin:2016urg}.

The symmetry restoration at high temperature implies that
conventional perturbation theory breaks down near the critical
temperature \cite{Quiros:1999jp}. As a matter of fact, if
perturbation theory were to remain valid, considering the
temperature-independence of the tree-level potential,
temperature-dependent radiative corrections should be incapable of
restoring the symmetry. In particular, it can be proved that the
one-loop approximation in terms of small coupling constants,
breaks down at a high temperature, due to the appearance of
infrared divergences for the zero Matsubara modes of bosonic
degrees of freedom \cite{Senaha:2020mop}. Therefore, higher loop
corrections that are contained in the daisy diagrams have to be
resummed in the perturbative expansion. In principle, the
so-called daisy resummation is a method aiming to sum all powers
of the coupling constant resulting in a theory where the effective
mass is \(m^2_i \to M^2_i = m^2_i + \Pi_i (T) \), which is called
the thermal mass, and \(\Pi_i(T)\) is the temperature-dependent
self-energy corresponding to the one-loop resummed diagrams to
leading powers of the temperature. This replacement in the full
temperature-dependent effective potential is usually described by
the Parwani scheme \cite{Parwani:1991gq} and leads to,
\begin{equation}\label{Parwani}
    V_{\text{eff}} (h,\phi,T) = V_0 (h,\phi) + \sum_{i} \left[ V^i_1 \left(m^2_i(h,\phi) + \Pi_i(T)\right) + V^i_T \left(m^2_i(h,\phi)+ \Pi_i(T),T \right) \right].
\end{equation}
A similar method, developed by Arnold and Espinosa
\cite{Arnold:1992rz}, resums only the zero Matsubara modes that
cause the infrared divergence. In the Arnold-Espinosa scheme, the
resummed full effective potential is written as follows,
\begin{equation}\label{Arnold-Espinosa}
     V_{\text{eff}} (h,\phi,T) = V_0 (h,\phi) + \sum_{i} \left[ V^i_1 \left(m^2_i(h,\phi)\right) + V^i_T \left(m^2_i(h,\phi),T \right) + V^i_{ring} \left(m^2_i(h,\phi),T\right) \right].
\end{equation}
The last term is added to incorporate the daisy
resummation\footnote{This scheme is equivalent to the Parwani
scheme in the high-temperature limit \cite{Curtin:2016urg}.} and
it is given by,
\begin{equation}\label{rings}
    V^i_{ring} \left(m^2_i(h,\phi),T \right) = \frac{\overline{n}_i T}{12\pi} \left[m^3_i(h,\phi) - \left(m^2_i(h,\phi) + \Pi_i(T) \right)^{3/2} \right],
\end{equation}
where \(i = \{h, \chi, \phi, W, Z, \gamma\}\) and
\(\overline{n}_i = \{1, 3, 1, 2, 1, 1\}\) is the modified number
of degrees of freedom, which takes into account that only the
longitudinal polarizations of the gauge bosons contribute to the
temperature-dependent self-energy
\cite{Arnold:1992rz,Carrington:1991hz}. A comparison between these
schemes is presented in Refs. \cite{Athron:2022jyi,
Croon:2020cgk}, where it is apparently shown that the numerical
differences between the two schemes regarding the critical
temperature and the ratio \(\upsilon_c/T_c\) are relatively small.
Finally, the Arnold-Espinosa scheme is widely preferred in beyond
the Standard Model physics and it will be adopted throughout this
paper.

In this model, the temperature-dependent self-energy for the
scalar fields is computed as,
\begin{equation}\label{T-higgs}
    \Pi_h (T) = \Pi_{\chi} (T) = \left(\frac{3g^2 }{16} + \frac{g^{\prime 2}}{16}  +\frac{y^2_t }{4} + \frac{\lambda_H}{2} + \frac{\lambda_{HS}}{12}\right) T^2,
\end{equation}
\begin{equation}\label{T-singlet}
    \Pi_S (T) = \left( \frac{\lambda_S}{4} + \frac{\lambda_{HS}}{3} +  \frac{\lambda \upsilon^2}{2M^2} \right) T^2.
\end{equation}
The thermal corrections of the gauge bosons require special
treatment as the transverse gauge fields have zero thermal
corrections \cite{Carrington:1991hz}. According to Ref.
\cite{Carrington:1991hz}, we compute the temperature-dependent
self-energy of the longitudinal gauge bosons in the high-temperature limit,
\begin{equation}\label{T-W}
    \Pi_{W_L} (T) = \frac{11}{6}g^2 T^2,
\end{equation}
\begin{equation}\label{T-Z}
    \Pi_{Z_L} (h, T) = \frac{11}{6} g^2T^2 - \frac{g^{\prime 2}}{4} h^2,
\end{equation}
\begin{equation}\label{T-Photon}
    \Pi_{\gamma_L} (h,T) = \frac{11}{6}g^{\prime 2} T^2 + \frac{g^{\prime 2}}{4} h^2,
\end{equation}
where the gauge boson mass matrix in the basis \((A^{1}_{\mu}, A^{2}_{\mu}, A^{3}_{\mu}, B_{\mu})\) is non-diagonal
\begin{equation}\label{gaugebosonmassmatrix}
    M^2_{GB} (h, T) = \begin{pmatrix}
                \frac{1}{4}g^2 h^2 + \frac{11}{6}g^2 T^2 & 0 & 0 & 0 \\
                0 & \frac{1}{4}g^2 h^2 + \frac{11}{6}g^2 T^2 & 0 & 0 \\
                0 & 0 & \frac{1}{4}g^2 h^2 + \frac{11}{6}g^2 T^2 & -\frac{1}{4}gg^{\prime}h^2 \\
                0 & 0 & -\frac{1}{4}gg^{\prime}h^2  & \frac{1}{4}g^{\prime 2} h^2 + \frac{11}{6}g^{\prime} T^2 \\
                    \end{pmatrix}.
\end{equation}
Nevertheless, the photon and the \(Z\) boson are not mass eigenstates
at high temperature, since there is an additional mixing term
between the \(Z\) boson and the photon \cite{Croon:2020cgk}.
Therefore, the gauge boson mass matrix has the following
eigenvalues for the longitudinal photon and \(Z\) boson:
\begin{equation}\label{Z-thermalmass}
    M^2_{Z_L} = \frac{1}{2} \left[ \frac{1}{4} \left(g^2 + g^{\prime 2}\right) h^2 + \frac{11}{6} \left(g^2 + g^{\prime 2} \right) T^2 + \sqrt{\left(g^2 - g^{\prime 2} \right)^2 \left( \frac{1}{4} h^2 + \frac{11}{6}  T^2 \right)^2  + \frac{g^2 g^{\prime 2}}{4} h^4 }\right],
\end{equation}
\begin{equation}\label{Photon-thermalmass}
    M^2_{\gamma_L} = \frac{1}{2} \left[ \frac{1}{4} \left(g^2 + g^{\prime 2}\right) h^2 + \frac{11}{6} \left(g^2 + g^{\prime 2} \right) T^2 - \sqrt{\left(g^2 - g^{\prime 2} \right)^2 \left( \frac{1}{4} h^2 + \frac{11}{6}  T^2 \right)^2  + \frac{g^2 g^{\prime 2}}{4} h^4 }\right].
\end{equation}
In the limit \(m_W^2 (h)/T^2 \ll 1 \), the full resummed mass
eigenvalues (\ref{Z-thermalmass}) and (\ref{Photon-thermalmass})
can be effectively approximated by those in Eqs. (\ref{T-Z}) and
(\ref{T-Photon}). The numerical difference between these
expressions is small, which indicates that one could treat the
photon and \(Z\) boson as mass eigenstates \cite{Croon:2020cgk}.

Finally, it is essential to mention that the full effective
potential can have imaginary contributions in the case of negative
squared effective masses. This holds true especially for the
scalar fields due to the logarithmic and cubic terms in the
thermal functions (\ref{bosonthermalfunction}) and
(\ref{fermionthermalfunction}), since the gauge bosons and the top
quark have always a positive squared effective mass. In general,
the effective mass in the logarithmic term is cancelled by its
counterpart in the one-loop zero-temperature correction. Moreover,
the daisy resummation could potentially cure the imaginary part
originating from the cubic term which signals the breakdown of the
perturbative expansion\footnote{Weinberg and Wu claimed that the
imaginary part of the effective potential could also be
interpreted as a decay rate of a state of the scalar fields
\cite{Weinberg:1987vp}.}. Nevertheless, the \(m^2_i(h,\phi) +
\Pi_i(T)\) in the ring correction can be negative for certain
temperatures and field values. As a result, the effective
potential is still complex and we consider only the real part of
the full effective potential and ensure the field's stability
during the phase transition, provided its imaginary counterpart
remains sufficiently insignificant. A relevant discussion can be
found in Ref. \cite{Delaunay:2007wb}.

To sum up, in the Arnold-Espinosa scheme, the full effective
potential that describes the dynamics of the phase transition of
our singlet-extended SM reads,
\begin{equation}\label{generalfullpotential}
\begin{split}
    V_{\text{eff}} (h, \phi, T)  = & - \frac{\mu^{2}_H}{2} h^2 + \frac{\lambda_H}{4} h^4 - \frac{\mu^2_S}{2}  \phi^2 + \frac{\lambda_S}{4} \phi^4 + \frac{\lambda_{HS}}{2} h^2 \phi^2 + \lambda \frac{h^2 \phi^4}{2M^2} \\
    & + \sum_{i}  \frac{n_i m^4_{i}(h,\phi)}{64 \pi^2}\left[ \ln \left( \frac{m^2_{i}(h,\phi)}{\mu^2_R}\right) - C_i \right] -  \frac{n_t m^4_{t}(h,\phi)}{64 \pi^2}\left[ \ln \left( \frac{m^2_{t}(h,\phi)}{\mu^2_R}\right) - C_t \right] \\
    & + \sum_{i} \frac{n_iT^4}{2 \pi^2} J_{B} \left(\frac{m^2_i (h,\phi)}{T^2}\right) -  \frac{n_tT^4}{2 \pi^2} J_{F} \left(\frac{m^2_t (h)}{T^2}\right) \\
    & + \sum_{i} \frac{\overline{n}_i T}{12\pi} \left[m^3_i(h,\phi) - \left(M^2_i(h,\phi, T) \right)^{3/2} \right],
\end{split}
\end{equation}
where \(i = \{h, \phi, \chi, W, Z, \gamma \}\) corresponds to the bosons in the extended SM.

Our main results are based on the high-temperature expansion
(\ref{bosonthermalfunction}) and (\ref{fermionthermalfunction})
and its validity is primarily checked by the condition on the
value of the ratio \(M_i/T\), which was previously explained. If
these conditions are violated, we numerically compute the full
effective potential (\ref{generalfullpotential}). Therefore,
according to the high-temperature expansion, the full effective
potential (\ref{generalfullpotential}) becomes,
\begin{equation}\label{HT-fullpotential}
\begin{split}
    V^{\text{HT}}_{\text{eff}} (h, \phi, T)  = & - \frac{\mu^{2}_H}{2} h^2 + \frac{\lambda_H}{4} h^4 - \frac{\mu^2_S}{2}  \phi^2 + \frac{\lambda_S}{4} \phi^4 + \frac{\lambda_{HS}}{2} h^2 \phi^2 +  \frac{\lambda}{2M^2}h^2 \phi^4 \\
    & + \frac{m^2_h (h,\phi)}{24}T^2 - \frac{T}{12 \pi} \left[M^2_h (h,\phi, T) \right]^{3/2} + \frac{m^4_h(h,\phi)}{64 \pi^2} \left[\ln \left(\frac{a_b T^2}{\mu^2_R}\right) -\frac{3}{2} \right] \\
    & + \frac{3 m^2_{\chi} (h,\phi)}{24}T^2 - \frac{3T}{12 \pi} \left[ M^2_{\chi} (h,\phi, T) \right]^{3/2} + \frac{ 3 m^4_{\chi} (h,\phi)}{64 \pi^2} \left[\ln \left(\frac{a_b T^2}{\mu^2_R}\right) -\frac{3}{2} \right] \\
    & + \frac{m^2_{\phi} (h,\phi)}{24}T^2 - \frac{T}{12 \pi} \left[ M^2_{\phi} (h,\phi, T) \right]^{3/2} + \frac{ m^4_{\phi} (h,\phi)}{64 \pi^2} \left[\ln \left(\frac{a_b T^2}{\mu^2_R}\right) -\frac{3}{2} \right]  \\
    &  + \frac{6 m^2_{W} (h)}{24}T^2 - \frac{4T}{12 \pi} m^3_{W} (h) - \frac{2T}{12 \pi}\left[ M^2_{W_{L}} (h, T)  \right]^{3/2} + \frac{ 6 m^4_{W} (h)}{64 \pi^2} \left[\ln \left(\frac{a_b T^2}{\mu^2_R}\right) -\frac{5}{6} \right]  \\
    & + \frac{3 m^2_{Z} (h)}{24}T^2 - \frac{2T}{12 \pi} m^3_{Z} (h) - \frac{T}{12 \pi} \left[  M^2_{Z_{L}} (h, T) \right]^{3/2} + \frac{ 3 m^4_{Z} (h)}{64 \pi^2} \left[\ln \left(\frac{a_b T^2}{\mu^2_R}\right) -\frac{5}{6} \right]   \\
    & + \frac{12 m^2_{t} (h)}{48}T^2 - \frac{ 12 m^4_{t} (h)}{64 \pi^2} \left[\ln \left(\frac{a_f T^2}{\mu^2_R}\right) -\frac{3}{2} \right] - \frac{T}{12 \pi}\left[ M^2_{\gamma_{L}} (h, T) \right]^{3/2},
\end{split}
\end{equation}
where the effective and thermal masses are given above. It is
interesting to notice that only the cubic terms of the gauge
bosons do not cancel each other in the high-temperature expansion
of the effective potential and the ring corrections, in contrast
with the rest of the fields which have the same degrees of freedom
in the ring correction. Our main aim is to analyze in detail the
phase transitions that occur at finite temperature for the
effective potential (\ref{HT-fullpotential}). This is a cumbersome
procedure though, since the parameter space is highly constrained
from various aspects and several limitations and constraints apply
that reduce significantly the range of the values of the free
parameters. Thus before we actually study the phase transition, in
the next few sections we shall narrow down the values of the free
parameters that are phenomenologically acceptable and comply with
all the experimental and theoretical constraints.

\section{Electroweak Baryogenesis Constraints}

Initially, the vacuum of the Universe was stabilized
at the origin \((h, \phi) = (0, 0)\). At a high temperature \(T_s\), a phase transition
starts in the \(\phi\) direction from the origin to a non-zero vacuum expectation value (VEV). This phase transition can be either first
or second-order, depending on the singlet extension model's
parameters. As the Universe cools down, a second minimum
appears in the full effective potential, where a barrier is
gradually formed between the two minima. At the temperature
\(T_c\), the electroweak phase transition occurs as a first-order
phase transition, when the two local minima \((0,
\upsilon^{\prime}_{s})\) and \((\upsilon_c, 0)\) are degenerate as
it is illustrated in Figs. \ref{T-potential_2} and
\ref{T-potential_1}. Therefore, the transition from the false vacuum to
the true vacuum proceeds via thermal tunneling, when the bubbles
of the broken phase nucleate within the surrounding plasma of the
phase with the false vacuum.

The observed baryon asymmetry of the Universe can be attributed to
the electroweak baryogenesis (EWBG). This is a physical mechanism
in the early Universe that generates an asymmetry between baryons
and anti-baryons in the electroweak phase transition. This
asymmetry is established if the so-called Sakharov criteria
\cite{Sakharov:1967dj} are fulfilled, which namely are: (i) baryon
number violation, (ii) \(C\)-\(CP\) violation, and (iii) departure from
thermal equilibrium. The EWBG satisfies these criteria requiring a
first-order phase transition \cite{Quiros:1999jp, Trodden:1998ym,
Riotto:1998bt, Riotto:1999yt}. Baryon creation in EWBG occurs at
the vicinity of the expanding bubble walls. More specifically,
ahead of the bubble wall, \(CP\) and \(C\) asymmetries in particle number
densities can be produced by \(CP\)-violating interactions of the
plasma with the expanding bubble wall of the true vacuum
\cite{Morrissey:2012db}. Then, these asymmetries diffuse into the
symmetric phase in front of the bubble wall, where they are
converted to baryons by electroweak sphalerons
\cite{Cohen:1993nk}. In the broken phase, the rate of sphaleron
transitions can be strongly suppressed to avoid washing out the
generated baryons. Hence, it is necessary for a successful EWBG
scenario, that the baryon asymmetry generated at the expanding
bubble wall is not washed out by sphalerons within the broken
phase. In particular, EWBG requires a strong first-order
electroweak phase transition, characterized by the following
condition, which provides an approximation to a factor in the rate
of sphaleron transitions in the broken phase,
\begin{equation}\label{sphaleron_rate}
    \frac{\upsilon_c}{T_c} > 0.6 - 1.4.
\end{equation}
The lower bound of this so-called sphaleron rate
criterion\footnote{This ratio does not respect gauge invariance
and special treatment is required to derive the gauge invariant
form. However, this approach affects the numerical results much
less than the two-loop approximation \cite{Patel:2011th}.} varies
between 0.6 and 1.4 due to some uncertainties in the precise
calculation, but it is conventionally taken to be one
\cite{Patel:2011th, Fuyuto:2014yia}. In our study, we also
consider the lower limit in this range for the sake of
completeness. The critical temperature \(T_c\) and the Higgs VEV \(\upsilon_c\) are computed numerically
in this paper.

In fact, the three Sakharov criteria are satisfied in the SM
\cite{Quiros:1999jp}: the baryon number is violated by
non-perturbative effects, \(CP\) violation can be induced by the CKM
phase in the fermion mass matrix, and the Universe is out of
equilibrium due to the electroweak phase transition. However, \(CP\)
violation is insufficient to produce large enough chiral
asymmetries to explain the observed baryon-to-entropy ratio.
Furthermore, the electroweak phase transition should be strong
enough first-order, a requirement that is not met by the SM as demonstrated in Appendix A, thus
the motivation for singlet extensions is clear. These drawbacks in
SM motivate beyond the Standard Model physics which should suggest
an extension of the Higgs potential to realize a strong
first-order electroweak phase transition and an extra source of \(CP\)
violation. The former is successfully achieved by singlet scalar
extensions of SM, two-Higgs doublet models and higher order
operators to SM \cite{Athron:2023xlk} and this occurs in our case
too.

\(CP\)-violating sources in EWBG can be obtained by a plethora of
models such as real and complex singlet scalar extensions
\cite{Cline:2012hg, Vaskonen:2016yiu, Huang:2018aja,
Jiang:2015cwa, Grzadkowski:2018nbc}, two-Higgs doublet models
\cite{Cline:1995dg}, and composite Higgs models
\cite{Espinosa:2011eu}. A promising \(CP\)-violating source could be a
dimension-six operator, which is introduced as an effective field
theory \cite{Cline:2012hg}. This extension couples the singlet with
the top quark's mass, considering the singlet particle as a dark
matter candidate and modifying top quark's mass for non-zero
values of the singlet scalar field so that it is given by \(\lambda_{HS}\)
\begin{equation}\label{CP_source}
    y_t \bar{Q}_L H \left( 1 + \frac{\eta}{\Lambda^2} \phi^2 \right) t_R + \text{h.c.},
\end{equation}
where \(\eta\) is the complex phase and \(\Lambda\) is the energy
scale of the effective field theory. Consequently, the top quark
acquires a complex phase that varies in space along the profile of
the bubble wall, offering the crucial source of \(CP\) violation
needed to explain baryon asymmetry.

This higher dimension operator does not affect the critical
temperature and VEVs as it vanishes at both the Higgs and singlet
direction. This operator mostly raises the height of the barrier at
temperature \(T_c\), which leads to a thinner bubble wall. It is
additionally assumed that \(\phi/\Lambda\) remains very small
resulting in a negligible contribution to the effective potential
which does not significantly affect the dynamics of the phase
transition and it is not included in our analysis of the effective
potential, see also Refs. \cite{Cline:2012hg, Vaskonen:2016yiu}.
However, a reasonable claim could be that, the fact that this
higher dimension operator does not affect the critical temperature
and VEVs since it vanishes at both the Higgs and singlet direction
could lead to a deeper minimum away from these points? This could
be possible, but it is clearly mentioned that the ratio
\(\phi/\Lambda\) remains very small resulting in a negligible
contribution to the effective potential and a deeper minimum
cannot be generated in the context of a valid perturbative
analysis. The choice of the phase \(\eta\)
is not restricted by any additional experimental constraints due
to the forbidden Higgs-singlet mixing and the behavior of the
singlet at high temperatures. As a result, this phase can be
chosen to be maximally \(CP\)-violating and it easily generates
sufficient baryon asymmetry for EWBG \cite{Cline:2012hg}. However,
potential loop contributions to electric dipole moments might
exist. This is an important aspect that we did not take into
account in this work to simplify the discussion. The two-loop
Barr-Zee contributions to the electric dipole moments are
discussed in Ref. \cite{Espinosa:2011eu}. However, this would require
Higgs-singlet mixing, which is not the case in our model. In that scenario \(CP\) would be broken spontaneously at high temperature.

\section{Allowed Parameter Space for the Model and Phenomenological Constraints}

The parameter space of the model is generally described by the
three parameters of the real singlet extension \(\mu_S\),
\(\lambda_S\), and \(\lambda_{HS}\) and the Wilson
coefficient\footnote{The effective field theory is considered
active at the scale \(M = 15\) TeV in the following sections.}
\(\lambda\). In addition, a strong electroweak phase transition can be realized by a
parameter space that is mainly restricted by the sphaleron rate
criterion, the vacuum structure, the perturbativity of the couplings, and the
invisible Higgs decay width.

Firstly, we proceed on the premise that the current vacuum state
of the Universe is described by the Higgs VEV at zero temperature
with broken \(SU(2) \times U(1)_Y\) symmetry, while the singlet
has a zero VEV with unbroken \(\mathbb{Z}_2\) symmetry. As a
result, we demand that the Higgs minimum is the global minimum of
the effective potential at zero temperature by requiring,
\begin{equation}\label{global_min}
    V_0 (\upsilon, 0) < V_0 (0, \upsilon_s) \Rightarrow \lambda_S > \lambda_H \frac{\mu^4_S}{\mu^4_H},
\end{equation}
where \(\upsilon_s\) is the zero-temperature VEV of the singlet
scalar field in the \(\phi\) direction and it is assumed that
\(\mu^2_S > 0\) for a two-step electroweak phase
transition\footnote{See section VI.}. We additionally demand that
the tree-level potential is bounded from below, which leads to
\(\lambda_S > 0\) and \(\lambda >0\). This vacuum structure further implies that the singlet
scalar field acquires the following mass squared at zero
temperature, which is required to be positive,
\begin{equation}\label{singlet_mass}
    m^2_{S} = - \mu^2_S + \lambda_{HS} \upsilon^2 > 0.
\end{equation}
An expression for the minimum value of \(\lambda_S\) can then be
derived by equations (\ref{global_min}) and (\ref{singlet_mass}),
and it is,
\begin{equation}\label{lambda_min}
    \lambda^{min}_S = \frac{\lambda_H}{\mu^4_H} \left(m^2_S - \lambda_{HS} \upsilon^2 \right)^2.
\end{equation}
Accordingly, the quartic coupling can be cast into the form,
\begin{equation}\label{lambda_definition}
    \lambda_S = \lambda^{min}_S + a,
\end{equation}
where \(a\) is a positive parameter\footnote{It is commonly set to
\(a = 0.1\) for \(m_S \geq m_H/2\) \cite{Curtin:2014jma,
Chiang:2018gsn, Kurup:2017dzf, Jain:2017sqm, Senaha:2020mop}.}.
Consequently, it is more convenient and suitable to study the
parameter space in terms of \(m_S\), \(\lambda_{HS}\), and
\(\lambda\) for a given parameter \(a\). However, it is worth
mentioning that Eq. (\ref{global_min}) could be rewritten in the
equivalent form in terms of the other parameters,
\begin{equation}
    \mu^2_S < \upsilon^2 \sqrt{\lambda_H \lambda_S} \Rightarrow \lambda_{HS} < \frac{m^2_S}{\upsilon^2} + \sqrt{\lambda_H \lambda_S}.
\end{equation}
Secondly, the validity of the one-loop approximation can be
violated by large coupling constants. Thus, we impose the
perturbativity of the couplings in high energy scales, while the
Renormalization Group Equations (RGE) can be solved at one loop
for the couplings \(\lambda_{HS}\), \(\lambda_H\) and
\(\lambda_S\) (see Appendix B). The RGE evolution of the gauge
couplings and the top quark Yukawa coupling remains the same as in
the case of the SM. Moreover, the contribution of the higher order
operator to the RGEs is omitted since the effective field theory
is a weakly coupled theory with \(\lambda/M^2 < 10^{-4}\)
GeV\(^{-2}\). Following Ref. \cite{Curtin:2014jma}, the RGE
evolution shows that the model remains perturbative up to scales
\(10 - 100\) TeV depending on the Higgs-singlet coupling, as well
as \(\lambda^{min}_S < 8\) is imposed for a reliable perturbative
analysis which is equivalent to \(\lambda_{HS} < 5\) with the
singlet mass ranging from \(0 - 550\) GeV.

Moving on, the tree-level potential has an inherent
\(\mathbb{Z}_2\) symmetry, allowing the singlet scalar to act as a
dark matter candidate \cite{Alanne:2014bra, Cline:2013gha,
Ghorbani:2018yfr, Ghorbani:2020xqv, Cline:2012hg,
GAMBIT:2017gge,Athron:2018ipf,Feng:2014vea, Beniwal:2017eik}.
Hence, the unbroken \(\mathbb{Z}_2\) symmetry at zero temperature
ensures the stability of the dark matter particle and prohibits
the mixing between the Higgs field and the singlet scalar field.
As a consequence, the couplings of the Higgs boson to fermions and
gauge bosons remain identical to the SM, as well as further
constraints from electroweak precision tests and Higgs coupling
modifications are ruled out due to the forbidden Higgs-singlet
mixing.

Lastly, the EWBG can be probed by gravitational wave and collider
measurements \cite{Curtin:2014jma, Beniwal:2017eik, Alves:2018jsw,
Athron:2023xlk, Chala:2018ari}. In the case of \(m_S > m_H/2\),
the number of potential collider signatures are highly restricted
due to the forbidden Higgs-singlet mixing. However, future
colliders, such as a \(100\) TeV hadron collider, could test this
scenario, probing the direct production of the singlet states, and
the modification in the triple Higgs couplings and \(Zh\) cross
section \cite{Curtin:2014jma}. Regarding the gravitational
waves, their signals from an electroweak first-order phase
transition can be detected in the future observatories \cite{Beniwal:2017eik, Alves:2018jsw, Athron:2023xlk}, such as LISA.

\section{Constraints from Invisible Higgs Decay}

If \(m_S < m_H/2\), the decay \(h \to \phi \phi\) is kinematically
allowed and it contributes to the invisible decay width of the
Higgs boson. The ATLAS and CMS experiments have reported numerous
results of searches for invisible Higgs decays over the last
years. The branching ratio of the Higgs to the invisible singlet
sector is set to \(BR_{inv} < 0.11 - 0.19\) at \(95 \% \) CL,
where the upper and lower limits of the range correspond to the
results from the ATLAS \cite{ATLAS:2020kdi} and the CMS
collaboration \cite{CMS:2018yfx}, respectively. In Ref.
\cite{ParticleDataGroup:2022pth}, those results are analyzed
further. However, the most recent result from combination of
searches, which is not included in Ref.
\cite{ParticleDataGroup:2022pth}, indicates \(BR_{inv} < 0.107\)
at \(95 \% \) CL \cite{ATLAS:2023tkt}.

The Higgs decay width to visible channels is \(\Gamma_{vis} =
4.07\) MeV for \(m_H = 125\) GeV. Thus, if the branching ratio of
the Higgs to the invisible particles is fixed at \(BR_{inv} <
0.19\), the upper bound on the invisible decay width of the Higgs
boson is,
\begin{equation}\label{decay_width_0.19}
    \Gamma (h \to \phi \phi) < 0.955 \text{ MeV},
\end{equation}
where this decay width is given by,
\begin{equation}\label{decay_width}
    \Gamma (h \to \phi \phi) = \frac{\lambda^2_{HS} \upsilon^2}{32 \pi m_H} \sqrt{1 - \frac{4 m^2_S}{m^2_H}}.
\end{equation}
Hence, this bound leads to an additional constraint for the interaction coupling,
\begin{equation}\label{special_condition_coupling}
    \lambda_{HS} < \sqrt{ \frac{32 \pi m_H}{\upsilon^2}\left({1 - \frac{4 m^2_S}{m^2_H}} \right)^{-1/2}\Gamma_{m} (h \to \phi \phi) },
\end{equation}
where \(\Gamma_m (h \to \phi \phi) \) is the upper bound on the invisible decay width of the Higgs boson. Requiring \(\mu^2_S \geq 0\) and imposing the previous constraint result
in,
\begin{equation}\label{condition_coupling}
    \frac{m_S^2}{\upsilon^2} < \lambda_{HS} < \sqrt{\frac{32 \pi m_H}{\upsilon^2}\left({1 - \frac{4 m^2_S}{m^2_H}} \right)^{-1/2} \Gamma_{m} (h \to \phi \phi) },
\end{equation}
which is completely independent of the coupling \(\lambda_S\) (or equivalently \(a\)) and the Wilson coefficient \(\lambda\).

\begin{figure}[h!]
\centering
\includegraphics[width=20.6pc]{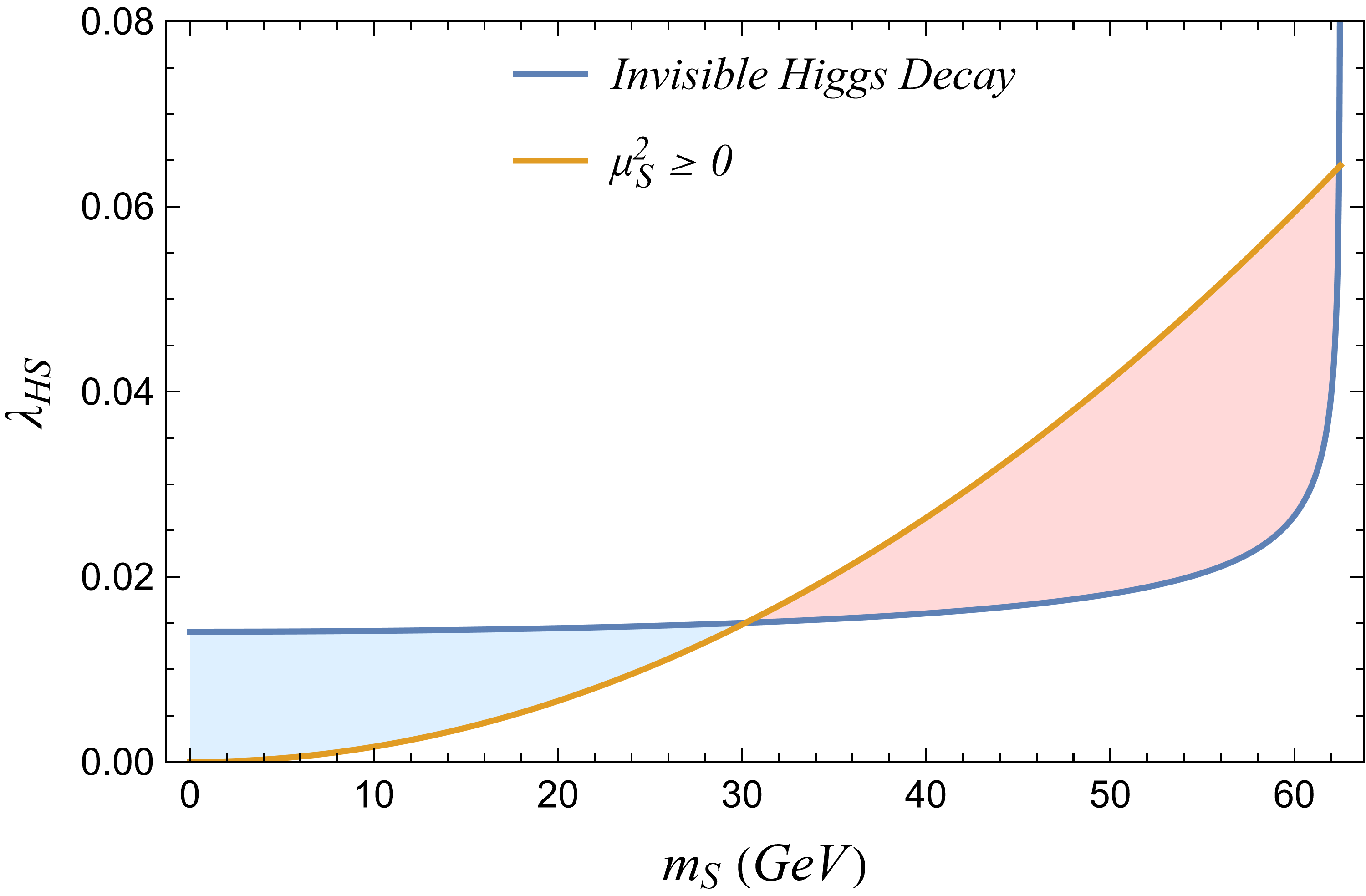}
\includegraphics[width=20.6pc]{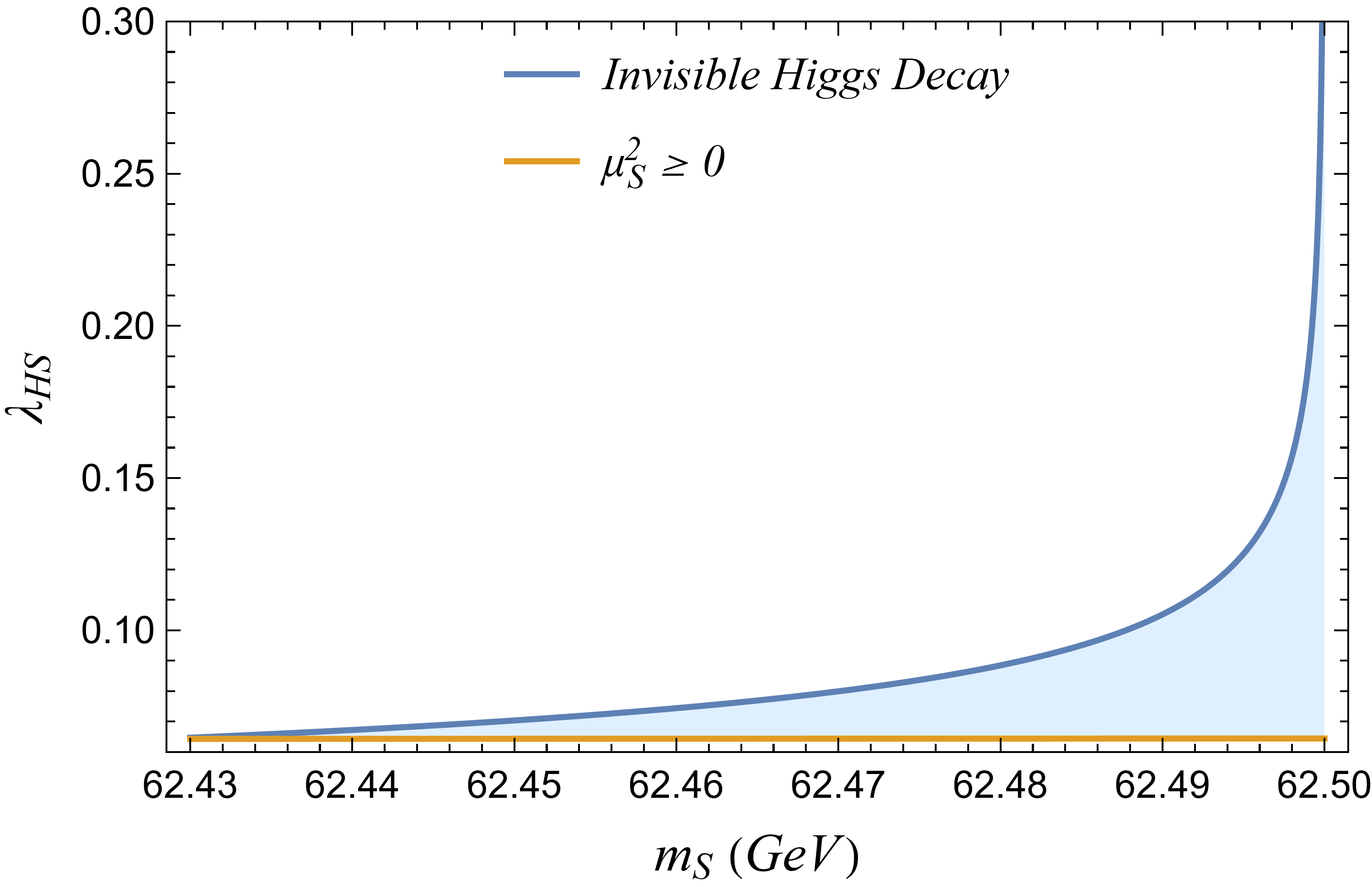}
\caption{\textbf{Left}: The parameter space (blue shaded) for
\(m_S < 62.5 \text{ GeV}\), which satisfies the invisible Higgs
decay constraint (\ref{decay_width_0.19}) and \(\mu_S^2 \geq 0\) setting \(BR_{inv} <
0.19\). \textbf{Right}: The higher mass region, \(m_S \geq 62.43\)
GeV, which satisfies the invisible Higgs decay constraint (\ref{decay_width_0.19}) and
\(\mu_S^2 \geq 0\) setting \(BR_{inv} < 0.19\).} \label{BR_0.19}
\end{figure}

Consequently, the allowed singlet masses are divided into two
regions,
\[m_S \leq 30.19 \text{ GeV} \quad \text{ and } \quad m_S \geq 62.43 \text{ GeV}.\]
where the two separated regions are merged only for \(BR_{inv} >
0.59\), which does not agree with the experimental data
\cite{ParticleDataGroup:2022pth}. We gathered the allowed singlet
masses in Fig. \ref{BR_0.19}. The lower mass region generally has
the upper bound of \(\lambda_{HS} = 0.014\), whereas the higher
mass region is characterized by a wider range of values for
\(\lambda_{HS}\), but it has almost a fixed singlet mass.

\par The invisible Higgs boson branching ratio can be also set to the
\(BR_{inv} < 0.11\) at \(95 \% \) CL in order to yield the
invisible decay width,
\begin{equation}\label{decay_width_0.11}
    \Gamma (h \to \phi \phi) < 0.503 \text{ MeV}
\end{equation}
with the following allowed mass regions which satisfy the condition (\ref{condition_coupling}),
\[m_S \leq 25.45 \text{ GeV} \quad \text{ and } \quad m_S \geq 62.48 \text{ GeV}.\]
It is immediately obvious that this upper bound further excludes
the low-mass region, because of the lower maximum value of the
coupling \(\lambda_{HS}\) compared to the previous case.
\begin{figure}[h!]
\centering
\includegraphics[width=20pc]{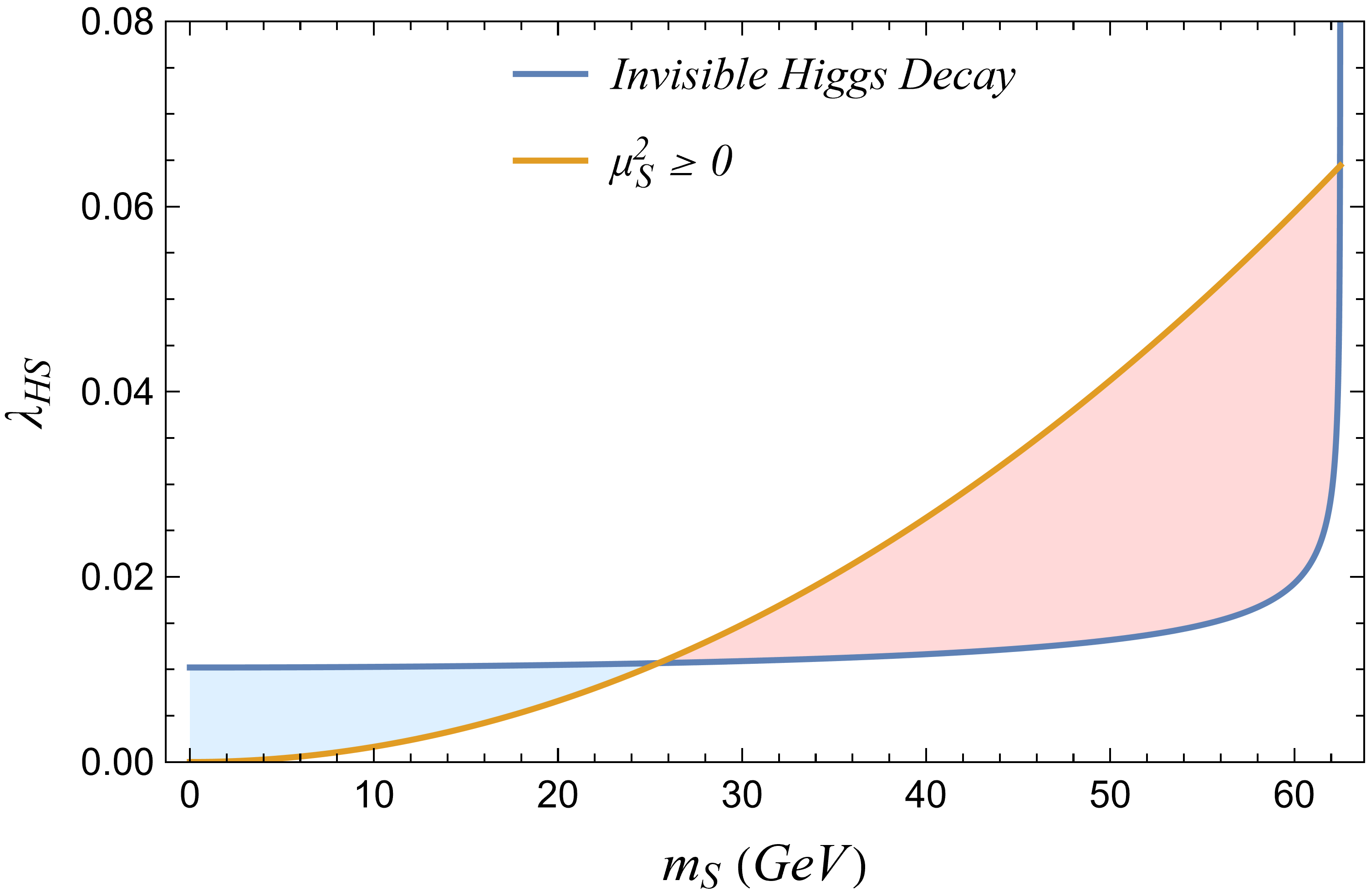}
\includegraphics[width=20pc]{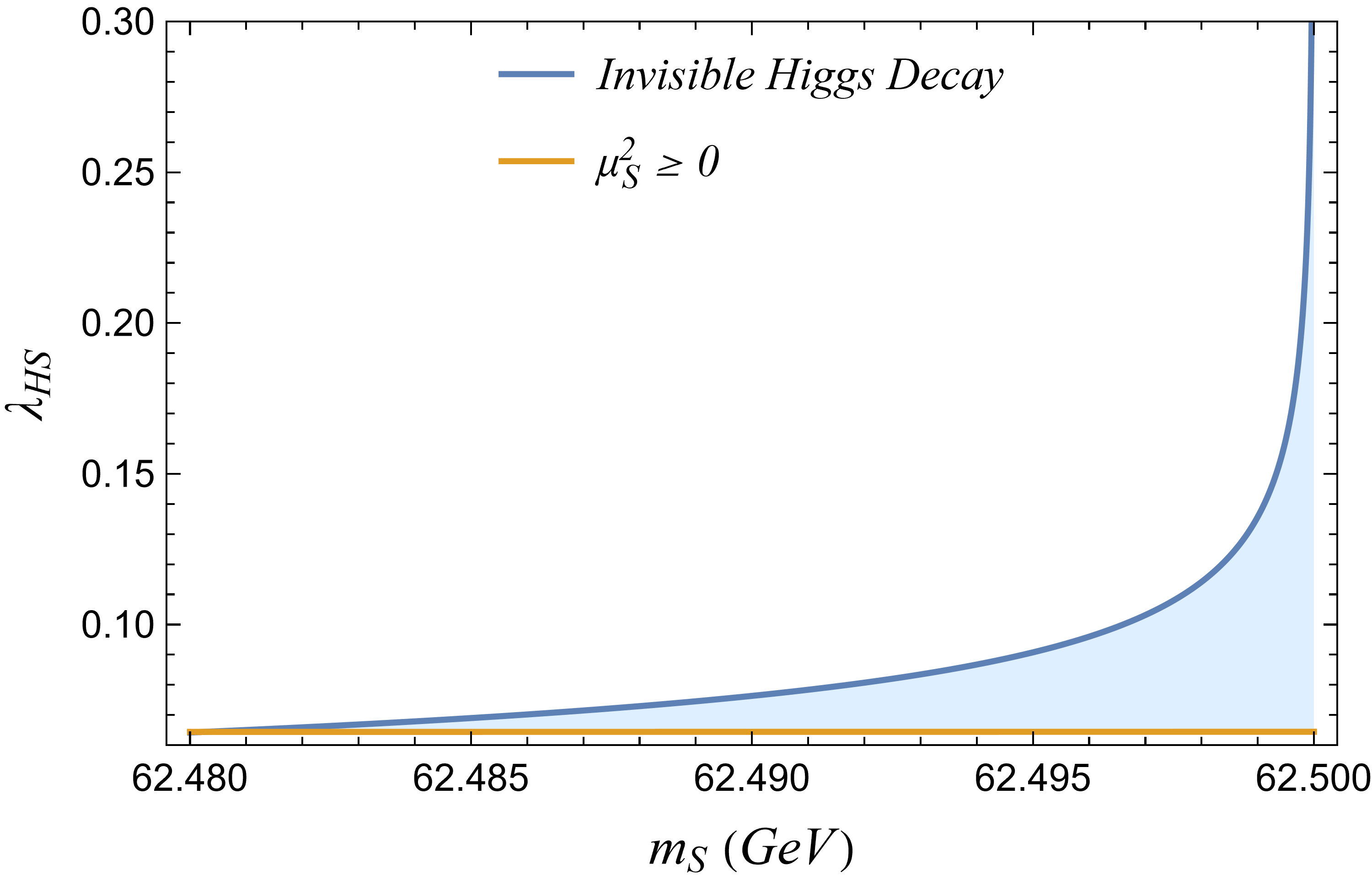}
\caption{\textbf{Left}: The parameter space (blue shaded) for
\(m_S < 62.5 \text{ GeV}\) which satisfies the invisible Higgs
decay constraint (\ref{decay_width_0.11}) and \(\mu_S^2 \geq 0\) setting \(BR_{inv} <
0.11\). \textbf{Right}: The higher mass region, \(m_S \geq 62.48\)
GeV, which satisfies the invisible Higgs decay constraint (\ref{decay_width_0.11}) and
\(\mu_S^2 \geq 0\) setting \(BR_{inv} < 0.11\).} \label{BR_0.11}
\end{figure}
\par Taking into account the sphaleron rate criterion
(\ref{sphaleron_rate}), the parameter space is eliminated further,
since very low values of the coupling \(\lambda_{HS}\) generate a
one-step or a two-step electroweak phase transition with a weak first-order phase
transition. Therefore, the lower bound on the Higgs-singlet
coupling can be higher than the one depicted in Figs. \ref{BR_0.19} and \ref{BR_0.11}.

\section{The Electroweak Phase Transition in the Singlet-extended SM: A Detailed Description}

\begin{figure}[h!]
\centering
\includegraphics[width=14pc]{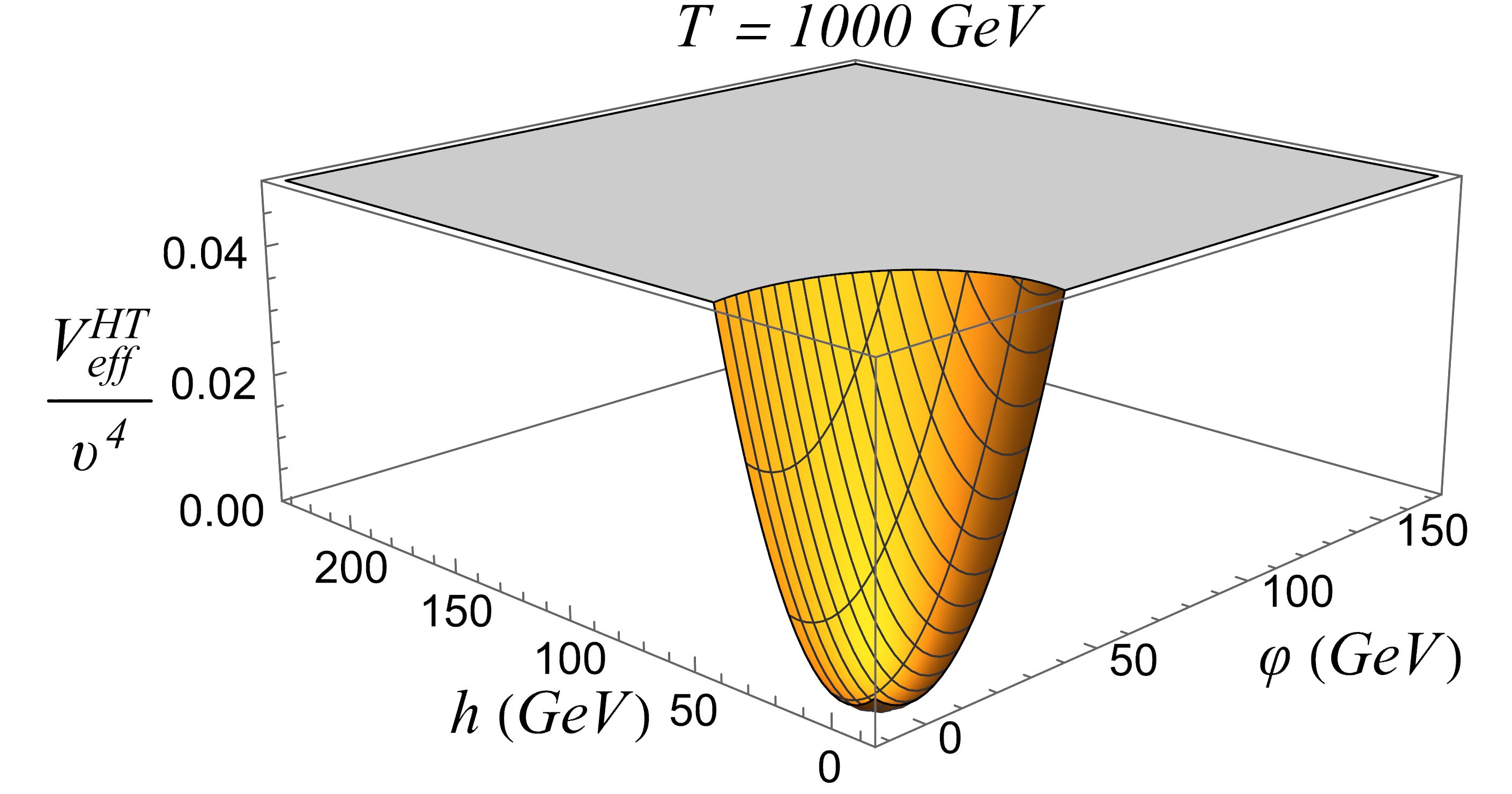}
\includegraphics[width=14pc]{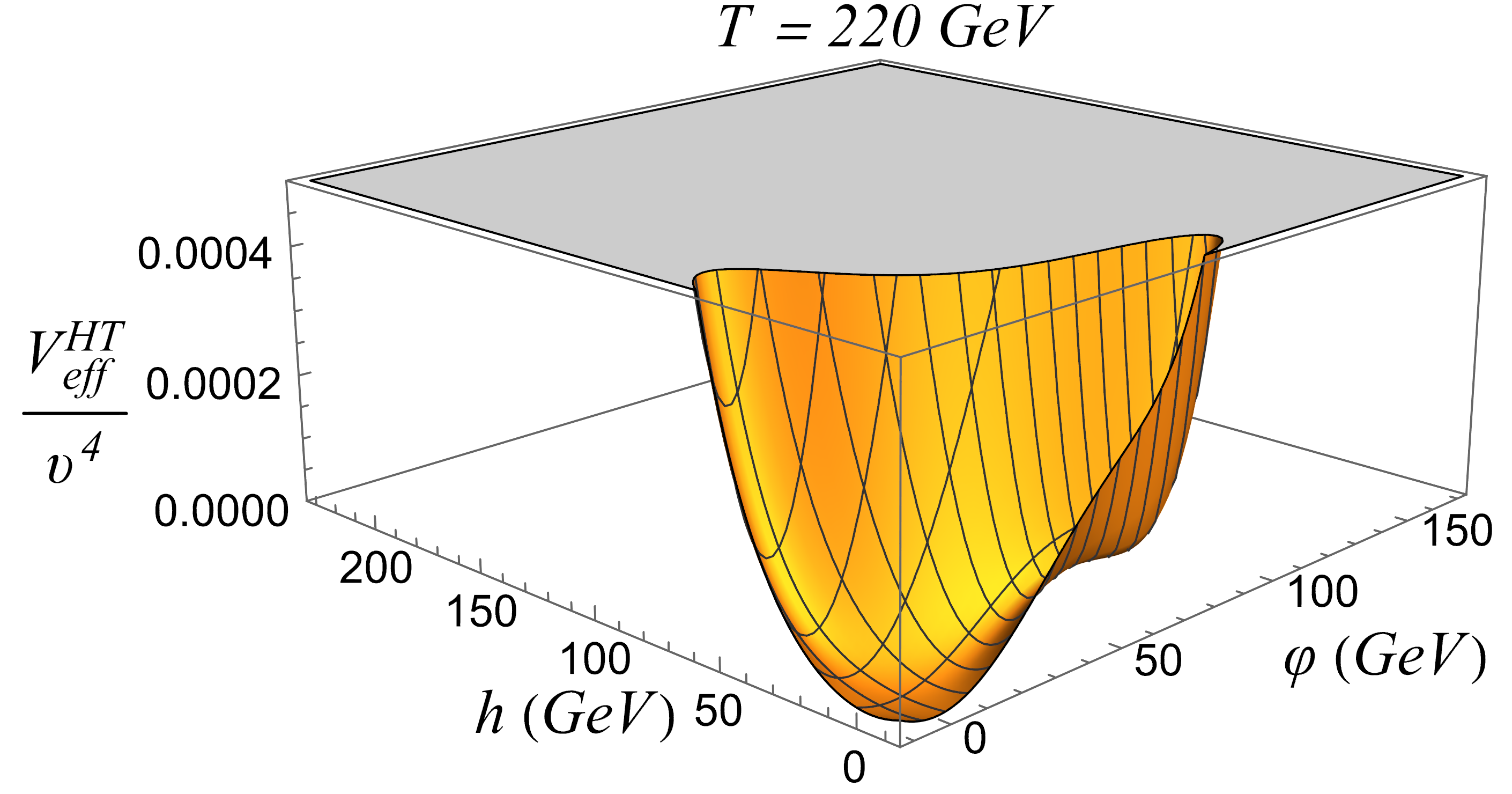}
\includegraphics[width=14pc]{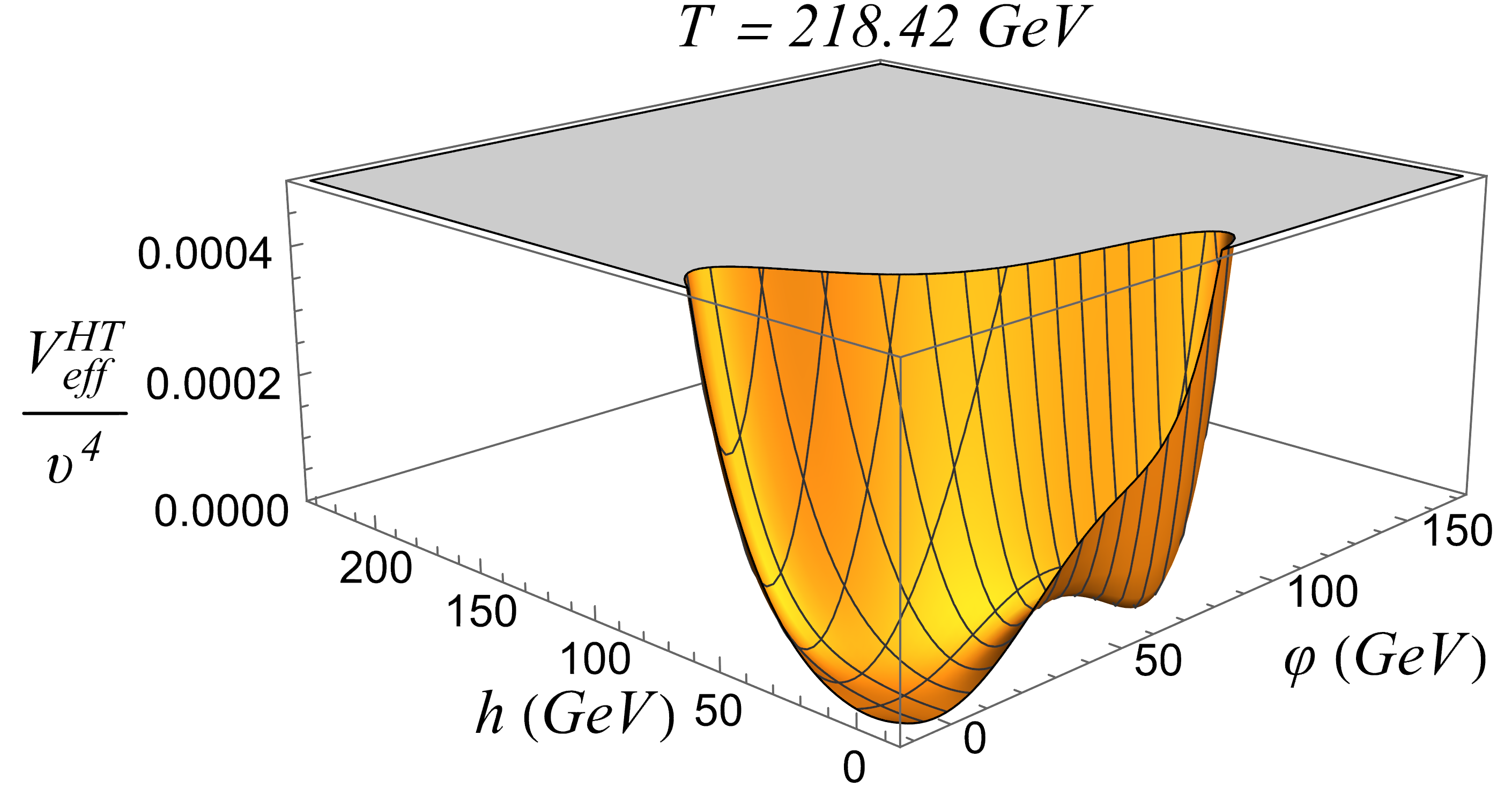}
\includegraphics[width=14pc]{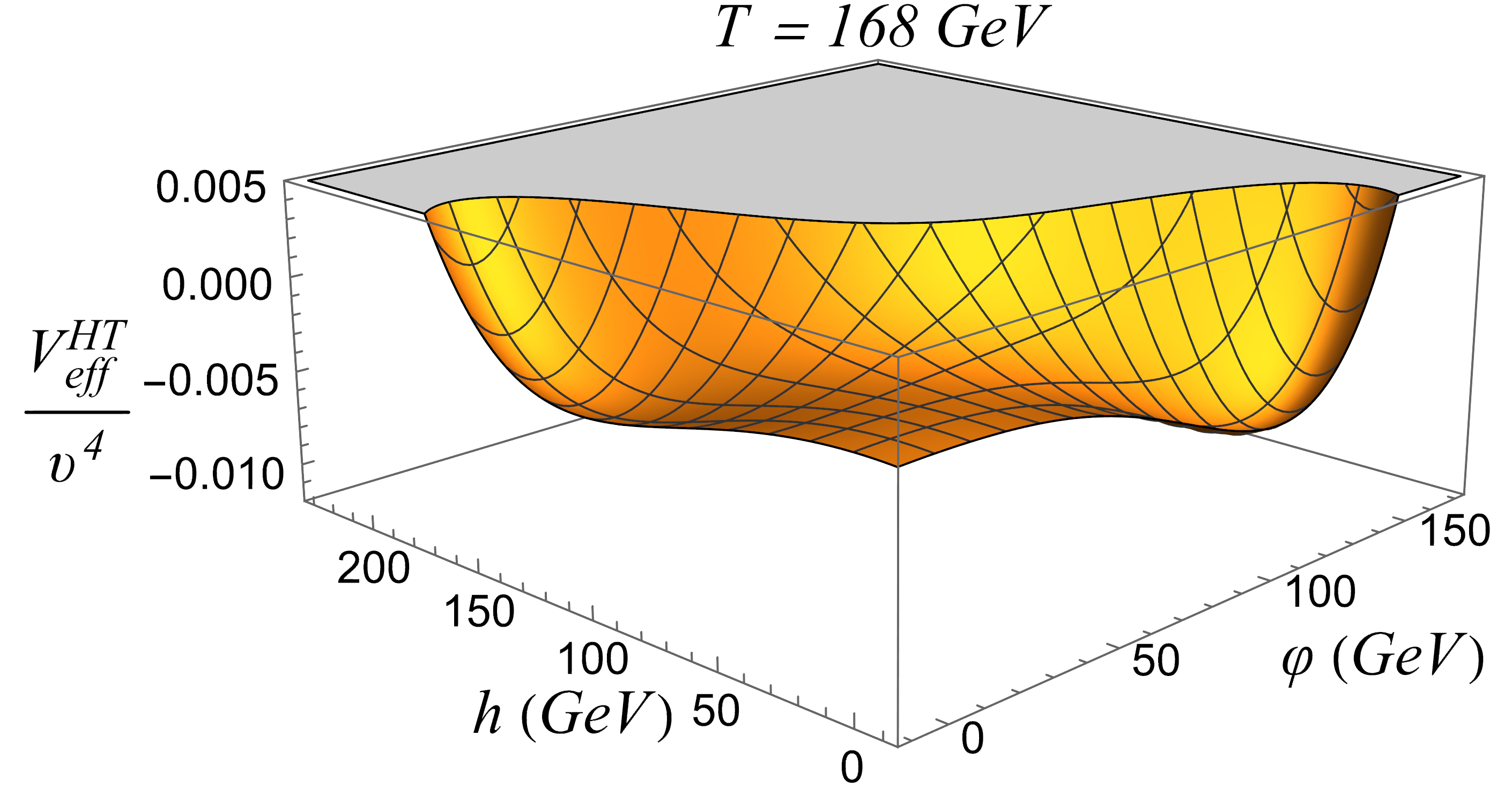}
\includegraphics[width=14pc]{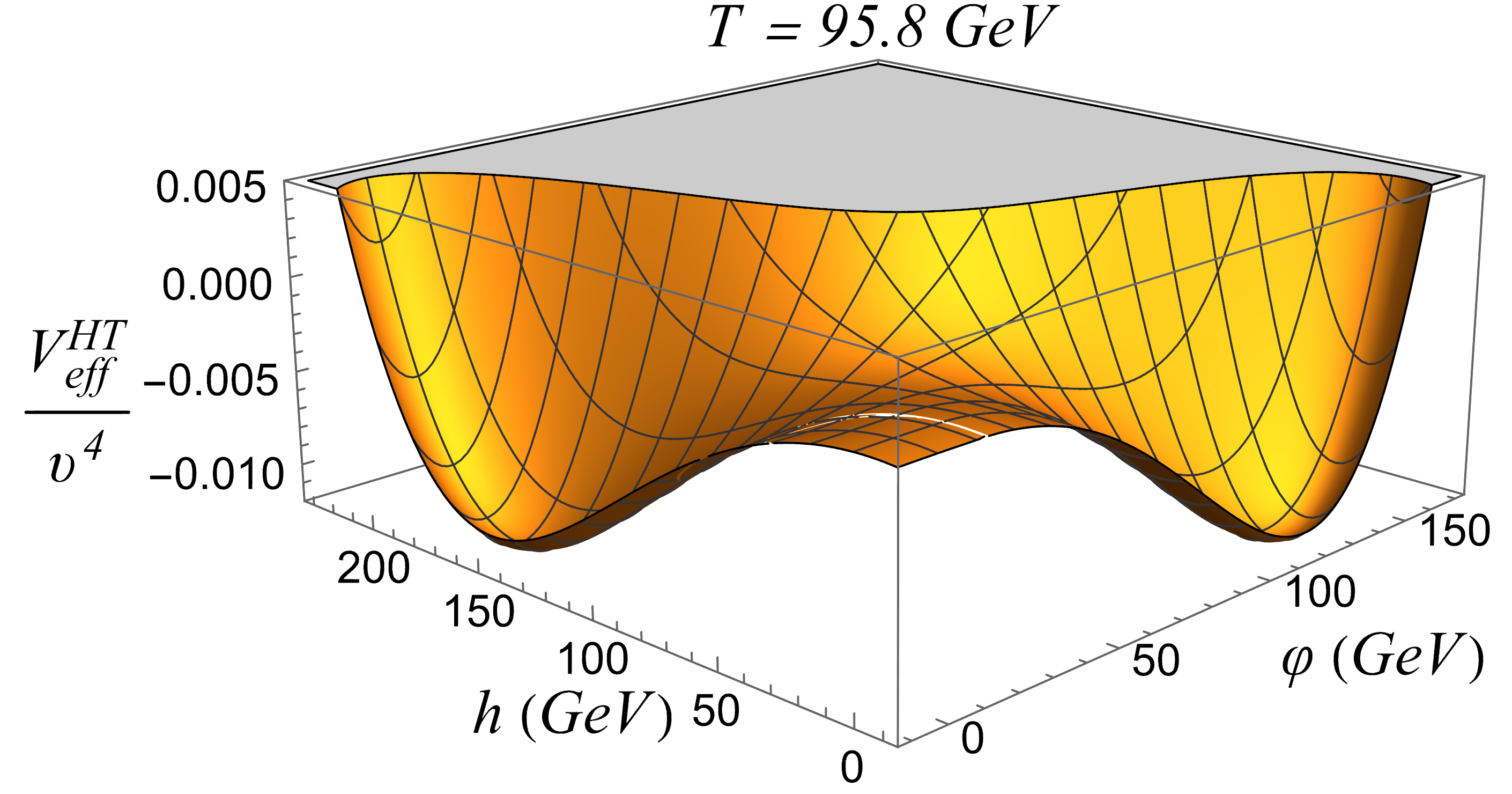}
\includegraphics[width=14pc]{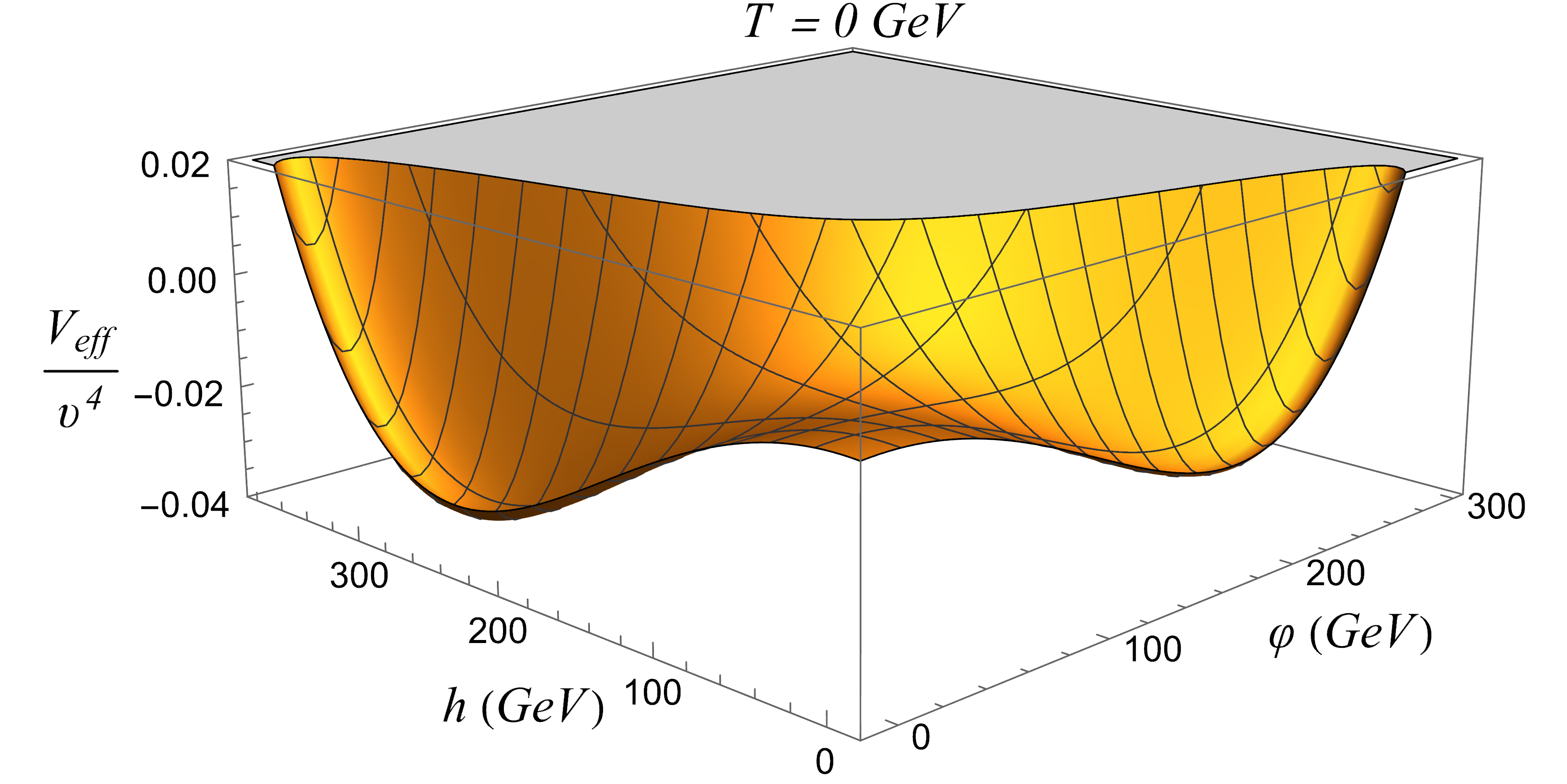}
\caption{The full effective potential during the strong
electroweak phase transition as the temperature decreases. In this
example, the singlet's phase transition is first-order using a
point of the parameter space with \(m_S = 500\) GeV,
\(\lambda_{HS} = 4.3\), \(\lambda/M^2 \simeq 2 \times 10^{-5}\)
GeV\(^{-2}\), and \(a = 0.1\).} \label{T-potential_2}
\end{figure}

\begin{figure}[h!]
\centering
\includegraphics[width=14pc]{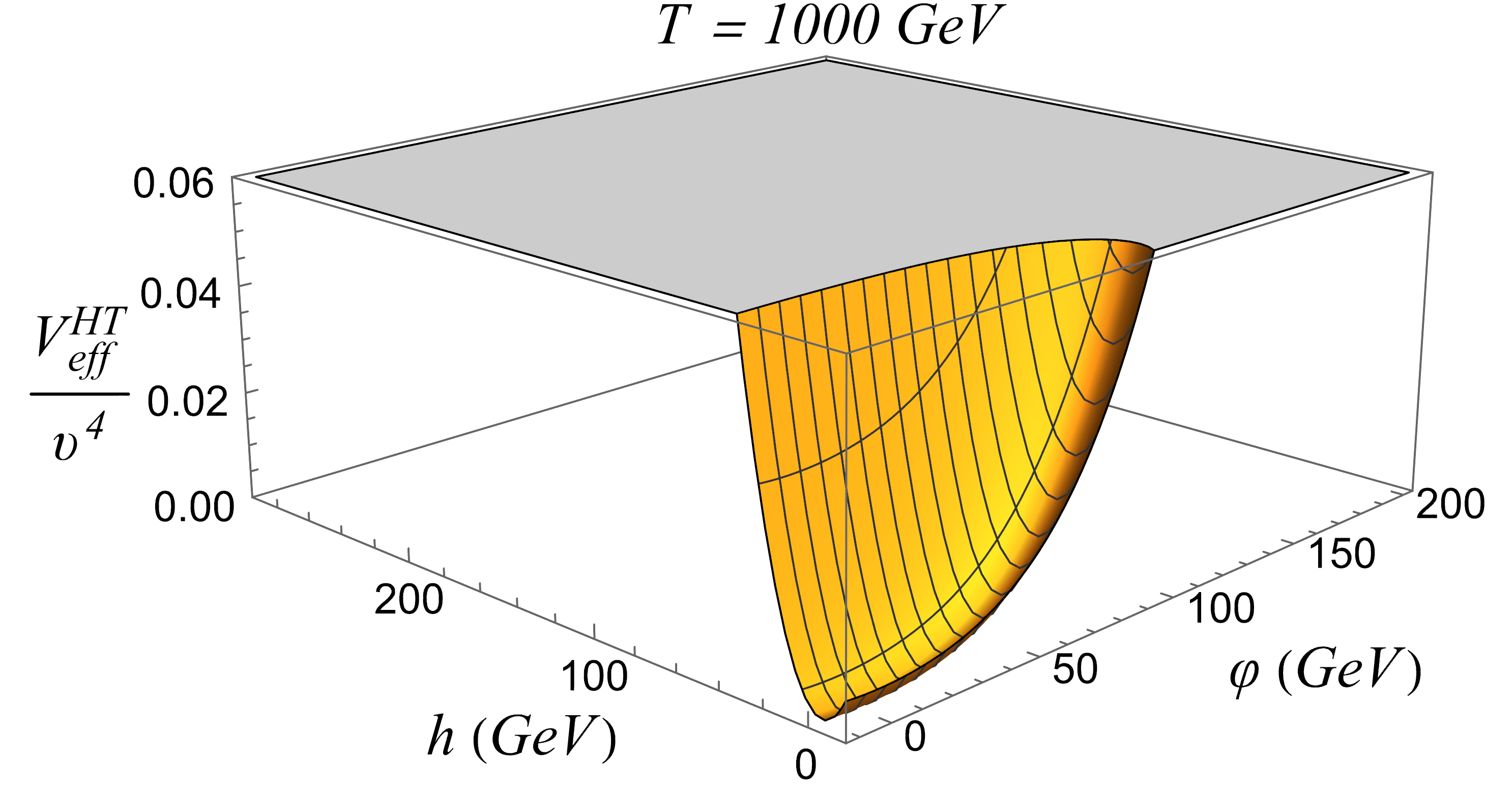}
\includegraphics[width=14pc]{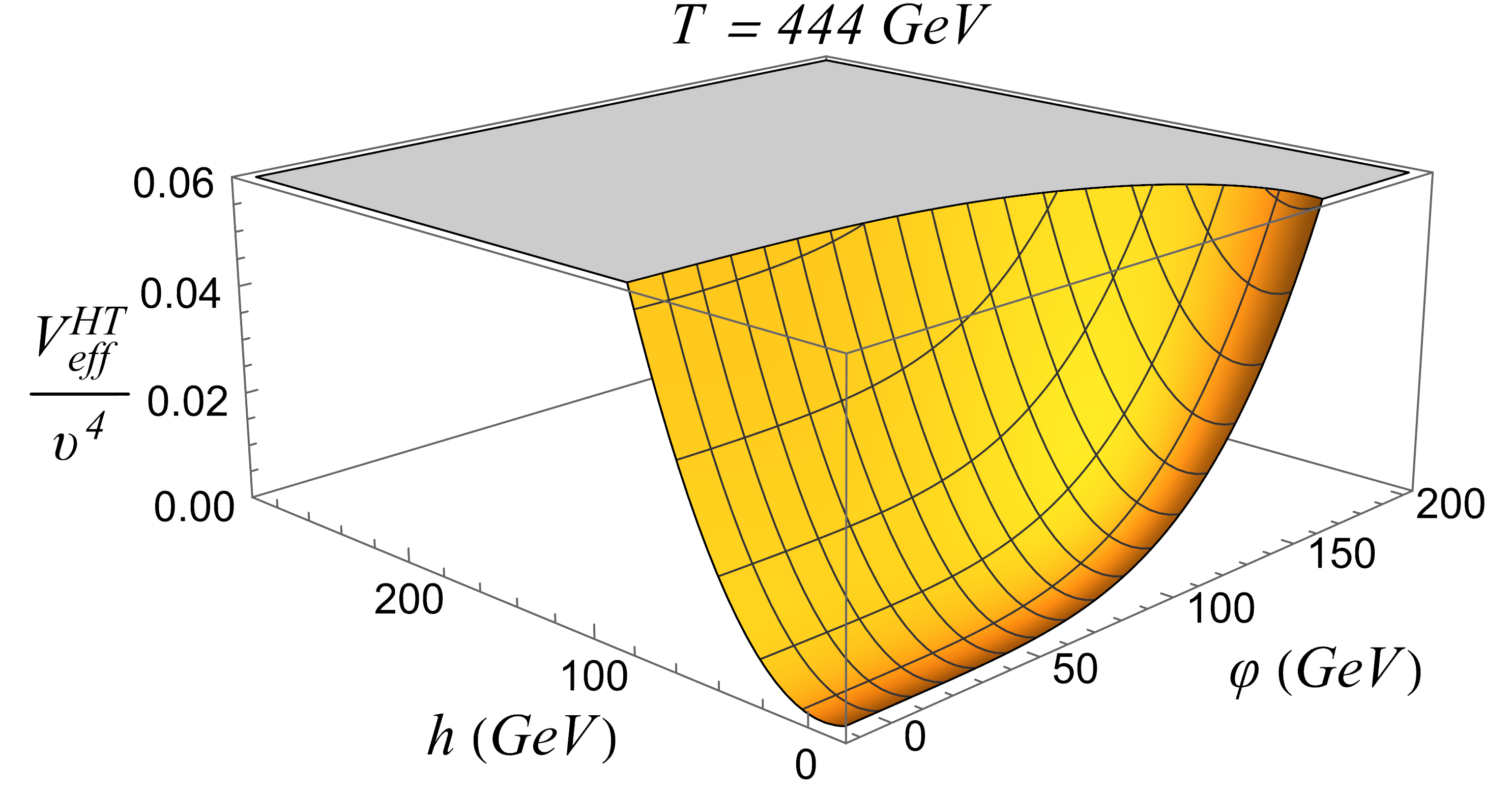}
\includegraphics[width=14pc]{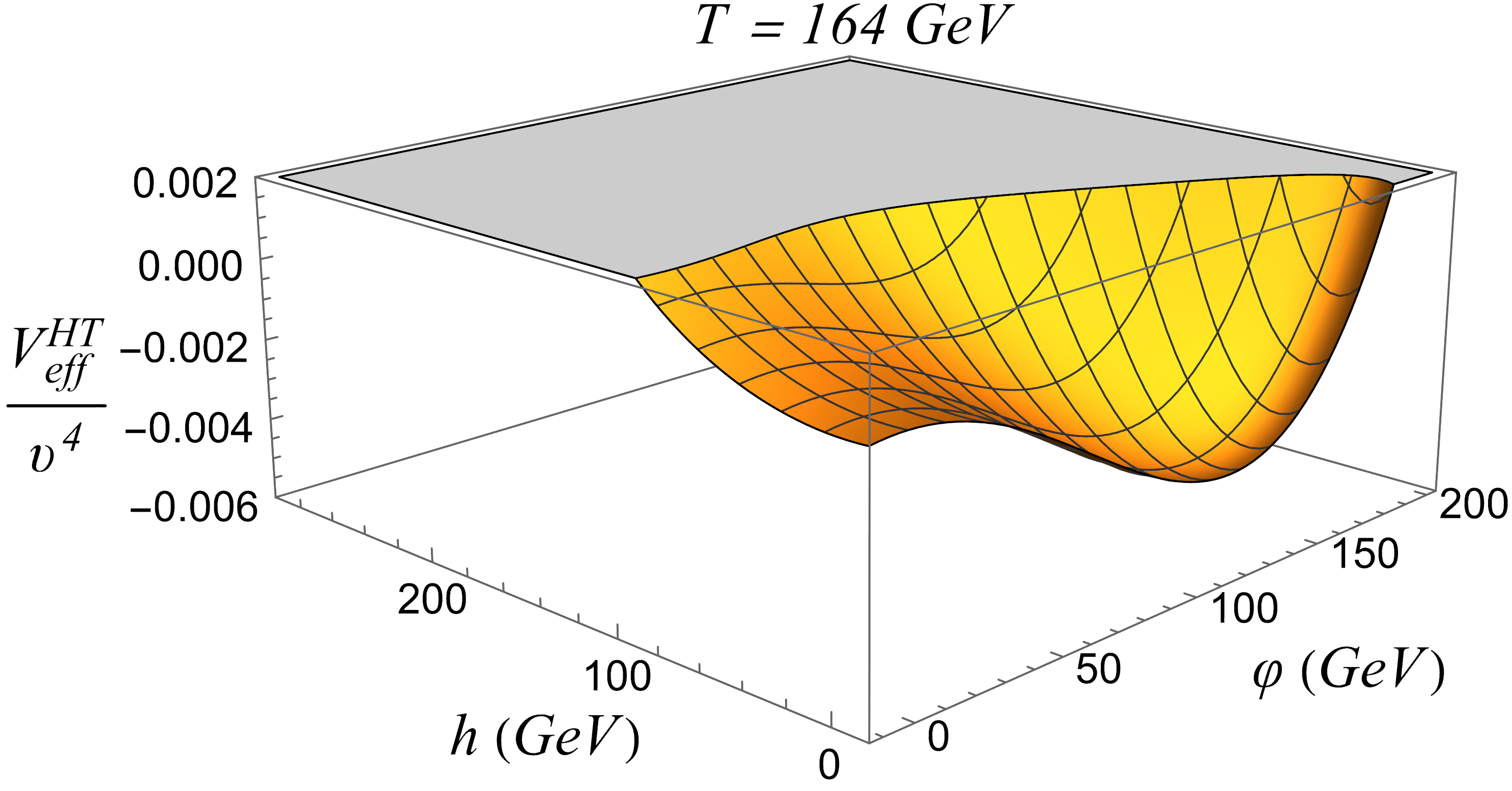}
\includegraphics[width=14pc]{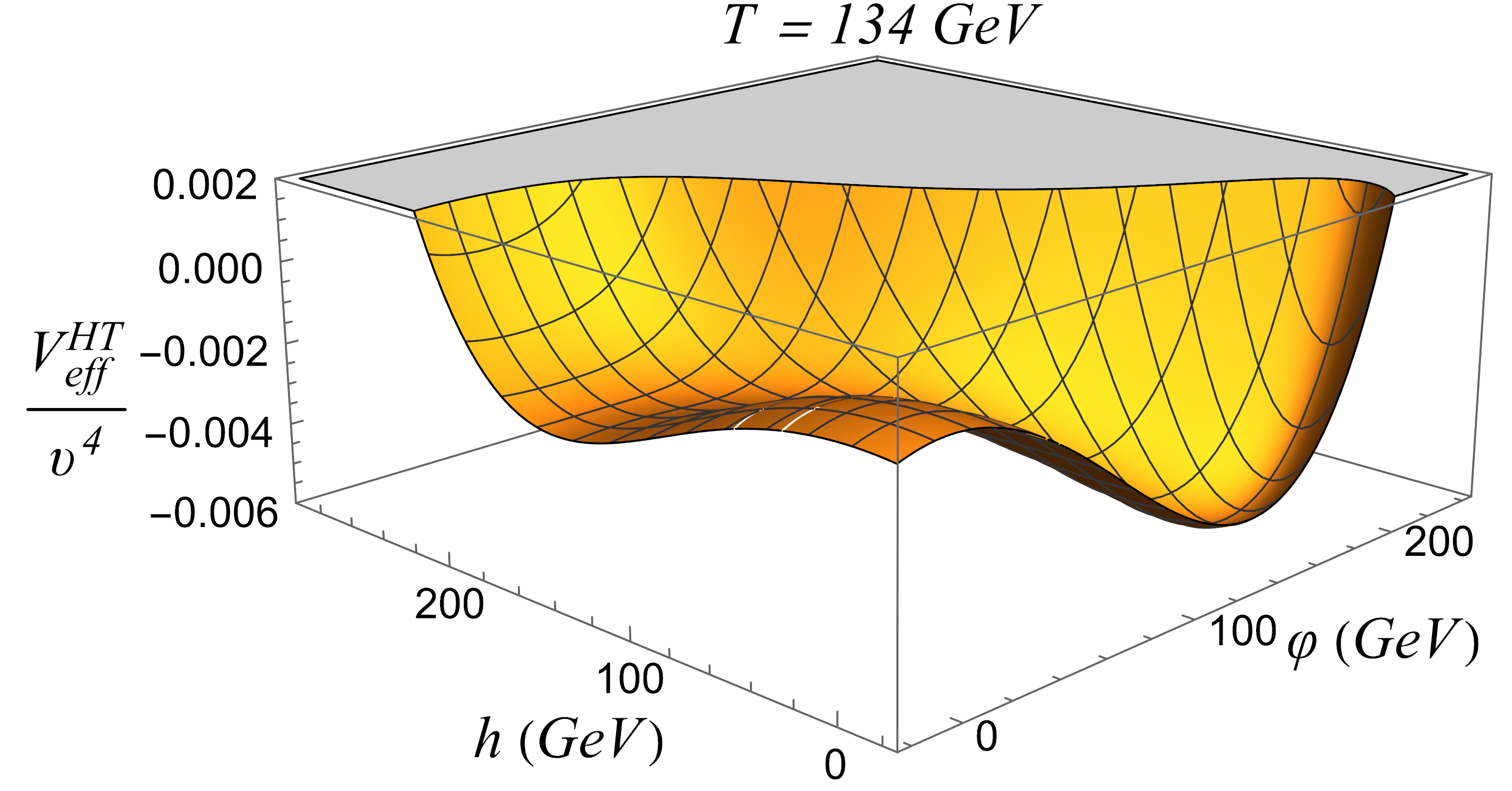}
\includegraphics[width=14pc]{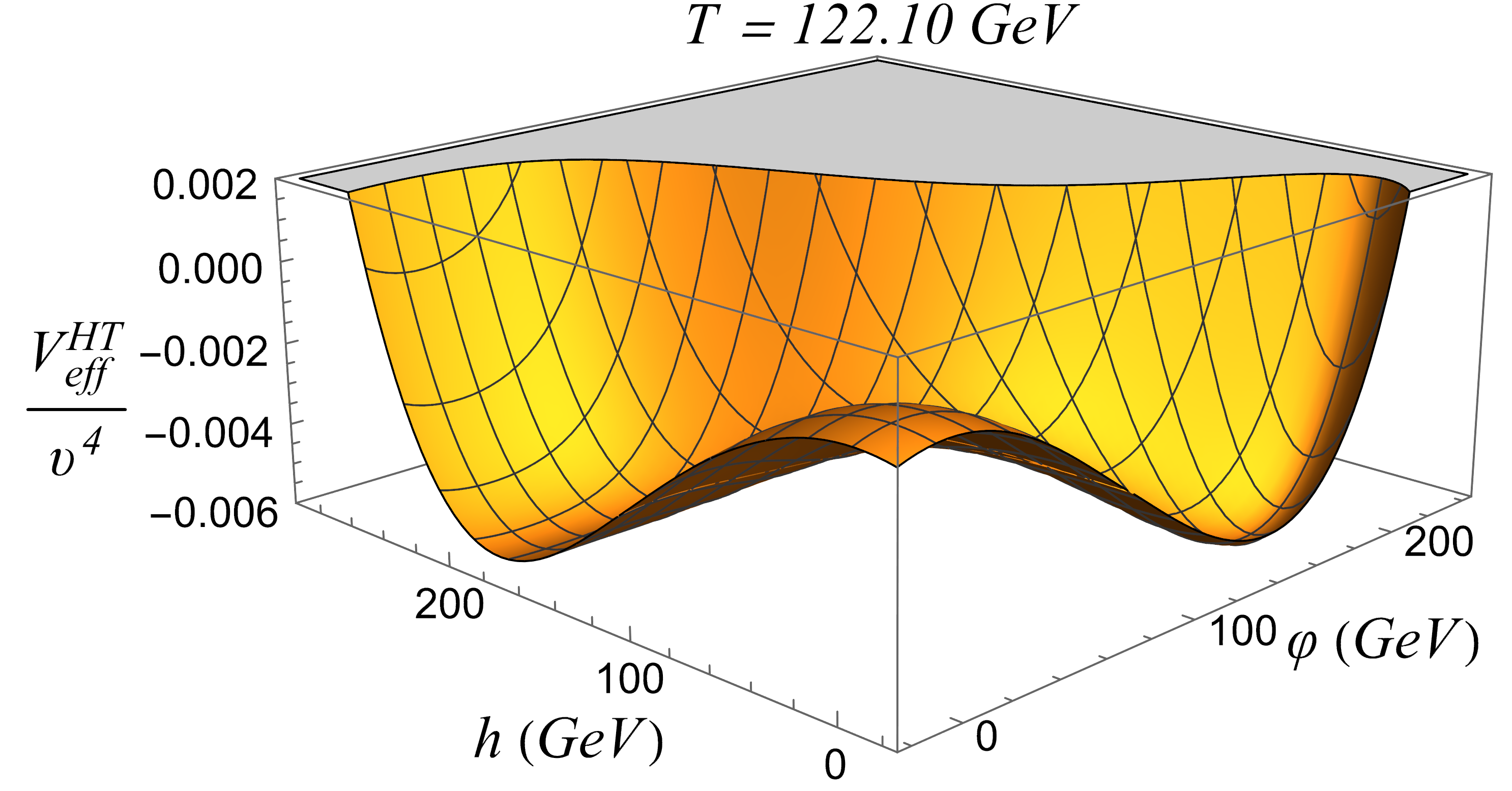}
\includegraphics[width=14pc]{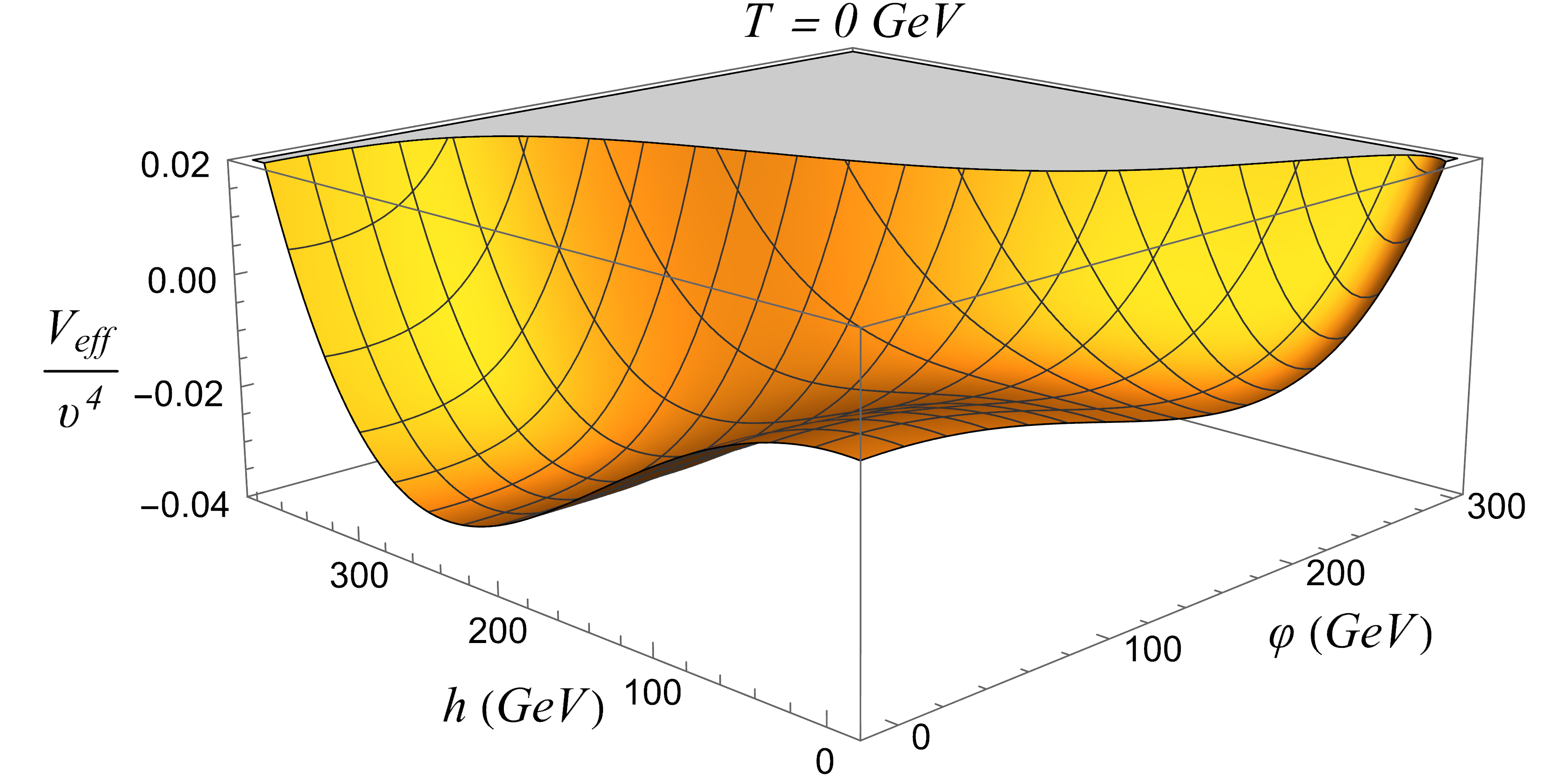}
\caption{The full effective potential during the strong
electroweak phase transition as the temperature decreases. In this
example, the singlet's phase transition is second-order using a
point of the parameter space with \(m_S = 62.5\) GeV,
\(\lambda_{HS} = 0.15\), \(\lambda/M^2 \simeq 2 \times 10^{-5}\)
GeV\(^{-2}\), and \(a = 0.1\).} \label{T-potential_1}
\end{figure}

Now we shall analyze the behavior of the singlet-extended
effective potential (\ref{HT-fullpotential}) in the full two-dimensional configuration space spanned by the Higgs and the
singlet scalar fields  \((h, \phi)\). We shall describe in some detail the behavior of the full effective potential at finite
temperature and in the next subsections we shall constrain the
free parameters of the model, in order to be phenomenologically
viable from various aspects. The results presented in Figs. \ref{T-potential_2} and \ref{T-potential_1} summarize some of the main results of this paper.

\par
In this work, the electroweak phase transition is considered as a
two-step phase transition which consists of a primary first-order
or second-order phase transition in the \(\phi\) direction from
the origin \((h, \phi) = (0, 0)\) to a non-zero VEV and a
subsequent first-order phase transition from \((0,
\upsilon^{\prime}_{s})\) to the Higgs vacuum \((\upsilon_c, 0)\).
In the early Universe, as the temperature decreases, the Higgs
minimum is formed, and a barrier is created between the Higgs and
the singlet minimum as shown in Figs. \ref{T-potential_2} and \ref{T-potential_1}. Then a strong first-order phase transition
signals the tunneling to the electroweak vacuum which remains the
global minimum at zero temperature. Instead of this cosmological
history, the electroweak symmetry breaking could alternatively
take place as a one-step strong first-order phase transition from
the origin (\(0, 0\)) to the electroweak vacuum \((\upsilon_c,
0)\) along the Higgs direction, while the singlet vacuum is
stabilized on the origin with zero VEV. This one-step scenario can
be driven by loop effects requiring \(\mu^2_S < 0\)
\cite{Chung:2012vg, Curtin:2014jma, Ghorbani:2020xqv}. A
comprehensive study on the classification of numerous models to
thermally driven, tree-level driven, and loop-driven is given in
Ref. \cite{Chung:2012vg}. Nevertheless, a one-step phase
transition can be also realized in the case of \(\mu^2_S > 0\) due
to thermal effects.

Furthermore, an instability can be developed along the singlet's
direction at extremely high temperatures for large values of the
couplings \(\lambda\), \(\lambda_{HS}\) and \(\lambda_{S}\). This
instability can also take place without the dimension-six operator
which was included in the Lagrangian, but in our model, it is
mainly caused by large \(\lambda\) and \(a\), which importantly
affect the singlet's effective potential. However, this behavior
does not alter the above discussion because the strong first-order
transition is still realized with \(T_c \ll T_s\). Moreover, we
need to stress that such an instability could lead to a high
temperature transition from the $\langle s \rangle=0$ minimum to
the $\langle s \rangle=v_{s'}$ vacuum, at a temperature way higher than the one corresponding to the electroweak phase transition.
This phase transition would generate defects, such as domain walls, and these could remain as remnants of this phase transition. 
This feature is mentionable and it occurs for the aforementioned values of the parameter, however, it relies on the fact that the temperature indeed reached such high values, which is not certain. We needed to report this because it is a peculiar and mentionable feature, but it must be treated with reluctancy since if it occurs, it occurs well before the electroweak phase transition. More importantly, from a physical point of view, perturbation theory does not break in the singlet sector, however, the ratio of the singlet mass over the corresponding temperature would be nearly zero, $m_S/T_s\sim 0$, thus normally, such high temperatures and the corresponding transitions in the effective potential should be disregarded, in the same way we disregarded the effect of small-mass quarks in the Higgs effective potential. 
We mention this issue for completeness though and in order to provide a spherical and complete view of the parameter space, for the convenience of the reader.

In addition, a common issue in this thermal history arises when
topological defects emerge due to the spontaneous symmetry
breaking of the discrete \(\mathbb{Z}_2\) symmetry, while the
singlet scalar field acquires a non-zero VEV at a high temperature
\cite{Zeldovich:1974uw}. Namely, electroweak baryogenesis can be
highly affected by the generated domain walls after the
high-temperature spontaneous symmetry breaking. Recently, it was
shown in Ref. \cite{Angelescu:2021pcd} that the inclusion of a
dimension-six singlet scalar field operator in the usual real
singlet extensions can resolve this problem, considering a scenario
in which the vacuum state never respected the \(\mathbb{Z}_2\)
symmetry. This scenario is not examined further in this study,
leaving such an analysis and the implications of topological
defects for future work. However, it is noticeable that in our
model, the \(\mathbb{Z}_2\) symmetry could be never restored at
high temperature due to the aforementioned instability developed
along the \(\phi\) direction in the effective potential. Namely,
one observes that the singlet VEV remains non-zero for large
\(\lambda\) and \(a\) at extremely high temperatures \(T > T_c\).
As a result, in these cases, the \(\mathbb{Z}_2\) symmetry is not
restored and the issue of the domain walls is avoided.

Finally, the influence of the higher order operator on the
electroweak phase transition can be comprehensively understood by
its mathematical aspects. In the Higgs direction, the non-zero
Wilson coefficient changes the thermal mass of the singlet
(\ref{T-singlet}), where \(\lambda\) is compared with the
couplings \(\lambda_{HS}\) and \(\lambda_S\), which is then
expressed in terms of singlet mass \(m_S\), the coupling
\(\lambda_{HS}\), and the parameter \(a\). This means that the
relation between their values can play an essential role for the
effect of the higher-order operator. In the \(\phi\) direction, on
the other hand, the Wilson coefficient mainly contributes to the
effective mass of the Higgs boson (\ref{effectivemasshiggs}), the
Goldstone bosons (\ref{effectivemasschi}), and the singlet
(\ref{effectivemasssinglet}), while the thermal mass of the
singlet remains the same as in the Higgs direction. As a
consequence, in the direction with \(h = 0\), the effect of the
Wilson coefficient is primarily determined by the coupling
\(\lambda_{HS}\) and the singlet mass \(m_S\). Therefore, it is
worth mentioning that the non-zero Wilson coefficient
significantly changes the effective potential
(\ref{HT-fullpotential}) in the \(\phi\) direction.

In the next subsections, the two-step strong electroweak phase transition is studied by dividing the parameter space of the singlet extension with the dimension-six operator into three regions: the low-mass region (\(m_S < m_H/2\)), the Higgs resonance region (\(m_S = m_H/2\)), and the high-mass region \((m_S > m_H/2\)).

\subsection{High-mass Region: \(m_S > m_H/2\)}

\begin{figure}[h!]
\centering
\includegraphics[width=30pc]{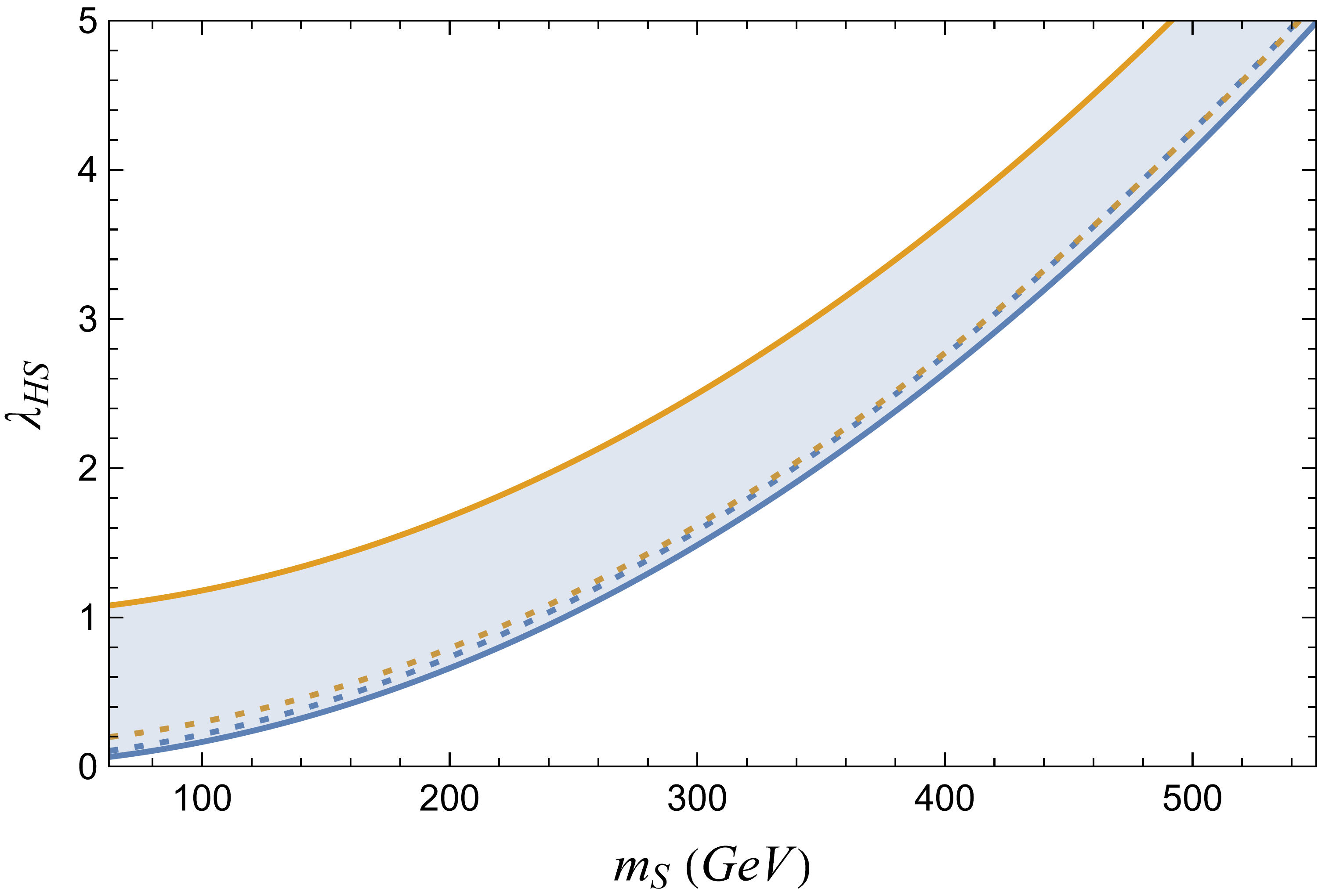}
\caption{The blue region is the parameter space of the singlet
model for a two-step electroweak phase transition with \(m_S >
m_H/2\), \(a = 0.1\) and \(\lambda = 0\). The orange dotted line
describes the constant \(\mu_S = 90\) GeV. The parameter space of
the strong two-step electroweak phase transition with
\(\upsilon_c/T_c > 1\) corresponds to the blue region which is
bounded from below by the blue dotted line. The critical
temperature in this parameter space varies from \(T_c \simeq 30 -
140\) GeV.}\label{ParameterSpace1}
\end{figure}

First of all, the two-step electroweak phase transition can be
described by the parameter space for \(m_H < 2 m_S\) in Fig.
\ref{ParameterSpace1} with zero Wilson coefficient, adopting the
common parametrization with \(a = 0.1\). As it has been previously
mentioned, this parameter space includes a number of small regions
with a one-step electroweak phase transition. In particular, it
was found that the parameter space with \(\mu_S \lesssim 90\) GeV
corresponds to a region with both scenarios. On the other hand,
imposing the criterion for a strong first-order phase transition,
reduces the generic two-step parameter space with \(\lambda = 0\).
More specifically, it is mainly changed for \(m_S > 200 \) GeV and the
low Higgs-singlet couplings as the criterion
(\ref{sphaleron_rate}) requires higher coupling\footnote{The
difference between the lowest \(\lambda_{HS}\) for satisfying the
sphaleron rate criterion with lower bound \(0.6\) and \(1\) is
negligible.} \(\lambda_{HS}\) in comparison with the generic case
of a two-step phase transition. Those higher values of the lower
bound for the coupling \(\lambda_{HS}\) vary with different
singlet masses and they are depicted by the blue dotted line in
Fig. \ref{ParameterSpace1}.
These results generally agree with similar studies
\cite{Curtin:2014jma, Beniwal:2017eik, Kurup:2017dzf,
Jain:2017sqm}. The renormalization scale dependence and other
theoretical uncertainties in the perturbative analysis (which are
discussed in Refs. \cite{Chiang:2018gsn,Athron:2022jyi,
Croon:2020cgk, Gould:2021oba}) may introduce small differences
between our results and the literature. Moreover, the authors of
Ref. \cite{Kurup:2017dzf} claimed that larger parts of the
two-step electroweak phase transition parameter space are ruled
out since relativistic speeds during the phase transition must not
be reached by the expanding bubble walls and it is required that
bubbles do nucleate at finite temperature. Nonetheless, we do not
consider these effects in the present study, leaving such
improvement for future work.

\begin{figure}[h!]
\centering
\includegraphics[width=30pc]{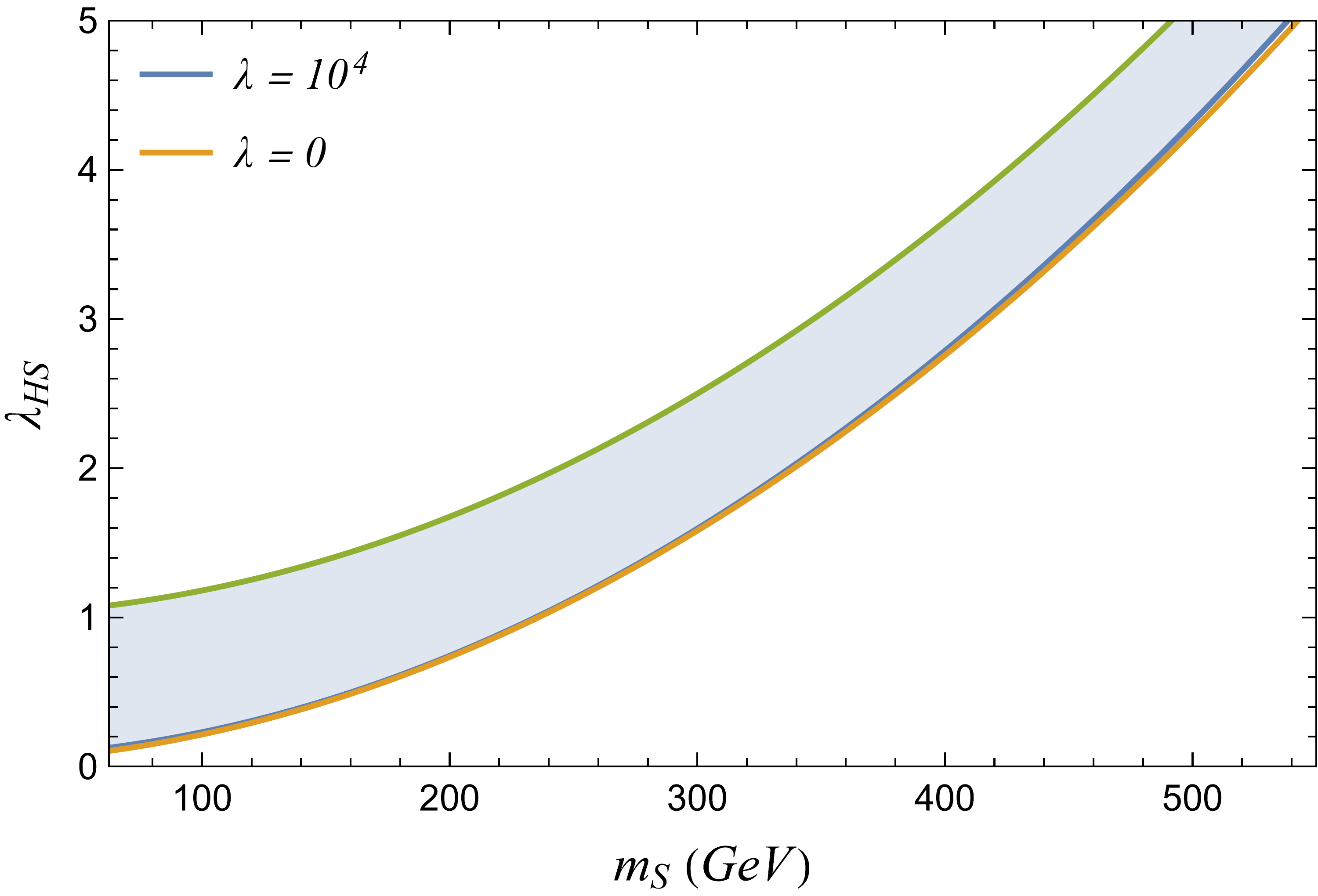}
\caption{The parameter space (blue region) of the singlet
extension with a dimension-six operator (\(\lambda = 10^4\)) in
order to realize a strong two-step electroweak phase transition
(\(\upsilon_c/T_c > 1\)). The orange line shows the lower bound of
the parameter space of the singlet extension which is represented
by a blue dotted line in Fig. \ref{ParameterSpace1}. In both
cases, we take \(a = 0.1\).}\label{ParameterSpace2}
\end{figure}

The presence of the higher order operator with \(\lambda <
10^{2}\) does not noticeably affect the aforementioned parameter
space of the strong electroweak phase transition. The lowest
coupling \(\lambda_{HS}\) for a strong phase transition, slightly
increases with \(\lambda \lesssim 10^{3}\) and the critical
temperature is significantly increased, whereas the ratio
\(\upsilon_c/T_c\) shows a downward trend. This effect is
strengthened by very high and very low singlet masses. On the
other hand, the maximum value of the coupling \(\lambda_{HS}\),
which is shown in Fig. \ref{ParameterSpace1}, is not changed by
the higher order operator. In this case, the effect of the
non-zero Wilson coefficient is highly weakened by lower singlet
masses. The comparison between the parameter space of the singlet
extension with \(\lambda = 0\) and \(\lambda = 10^4\) is presented
in Fig. \ref{ParameterSpace2}. It is clearly seen that the
influence of the dimension-six operator is not crucial for \(a =
0.1\). On the contrary, this behavior dramatically alters for
higher values of the parameter \(a\) and it is extensively studied
for lower singlet masses.

\par The behavior of the full effective potential at finite temperature
is depicted in Fig. \ref{T-potential_2} for the case at hand, by
taking into account all the phenomenological constraints obtained
in this subsection, and for various temperatures. In this example,
the singlet's phase transition is first-order using a point of the
parameter space with \(m_S = 500\) GeV, \(\lambda_{HS} = 4.3\),
\(\lambda/M^2 \simeq 2 \times 10^{-5}\) GeV\(^{-2}\) and \(a =
0.1\), but it can also be a second-order phase transition for
lower singlet masses and Higgs-singlet couplings. As it can be
seen, at high temperatures, the effective potential is symmetric,
and as the temperature decreases, the singlet vacuum with a
barrier between the second vacuum and the origin is developed. As the temperature decreases, the Higgs vacuum is generated with another barrier between the latter and the origin also existing. Accordingly, the Higgs vacuum is deeper than the singlet vacuum and the electroweak phase transition occurs.

\subsection{Higgs Resonance Region: \(m_S = m_H/2\)}

The Higgs resonance region is highly motivated by dark matter
physics  since the singlet should have a mass near \(m_S = 62.5\)
GeV as a viable dark matter candidate
\cite{GAMBIT:2017gge,Athron:2018ipf,Feng:2014vea} and it is
strongly restricted by the direct dark matter searches by LUX,
XENON1T, and XENONnT experiments \cite{Cline:2013gha, Cline:2012hg,
GAMBIT:2017gge,Athron:2018ipf,Feng:2014vea, Beniwal:2017eik}.
While in the majority of the parameter space without the
dimension-six operator, the assumption of the real singlet as a
dark matter candidate is inconsistent with a strong first-order
electroweak phase transition, it has been shown that in the Higgs
resonance region, the dark matter relic density can constitute an
important proportion of the total observed dark matter density
\cite{GAMBIT:2017gge,Athron:2018ipf,Feng:2014vea}.

Additionally, the full effective potential at finite temperature
is illustrated in Fig. \ref{T-potential_1} using a point of the
parameter space with \(\lambda_{HS} = 0.15\), \(\lambda/M^2 \simeq
2 \times 10^{-5}\) GeV\(^{-2}\), and \(a = 0.1\) and a similar
behavior is followed in the total parameter space with \(m_S =
m_H/2\). The singlet's phase transition is generally second-order
in this region in which the two-step electroweak phase transition
satisfies the sphaleron rate criterion.

\begin{figure}[h]
\centering
\includegraphics[width=30pc]{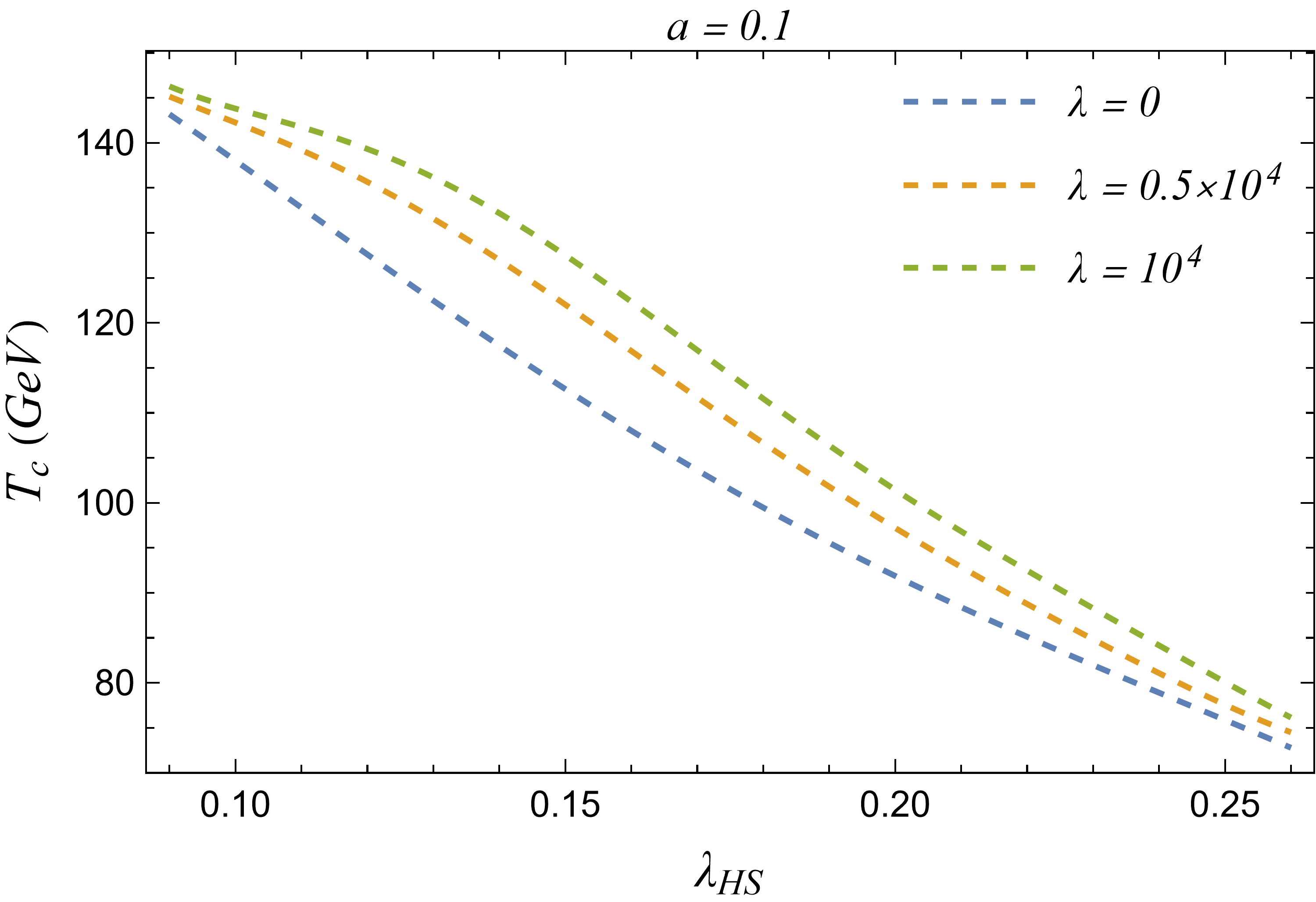}
\caption{The critical temperature (\(T_c\)) as a function of the
Higgs-singlet coupling (\(\lambda_{HS}\)) for $m_S = 62.5 $ GeV
and \(a = 0.1\).}\label{Tc_ms_625_a_01}
\end{figure}

\begin{figure}[h]
\centering
\includegraphics[width=20.5pc]{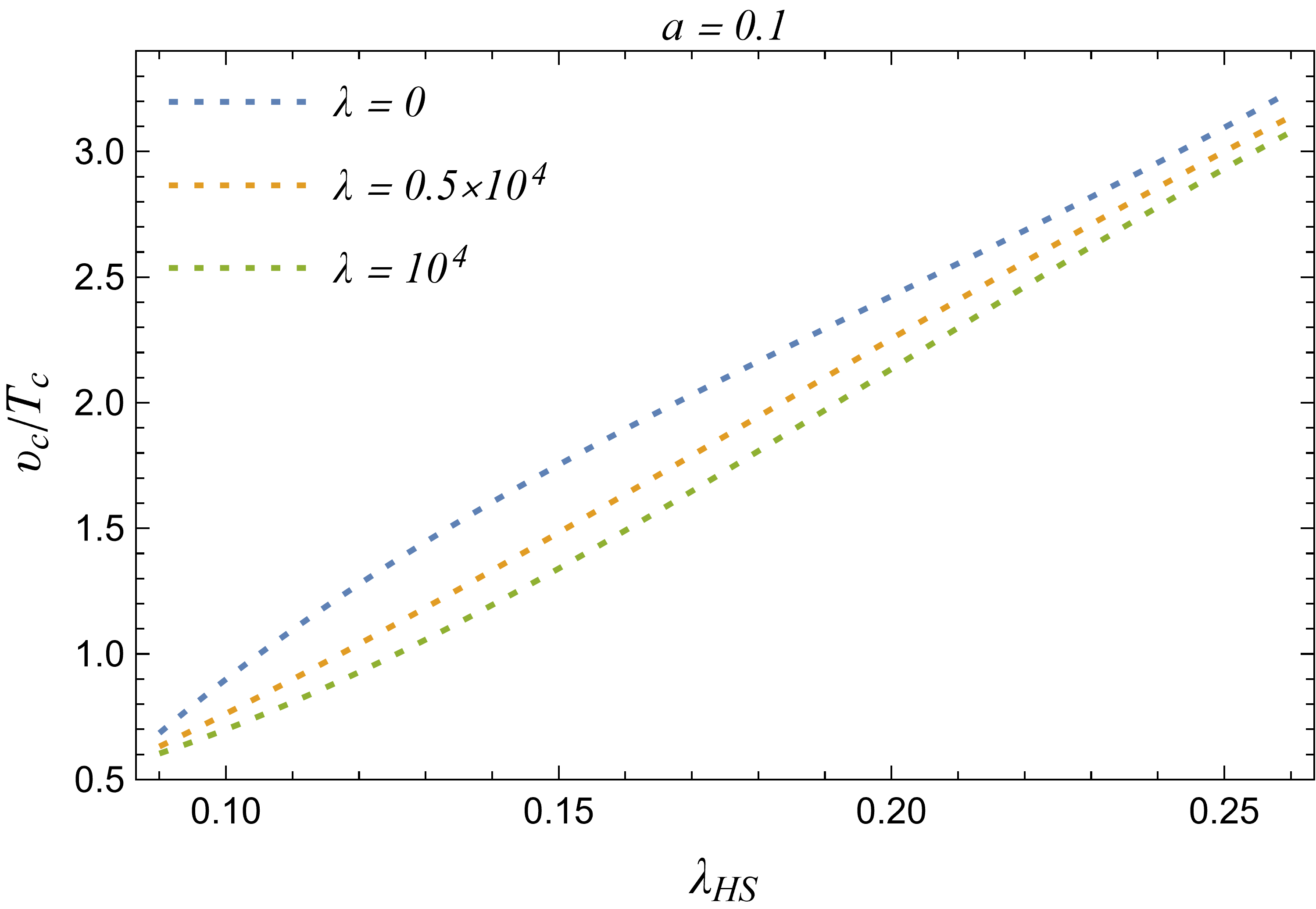}
\includegraphics[width=20.5pc]{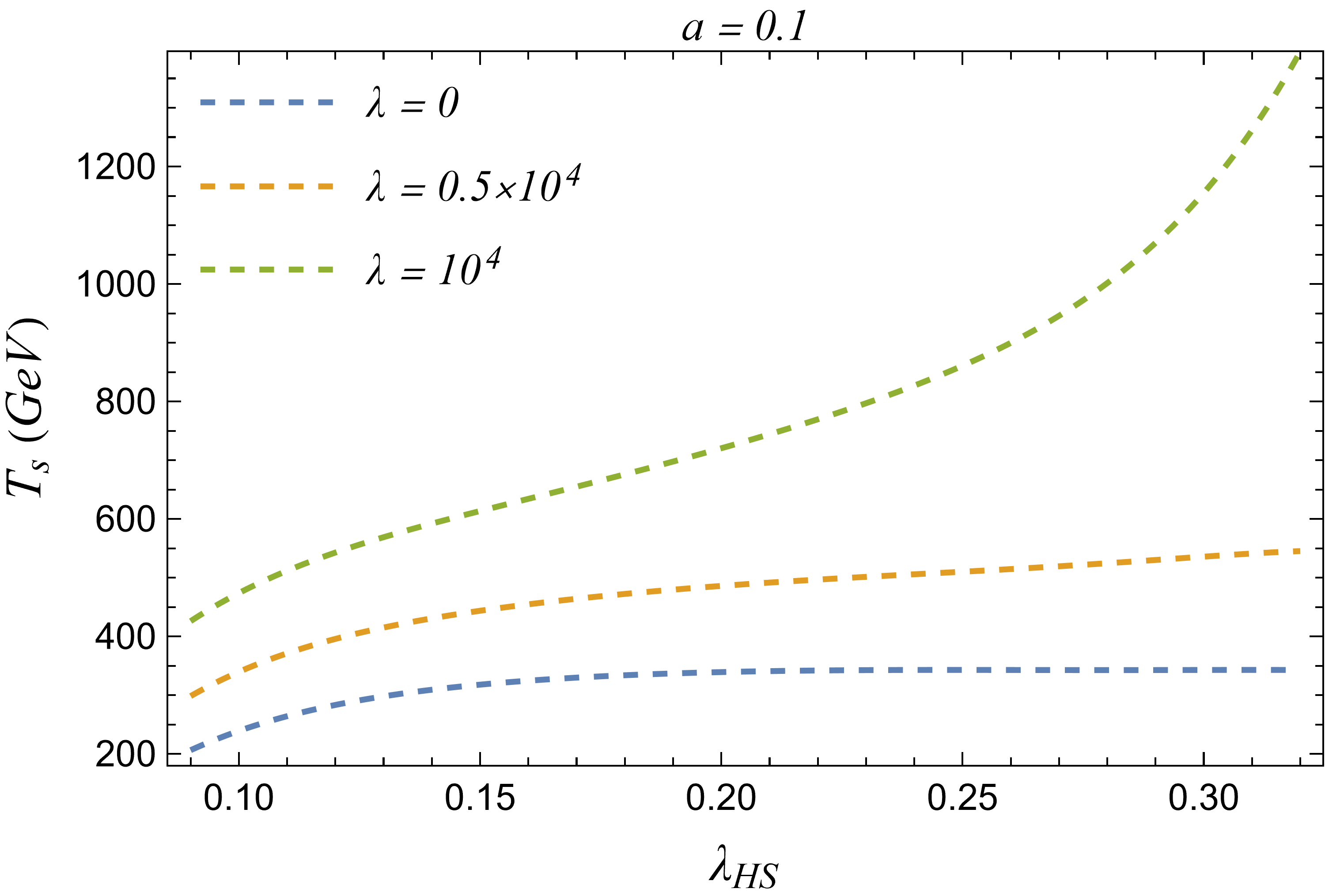}
\caption{\textbf{Left}: The sphaleron rate criterion as a function
of the coupling \(\lambda_{HS}\) for $m_S = 62.5 $ GeV and \(a =
0.1\) in the case of zero and non-zero Wilson coefficient.
\textbf{Right}: The critical temperature of the singlet's
second-order phase transition \(T_s\) as a function of the
coupling \(\lambda_{HS}\) for $m_S = 62.5 $ GeV and \(a =
0.1\).}\label{sr_Ts_ms_625_a_01}
\end{figure}

Firstly, the common value of \(a = 0.1\) is selected in order to
compute the critical temperature \(T_c\), the critical temperature
\(T_s\) of the singlet's phase transition and the sphaleron rate
criterion for various values of the Wilson coefficient and the
Higgs-singlet coupling which are presented in Figs. \ref{Tc_ms_625_a_01} and \ref{sr_Ts_ms_625_a_01}. The electroweak
phase transition is considerably affected by the higher order
operator for \(\lambda > 10^{3}\) and the parameter space is
restricted, as the ratio \(\upsilon_c/T_c\) drops for a non-zero
Wilson coefficient, and a higher \(\lambda_{HS}\) is necessary to
generate the strong electroweak phase transition. Moreover, the
phase transition in the \(\phi\) direction occurs at a very high
temperature for \(\lambda/M^2 \simeq 4 \times 10^{-5}\)
GeV\(^{-2}\), which shows the strong influence of the higher order
operator on the singlet's dynamics.

\par
\begin{figure}[h]
\centering
\includegraphics[width=20.5pc]{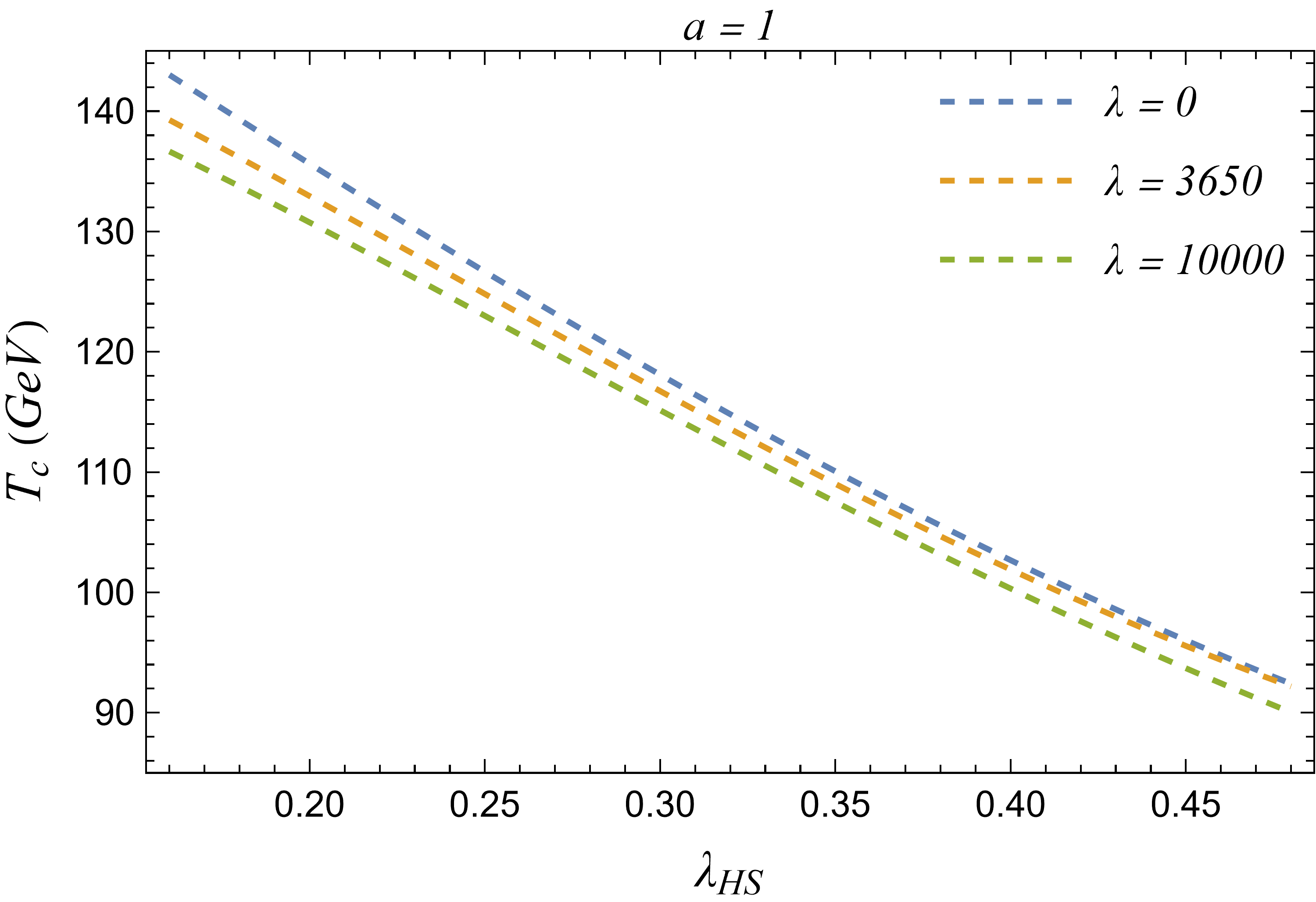}
\includegraphics[width=20.5pc]{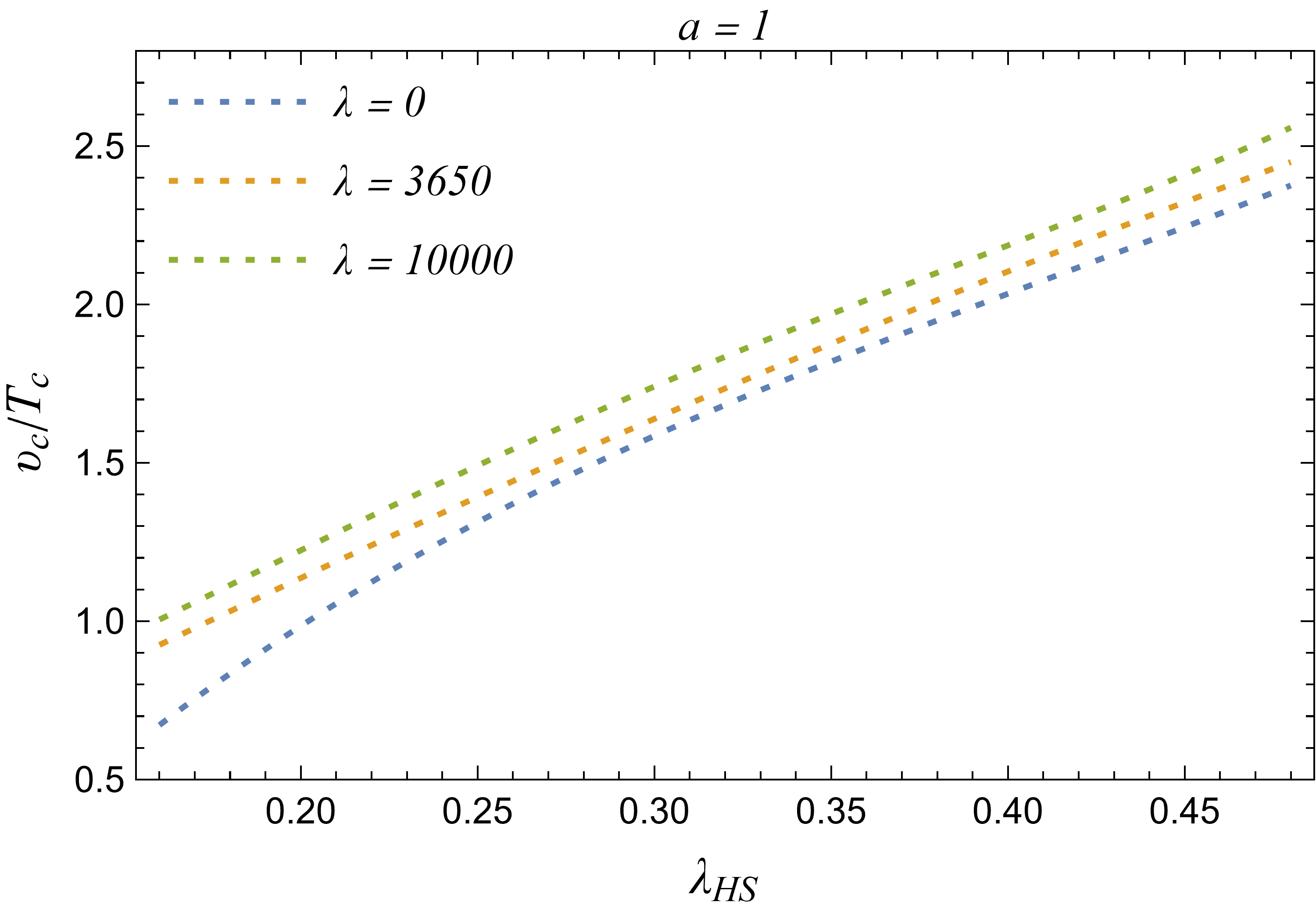}
\caption{The electroweak phase transition with $m_S = 62.5 $ GeV
and \(a = 1\): \textbf{Left}: The critical temperature (\(T_c\))
as a function of the Higgs-singlet coupling (\(\lambda_{HS}\)).
\textbf{Right}: The sphaleron rate criterion as a function of the
coupling \(\lambda_{HS}\) in the case of zero and non-zero Wilson
coefficient.}\label{Tc_sr_ms_625_a_1}
\end{figure}

 On the other hand, the parameter space of the strong electroweak
phase transition is expanded by taking \(a > 0.4\).  This is an
important feature of our model which we need to further highlight
at this point. This is clearly illustrated in the case of \(a =
1\) in Fig. \ref{Tc_sr_ms_625_a_1}, where the criterion
\(\upsilon_c/T_c > 1\) can be satisfied by much lower
\(\lambda_{HS}\) compared to the model with zero Wilson
coefficient. As a result, if \(\lambda/M^2 \gtrsim 10^{-4}\)
GeV\(^{-2}\), a strong electroweak phase transition can be
realized for every low Higgs-singlet coupling with \(\mu^2_S >
0\). Hence, the higher order operator could assist a strong
electroweak phase transition in regions of the parameter space
which were excluded in the previous singlet extensions of the SM.
\par
\begin{figure}[h!]
\centering
\includegraphics[width=30pc]{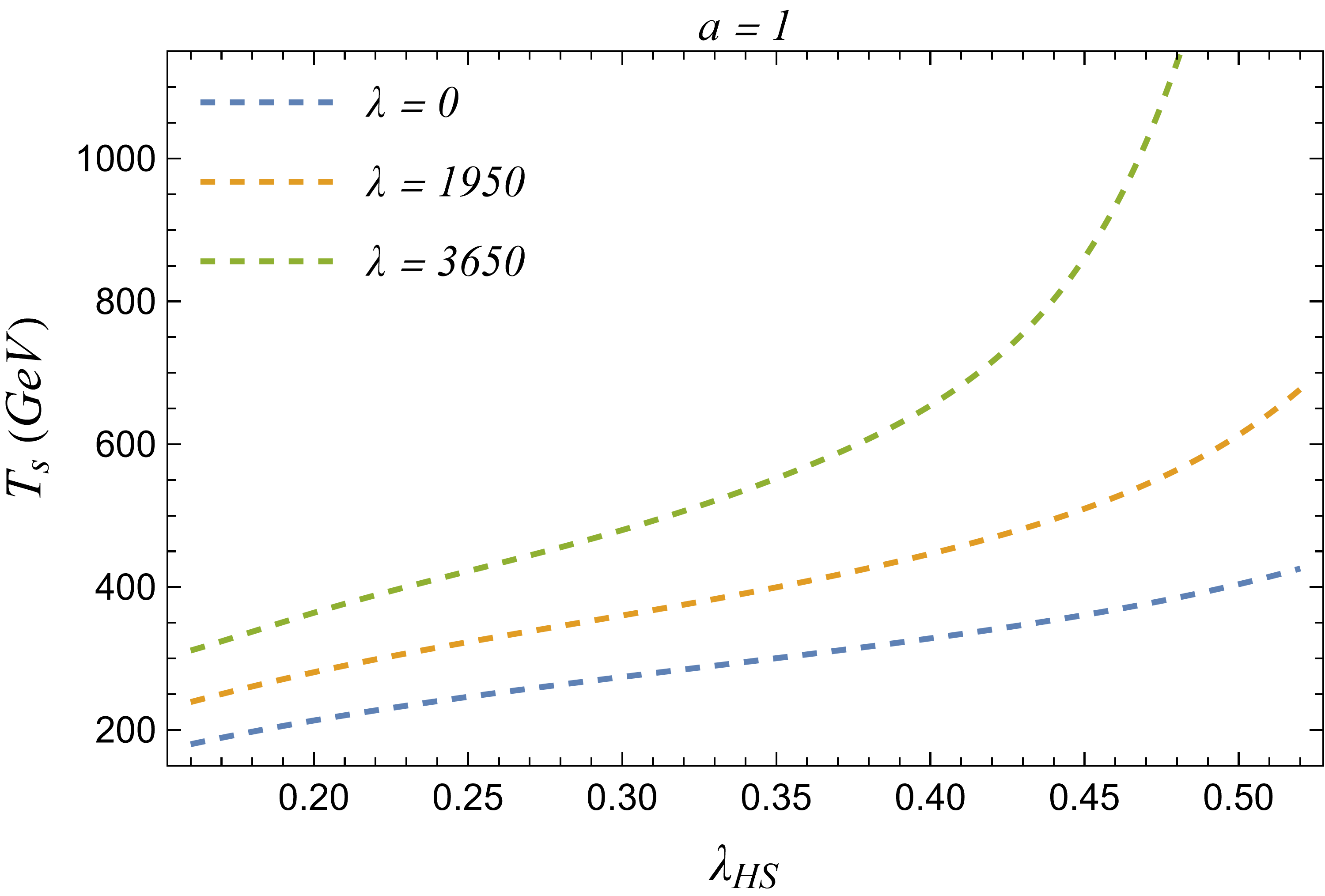}
\caption{The critical temperature of the singlet's second-order
phase transition \(T_s\) as a function of the coupling
\(\lambda_{HS}\) for $m_S = 62.5 $ GeV and \(a = 1\) in the
case of zero and non-zero Wilson
coefficient.}\label{Ts_ms_625_a_1}
\end{figure}

 Regarding the singlet's phase transition, the trend of the
critical temperature \(T_s\) (Fig. \ref{Ts_ms_625_a_1}) is similar to the case of \(a < 0.4\)
and \(T_s\) sharply increases as the Wilson coefficient
increases. In addition, the previously mentioned instability in
the \(\phi\) direction takes place for \(\lambda > 3650\) and
\(\lambda_{HS} > 0.52\) with \(a = 1\), which is also showcased by
the fact that \(\lambda_{HS} = 0.52\) leads to \(T_s \simeq 8768\)
GeV. It is noticeable that lower values of \(a\) lead to this
instability for higher values of the Wilson coefficient and the
Higgs-singlet coupling.
\par In Figs. \ref{Tc_ms_625_a_01} and \ref{Tc_sr_ms_625_a_1}, the
critical temperature with \(\lambda \ne 0\) approaches the
critical temperature with \(\lambda = 0\) for larger values of the
Higgs-singlet coupling. This behavior shows that the effect of the
higher order operator is stronger for low values of
\(\lambda_{HS}\). This is also apparent for the ratio in the
sphaleron rate criterion in Figs. \ref{sr_Ts_ms_625_a_01} and
\ref{Tc_sr_ms_625_a_1}. However, in the case of \(a = 0.1\), the
influence of the higher order operator is maximized for an
intermediate low value of the Higgs-singlet coupling and not for
the lower values.

Finally, it is essential to mention that if \(a <
0.1\),  the Higgs VEV \(\upsilon\) at zero temperature is not much
deeper than the singlet VEV \(\upsilon_s\). Namely, the condition
(\ref{global_min}) is satisfied for every \(a\) in Eq.
(\ref{lambda_definition}), but low values of \(a\) do not satisfy
the condition,
\begin{equation}
    V_0 (0, \upsilon_s) \gg V_0 (\upsilon, 0),
\end{equation}
which ensures that the Higgs vacuum is by far more energetically favorable compared to the singlet vacuum.

\subsection{Low-mass Region: \(m_S < m_H/2\)}

In the low mass region, the behavior of the full effective
potential remains the same as in the Higgs resonance region, while
the high-temperature phase transition in the \(\phi\) direction is
primarily second-order in the allowed parameter space.

\begin{figure}
\centering
\includegraphics[width=30pc]{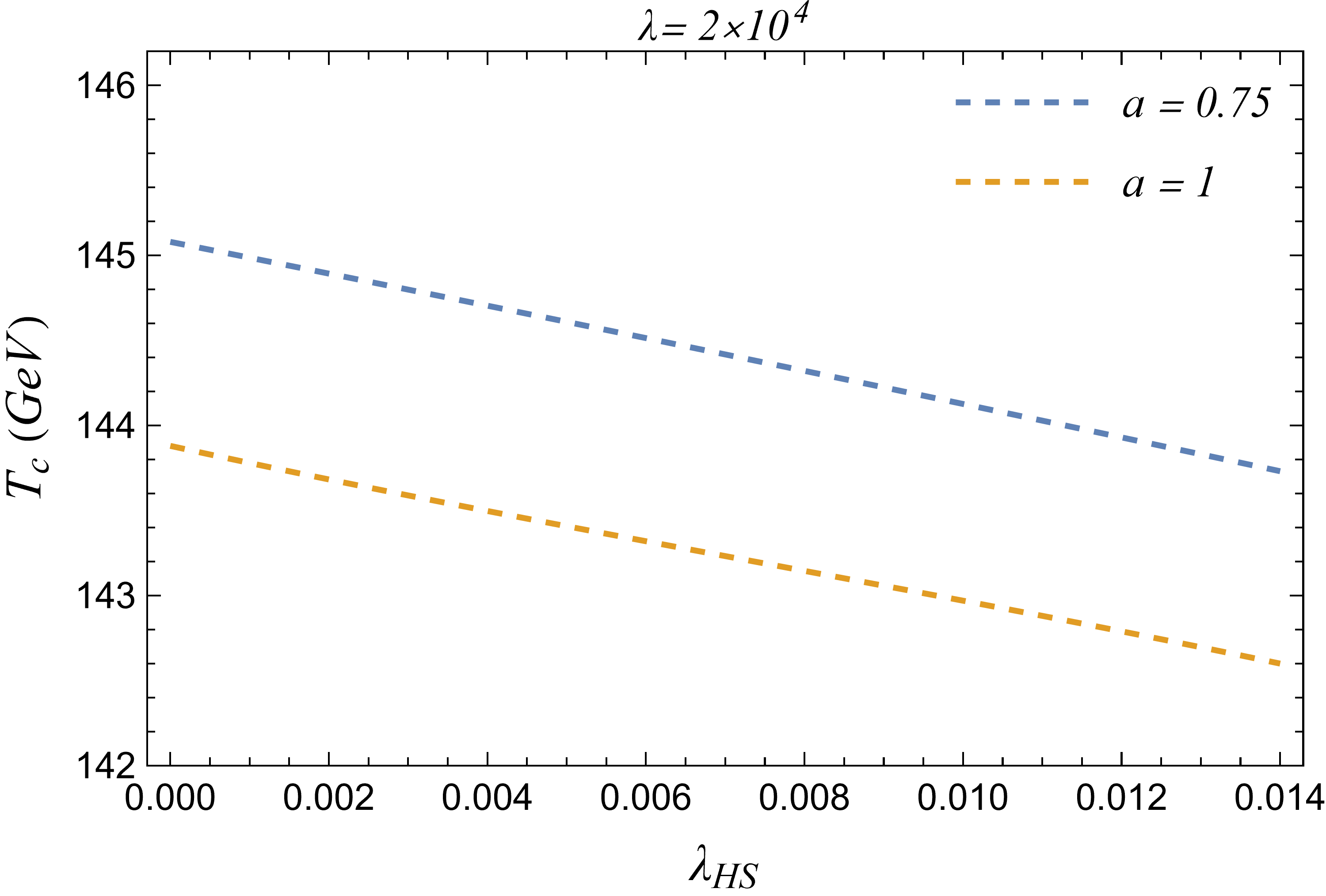}
\caption{The critical temperature as a function of the Higgs-singlet coupling for \(a = 0.75\) and \(1\) and \(m_S \leq 0.1\) GeV.}\label{T_c_a_1_0.75}
\end{figure}

It has been already shown that the parameter space for \(m_S <
m_H/2\) is severely eliminated by the invisible decay width of the
Higgs boson and the condition \(\mu^2_S > 0\). It is initially
assumed that the branching ratio of the Higgs boson to invisible
particles is set to \(BR_{inv} < 0.19\). However, the lower mass
region \(m_S \lesssim 30\) GeV, which is allowed by these constraints, is completely
excluded by the sphaleron rate criterion considering \(\lambda =
0\) and \(a \gtrsim 0.05 \) due to the low values of the
Higgs-singlet coupling. On the contrary, large values of \(\lambda
> 10^3\) and \(a \gtrsim 0.05 \) can easily assist the strong
phase transition for \(m_S \leq 1 \) GeV. This is clearly
illustrated if we take \(\lambda = 2 \times 10^4\) and \(a =
0.75\) as a strong electroweak phase transition (\(\upsilon_c/T_c
> 0.6\)) occurs for every coupling \(\lambda_{HS}\) and \(m_S <
10\) GeV. This is shown in Fig. \ref{T_c_a_1_0.75}. More
specifically, the lower bound of \(\lambda_{HS}\) with a singlet
mass \(m_S = 0.1\) GeV and \(\lambda = 2 \times 10^4\) can
generate a strong electroweak phase transition and it is at least
\(10^6\) times smaller than the corresponding value in the singlet
extension without the higher order operator. Therefore, the
presence of the higher order operator allows very low interaction
couplings which easily avoid the constraint imposed by the
branching ratio \(BR_{inv} < 0.10\).
\par
\begin{figure}
\centering
\includegraphics[width=30pc]{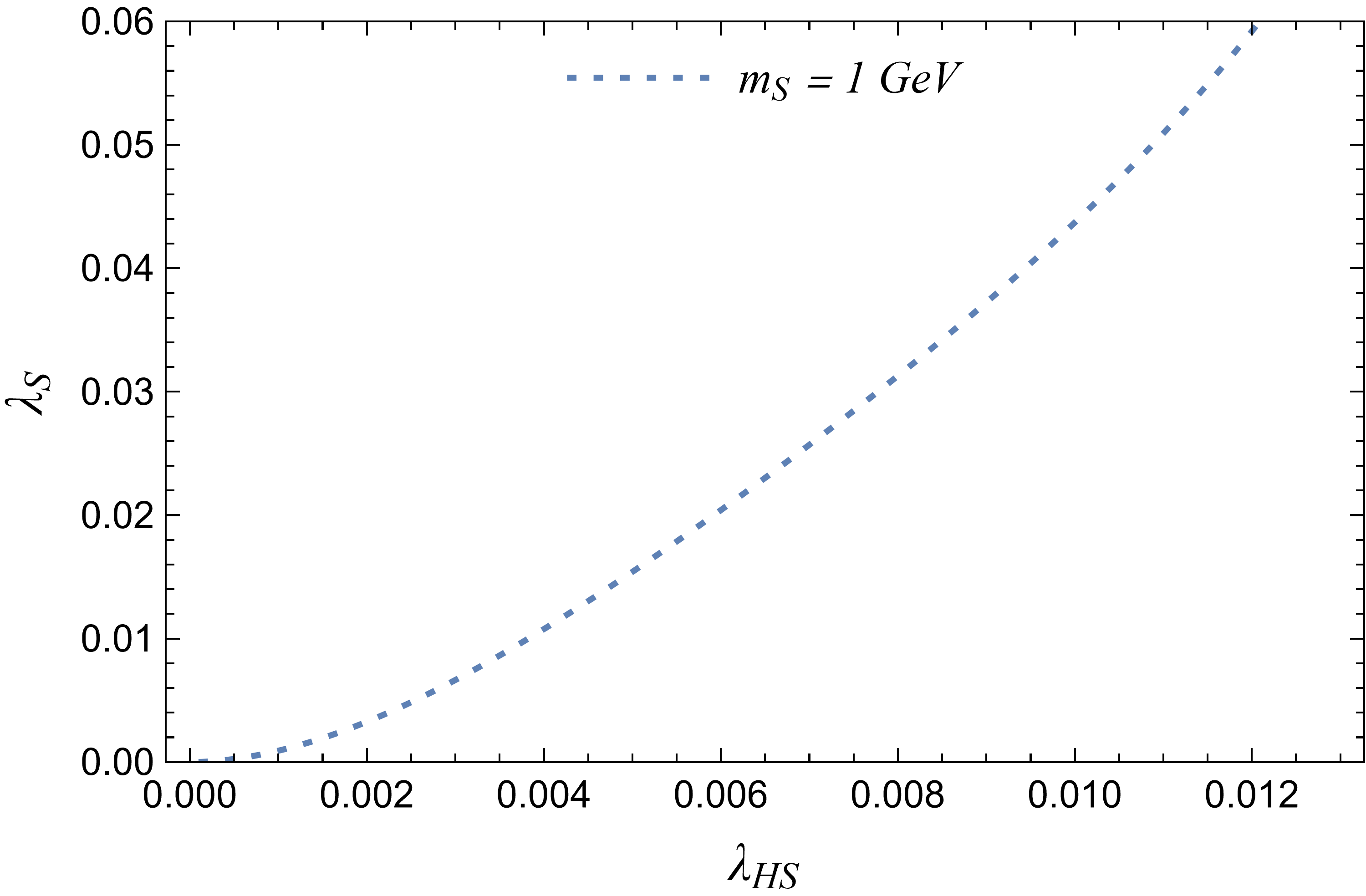}
\caption{The dependence of the lowest \(\lambda_{HS}\), which is
allowed by all the previous constraints imposing the criterion
\(\upsilon_c/T_c > 0.6\), on the parameter \(a\) for $m_S = 1 $ GeV
and \(\lambda = 0\). This dependence was numerically computed to
explain qualitatively that the parameter space expands its lower
bound for low values of the quartic coupling
\(\lambda_S\).}\label{lambda_s_k_lowest_ms_1_GeV}
\end{figure}

 In contrast, the previous behavior is changed for
values \(a \lesssim 0.05 \), as the zero Wilson coefficient can
lead to a strong first-order phase transition, whereas the
presence of the higher order operator weakens the electroweak
phase transition. The electroweak phase transition with zero
Wilson coefficient mainly occurs for \(m_S < 10 \) GeV in order to
have a two-step phase transition, but this upper limit on the mass
increases for very low values of the parameter \(a\) and
\(\lambda_{HS}\). This behavior is also illustrated in Fig.
\ref{lambda_s_k_lowest_ms_1_GeV} since it shows that the lowest
allowed value of \(\lambda_{HS}\) decreases as the parameter \(a\)
drops for a given singlet mass.
\par
\begin{figure}
\centering
\includegraphics[width=30pc]{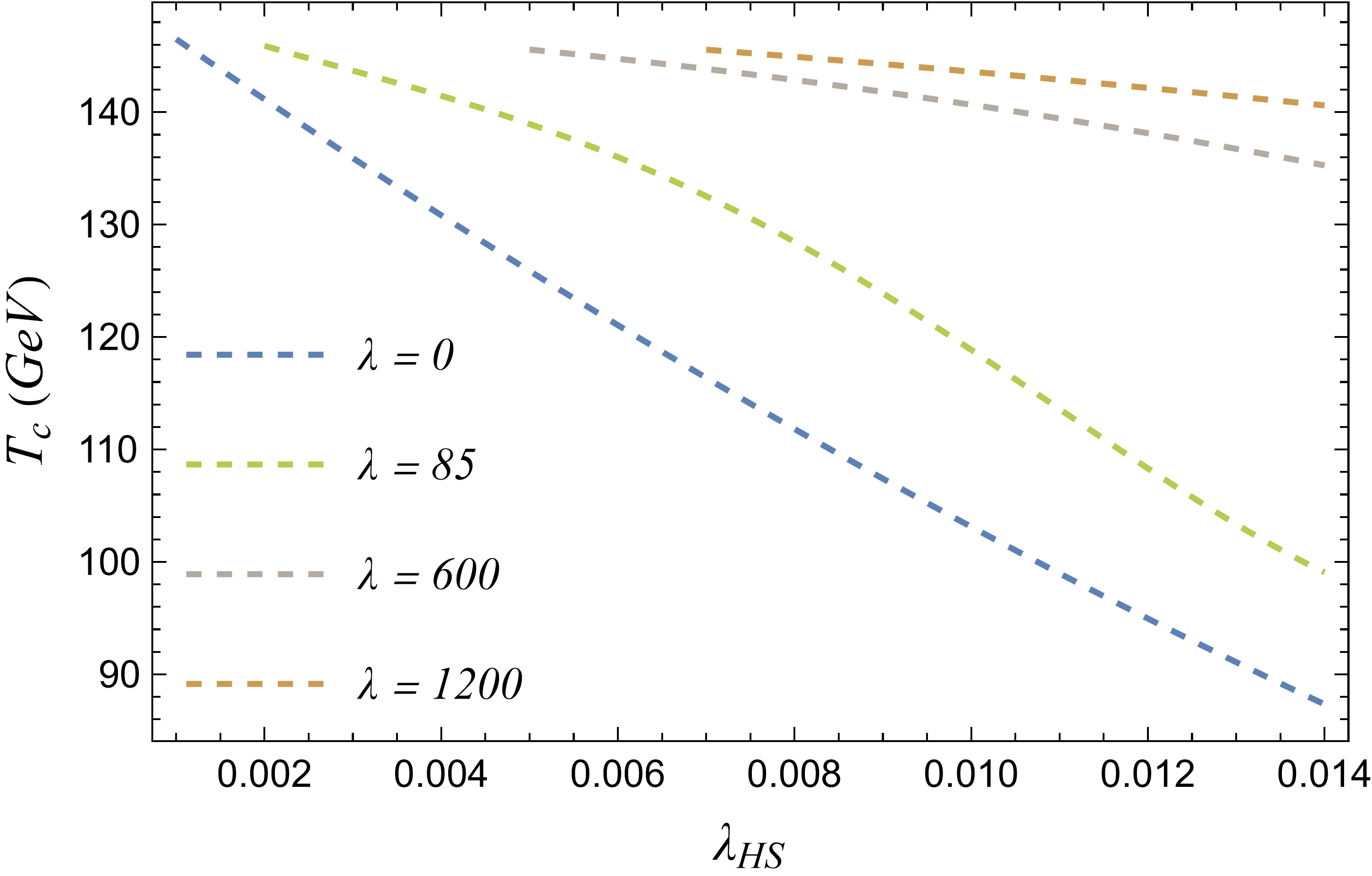}
\caption{The critical temperature as a function of the coupling
\(\lambda_{HS}\) for $m_S \leq 1 $GeV and \(a = 0.001\) in the
case of zero and non-zero Wilson coefficient \(\lambda = 0\)
(blue), \(85\) (green), \(600\) (gray), \(1200\)
(brown).}\label{critical_temperature_ms_0.1_a_0.001}
\end{figure}

\begin{figure}
\centering
\includegraphics[width=20.5pc]{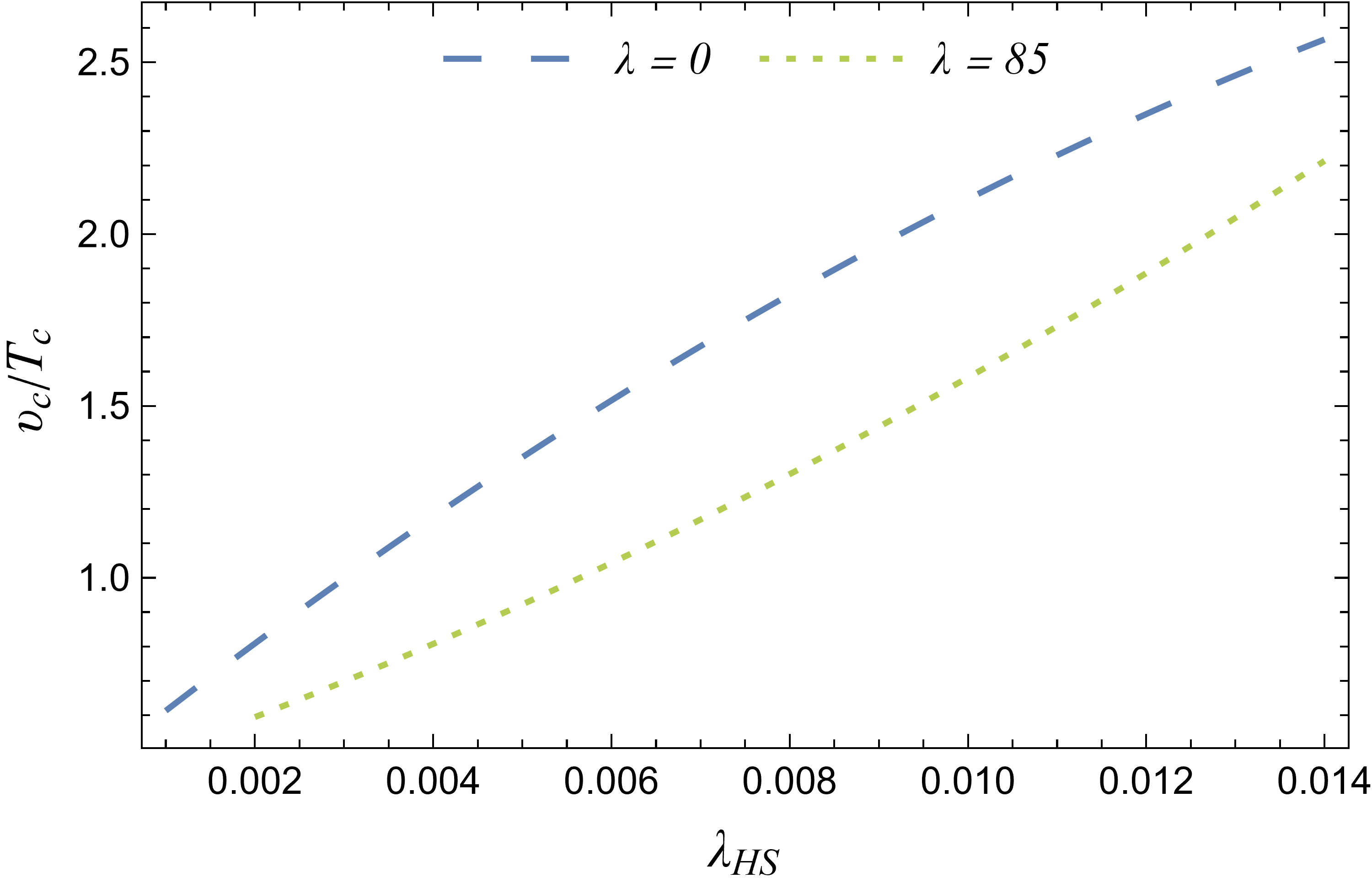}
\includegraphics[width=20.5pc]{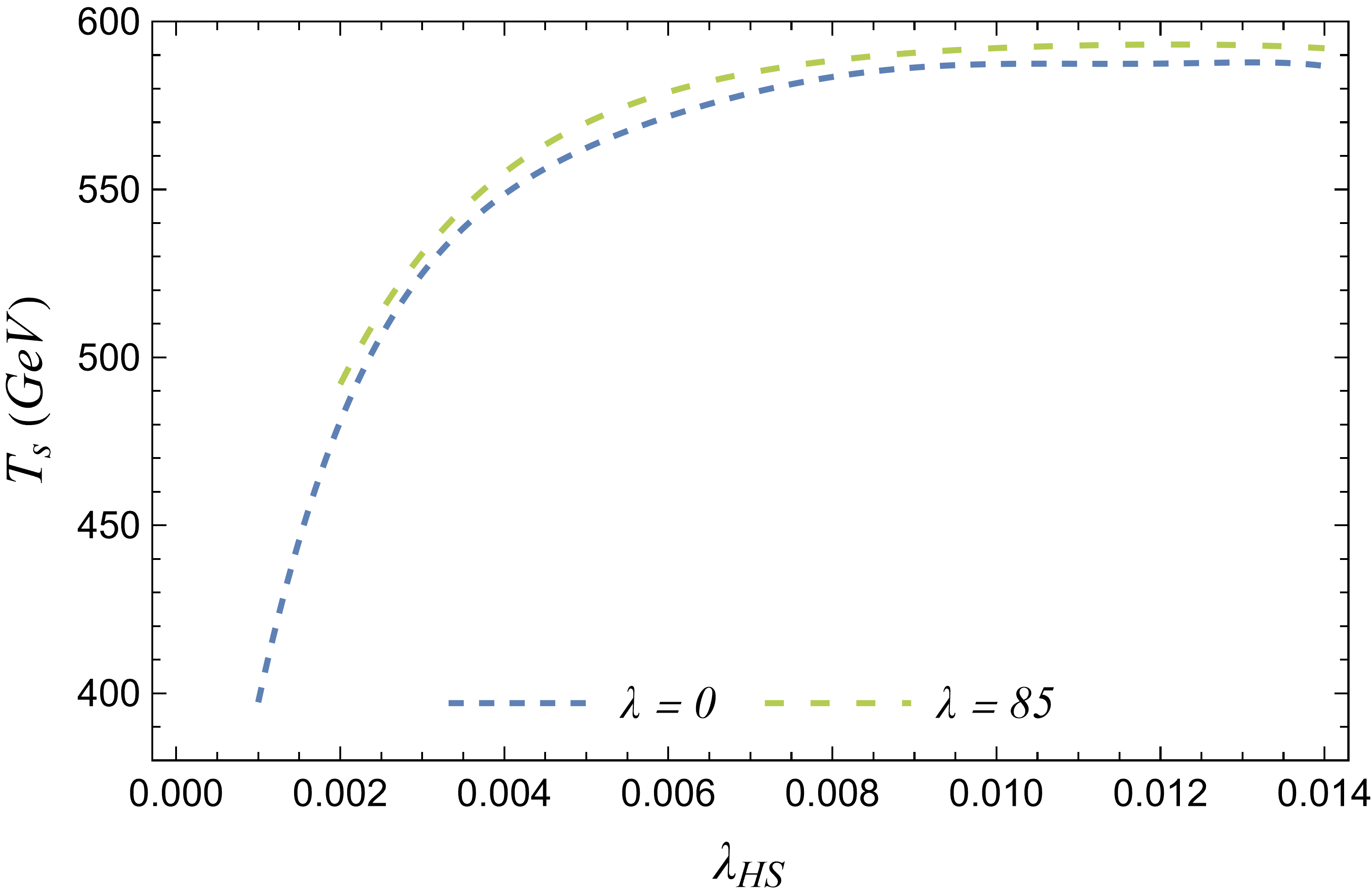}
\caption{\textbf{Left}: The sphaleron rate criterion as a function
of the coupling \(\lambda_{HS}\) for $m_S \leq 1 $GeV and \(a =
0.001\) in the case of zero and non-zero Wilson coefficients.
\textbf{Right}: The critical temperature of the singlet's
second-order phase transition \(T_s\) as a function of the
coupling \(\lambda_{HS}\) for $m_S \leq 0.1 $GeV and \(a = 0.001\)
in the case of zero and non-zero Wilson
coefficient.}\label{sr_Ts_ms_0.1_a_0.001}
\end{figure}

 Some typical results are presented for \(m_S \leq 1 \) GeV and \(a
= 0.001\) in Figs. \ref{critical_temperature_ms_0.1_a_0.001} and
\ref{sr_Ts_ms_0.1_a_0.001}. Firstly, it can be clearly seen that
the lower bound of \(\lambda_{HS}\) is strongly constrained by a
non-zero total Wilson coefficient \(\lambda/M^2 \simeq 10^{-7} -
10^{-5}\) GeV\(^{-2}\). In addition, the full effective potential around the critical temperature \(T_c\) remains the same for masses between \(0 - 1\) GeV with maximum deviations in the critical temperature of the order \(10^{-2}\). In Fig.
\ref{critical_temperature_ms_0.1_a_0.001}, the trend of critical
temperature \(T_c\) is similar to the one in the previous mass region,
whereas the temperature \(T_s\) in Fig. \ref{sr_Ts_ms_0.1_a_0.001}
stabilizes for large values of \(\lambda_{HS}\), in contrast to
the behavior described in the previous sections.
\par In the case of \(BR_{inv} < 0.11\), in the lower mass region, the previous conclusions do not change as this constraint only decreases the maximum allowed value of the Higgs-singlet coupling to \(\lambda^{max}_{HS} = 0.10\), which subsequently excludes some scenarios, especially for \(\lambda = 0\).

In the second region of the parameter space for \(m_S < m_H/2\),
the singlet mass is strictly restricted to almost the half
mass of the Higgs boson due to the constraints in Eq.
(\ref{condition_coupling}) (Figs. \ref{BR_0.19} and \ref{BR_0.11}).
As a result, there is no noticeable difference between this case
and the Higgs resonance regarding the second step
in the two-step electroweak phase transition, while the phase
transition in the \(\phi\) direction is slightly affected. Namely,
the results on the critical temperature \(T_c\) and the Higgs VEV
\(\upsilon_c\) remain the same as the results in the previous
section. Assuming \(BR_{inv} < 0.19\), the constraint
(\ref{condition_coupling}) leads to the singlet mass \(m_S = 62.48\)
GeV and coupling \(\lambda_{HS} = 0.88\) as the only strong
electroweak phase transition scenario\footnote{The sphaleron rate
criterion is satisfied: \(\frac{\upsilon_c}{T_c} \simeq 0.61\).}
for \(a = 0.1\) in the standard singlet extension. Thus, the
constraint (\ref{condition_coupling}) imposes that \( m_S \geq
62.48\) GeV and leads to a narrow parameter space\footnote{The
maximum value of \(\lambda_{HS}\) in Eq.
(\ref{special_condition_coupling}) sharply increases as it
approaches half of the Higgs mass. This maximum value for \(m_S =
62.49998\) GeV reaches nearly half of the upper bound of
\(\lambda_{HS}\) in the parameter space shown in Fig.
\ref{ParameterSpace2}.}. On the other hand, the higher order operator can decrease the lower bound on the Higgs-singlet coupling due to the sphaleron rate criterion as it was discussed earlier. It is remarkable that in the higher mass
region the coupling \(\lambda_{HS} > 0.065\) (allowed by Eq.
(\ref{condition_coupling})) generates a viable electroweak phase
transition for \(\lambda/M^2 \simeq 9 \times 10^{-5}\)
GeV\(^{-2}\) and \(a = 0.75\). Consequently, the inclusion of the
dimension-six operator can contribute to the occurrence of a strong
electroweak first-order phase transition in the higher mass region
of the parameter space for \(m_S < m_H/2\). This discussion is
analogous in the case of \(BR_{inv} < 0.11\) because the lower
bound of the Higgs-singlet coupling remains unaffected, whereas
the size of the parameter space differs, with the parameter space
being highly restricted for \(m_S < 62.499\) GeV.

\section{Conclusions and Discussion}

Our current perception for the Universe during its primordial
classical and quantum evolution stages is rather vague. Too many
questions have to be answered, and the current and future CMB and
gravitational wave experiments are expected to shed light on these
stages of our Universe's evolution. The mechanism that gave masses to
the matter particles is the electroweak symmetry breaking and this
is believed to have occurred via a first-order phase transition.
However, the electroweak phase transition in the SM is not a sufficiently strong first-order phase transition. To this end, in this article, we studied the electroweak phase transition in the context of the real singlet extensions of the SM, including dimension-six non-renormalizable operators that couple the singlet scalar field with the Higgs doublet. We showed that the electroweak phase transition occurs as
a two-step phase transition, which consists of the singlet's
phase transition at high temperature and a subsequent strong
first-order phase transition in the Higgs and SM sector. In
addition, considering a viable \(CP\)-violation source, such as a
dimension-six operator, which couples the singlet to the top-quark
mass, the electroweak baryogenesis can be realized to describe the
baryon asymmetry in the current Universe. This scenario is also compatible with assuming the singlet particle as a dark matter candidate with a mass nearly half of the Higgs mass, which is highly restricted by the direct dark matter searches. As we previously discussed, the
parameter space is majorly reduced by the sphaleron rate
criterion, the vacuum structure, the perturbativity of couplings,
and the invisible Higgs decay width. Consequently, we studied the
two-step strong electroweak phase transition, which respects those
constraints for \(m_S = 0 - 550\) GeV. In this model, the critical
temperature of the first-order phase transition varies from \(T_c
\simeq 30 - 200\) GeV, depending on the parameters of the model.

The impact of the dimension-six operator is summarized
as follows:
\begin{itemize}
    \item The presence of the dimension-six operator mainly modifies the thermal mass of the singlet in the Higgs direction, which alters the first-order phase transition, while the singlet's phase transition is strongly affected by the non-zero Wilson coefficient and its critical temperature increases for higher Wilson coefficients. In addition, the effect of the higher order operator is washed out by large values of the Higgs-singlet coupling and is weakened for coefficients \(\lambda < 10^3\) in most cases.
    \item The dimension-six operator can generally assist in generating the strong
    electroweak phase transition for low Higgs-singlet couplings which
    were not allowed in the standard singlet extensions of the SM.
    \item The parameter space of our model for \(m_S > m_H/2\) and \(a = 0.1\) remains
    approximately the same as in the standard singlet extension. In the Higgs resonance region, the singlet particle could be a dark matter candidate, and the parameter space for \(a > 0.4\) is expanded for large values of the Wilson coefficient, whereas this trend is reversed for \(a < 0.4\).
    \item The low-mass region is significantly excluded by the invisible Higgs
    decay width and the sphaleron rate criterion in the context of the
    standard singlet extensions. Nevertheless, a strong electroweak phase transition can be generated for low singlet masses \(m_S < m_H/2\) by including the dimension-six operator with \(\lambda/M^2 \gtrsim 5 \times 10^{-6}\) GeV\(^{-2}\) and \(a \gtrsim 0.05\).

\end{itemize}
\par A study we did not include in our analysis, is the study of the
stochastic gravitational wave background that corresponds to the
first-order phase transitions occurring in our model. It is
expected that the collisions between the bubbles of vacua during
the first-order phase transition can generate a stochastic
gravitational wave background in the frequencies probed by current
and future gravitational wave experiments. We aim to analyze the
stochastic gravitational wave background generated by our model in
a near future work.

\section*{Acknowledgments}

This research has been is funded by the Committee of Science of
the Ministry of Education and Science of the Republic of
Kazakhstan (V.K.O) (Grant No. AP19674478).

\section*{APPENDIX A: Electroweak Phase Transition in the SM}

The electroweak phase transition can be realized in the context of
the SM as a first-order phase transition \cite{Anderson:1991zb,
Quiros:1999jp,Arnold:1992rz,Carrington:1991hz,Morrissey:2012db,Dine:1992wr,Dolan:1973qd,Senaha:2020mop},
but it is insufficiently strong to generate the baryon asymmetry.
The dynamics of the phase transition are described by the
temperature-dependent effective potential of the Higgs field
including the dominant contributions of the gauge bosons, the top
quark and the Goldstone bosons. This effective potential is
explicitly presented in section II and in the Arnold-Espinosa scheme it is written as
\begin{equation}\label{eq:A1}
\begin{split}
    V^{SM}_{eff} (h, T) = & - \frac{\mu^{2}_H}{2} h^2 + \frac{\lambda_H}{4} h^4 + \sum_{i} (-1)^{F_i} n_i \frac{m^4_{i}(h)}{64 \pi^2}\left[ \ln \left( \frac{m^2_{i}(h)}{\mu^2_R}\right) - C_i \right] - \frac{n_t m^4_{t}(h)}{64 \pi^2}\left[ \ln \left( \frac{m^2_{t}(h)}{\mu^2_R}\right) - C_t \right] \\
    & + \sum_{i} \frac{n_iT^4}{2 \pi^2} J_{B} \left(\frac{m^2_i (h)}{T^2}\right) - \frac{n_t T^4}{2 \pi^2} J_{F} \left(\frac{m^2_t (h)}{T^2}\right) \\
    & + \sum_{i} \frac{\overline{n}_i T}{12\pi} \left[m^3_i(h) - \left(M^2_i(h,T) \right)^{3/2} \right],
\end{split}
\end{equation}
where \(i = \{h, \chi, W, Z, \gamma \}\) corresponds to the bosons
in the SM.

The high-temperature expansion of the thermal functions (\ref{bosonthermalfunction}) and (\ref{fermionthermalfunction}) for the temperature-dependent one-loop effective potential is valid with very high accuracy in the case of the phase transition. Hence, the high-temperature expansion is implemented in the full effective potential which reads
\begin{equation}\label{eq:A2}
\begin{split}
    V^{SM}_{eff} (h, T) = & - \frac{\mu^{2}_H}{2} h^2 + \frac{\lambda_H}{4} h^4 + \frac{m^2_h (h)}{24}T^2 - \frac{T}{12 \pi} \left[m^2_h (h) + \Pi_h (T)\right]^{3/2} + \frac{m^4_h(h)}{64 \pi^2} \left[\ln \left(\frac{a_b T^2}{\mu^2_R}\right) -\frac{3}{2} \right] \\
    & + \frac{3 m^2_{\chi} (h)}{24}T^2 - \frac{3T}{12 \pi} \left[ m^2_{\chi} (h) + \Pi_{\chi}(T) \right]^{3/2} + \frac{ 3 m^4_{\chi} (h)}{64 \pi^2} \left[\ln \left(\frac{a_b T^2}{\mu^2_R}\right) -\frac{3}{2} \right] \\
    & + \frac{6 m^2_{W} (h)}{24}T^2 - \frac{4T}{12 \pi} m^3_{W} (h) - \frac{2T}{12 \pi}\left[ m^2_{W} (h) + \Pi_{W_L} (T) \right]^{3/2} + \frac{ 6 m^4_{W} (h)}{64 \pi^2} \left[\ln \left(\frac{a_b T^2}{\mu^2_R}\right) -\frac{5}{6} \right] \\
    & + \frac{3 m^2_{Z} (h)}{24}T^2 - \frac{2T}{12 \pi} m^3_{Z} (h) - \frac{T}{12 \pi} \left[  M^2_{Z_{L}} (h, T) \right]^{3/2} + \frac{ 3 m^4_{Z} (h)}{64 \pi^2} \left[\ln \left(\frac{a_b T^2}{\mu^2_R}\right) -\frac{5}{6} \right] \\
    & + \frac{12 m^2_{t} (h)}{48}T^2 - \frac{ 12 m^4_{t} (h)}{64 \pi^2} \left[\ln \left(\frac{a_f T^2}{\mu^2_R}\right) -\frac{3}{2} \right] - \frac{T}{12 \pi}\left[ M^2_{\gamma_{L}} (h, T) \right]^{3/2},
\end{split}
\end{equation}
where the effective masses are given by
\begin{equation}\label{eq:A3}
    m^2_h (h) = - \mu^2_H + 3\lambda_H h^2,
\end{equation}
\begin{equation}
    m^2_{\chi} (h) = - \mu^2_H + \lambda_H h^2,
\end{equation}
and Eqs. (\ref{effectivemassW}), (\ref{effectivemassZ}), (\ref{effectivemasstop}), while the temperature-dependent self-energy of the Higgs and the Goldstone bosons reads
\begin{equation}\label{eq:A4}
    \Pi_h (T) = \Pi_{\chi} (T) = \left(\frac{3g^2 }{16} + \frac{g^{\prime 2}}{16}  +\frac{y^2_t }{4} + \frac{\lambda_H}{2}\right) T^2
\end{equation}
and the thermal masses of the gauge bosons are given by Eqs.  (\ref{T-W}), (\ref{Z-thermalmass}), and (\ref{Photon-thermalmass}).
\begin{figure}
\centering
\includegraphics[width=35pc]{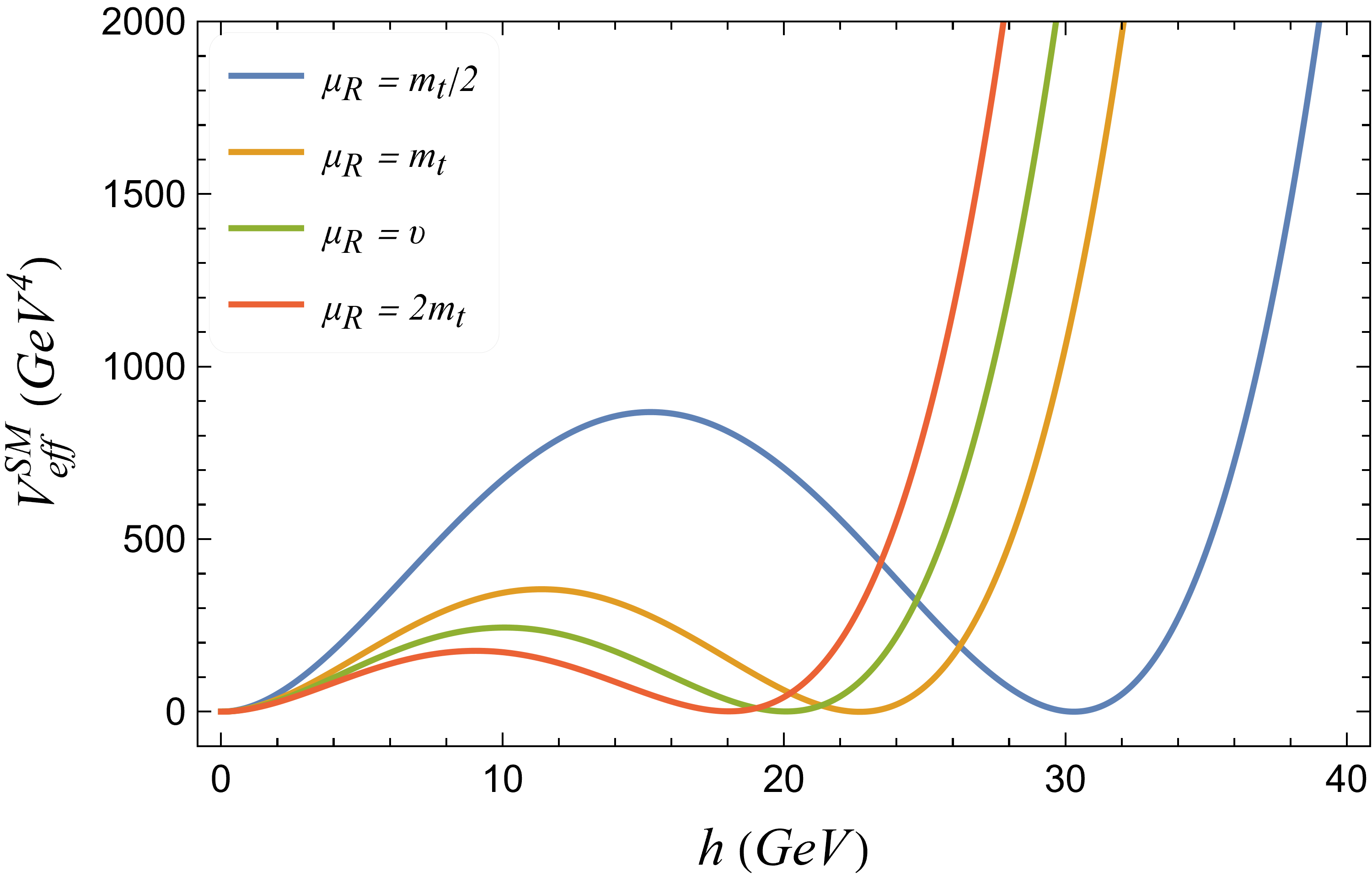}
\caption{The full effective potential of the SM at the critical
temperature for numerous renormalization scales \(\mu_R = m_t/2\)
(blue), \(m_t\) (orange), \(\upsilon\) (green) and \(2m_t\)
(red).}\label{electroweak phase transition_in_SM}
\end{figure}

\par The electroweak phase transition is studied by varying the
renormalization scale for different values. The full effective
potential (\ref{eq:A2}) is illustrated in Fig. \ref{electroweak phase transition_in_SM} for \(\mu_R =
m_t/2\) (blue), \(m_t\) (orange), \(\upsilon\) (green) and
\(2m_t\) (red).

\begin{table}[h!]
\centering
\begin{tabularx}{0.5\textwidth} {
  | >{\centering\arraybackslash}X
  | >{\centering\arraybackslash}X
  | >{\centering\arraybackslash}X| }
 \hline
 \(\mu_R\) (GeV) & \(T_c\) (GeV) & \(\upsilon_c\) (GeV)\\
 \hline
 86.5 & 148.987 &  30\\
 \hline
 173.0 & 148.326 & 23\\
 \hline
 246.2 & 148.002 & 20\\
 \hline
 346.0 & 147.693 & 18\\
 \hline
\end{tabularx}
\caption{The critical temperature of the electroweak phase
transition (\(T_c\)) and the Higgs VEV (\(\upsilon_c\)) for
different values of the renormalization scale.} \label{table:1}
\end{table}
It is immediately apparent that the sphaleron rate criterion is
not satisfied as well as the maximum ratio is achieved for \(\mu_R
= m_t/2\) with \(\upsilon_c/T_c \simeq 0.2 < 0.6\); namely, the
electroweak phase transition is not a strong enough first-order
phase transition\footnote{Likewise, the authors of Ref.
\cite{Anderson:1991zb} argued that the sphaleron rate criterion is
not satisfied for \(m_H = 125\) GeV in SM. In particular, they
calculated that \(\upsilon_c/ T_c \simeq 0.2\) for \(m_H = 120\)
GeV and \(m_t = 170\) GeV, considering the high-temperature expansion without the daisy
resummation.}. Similar calculations can be found in
Refs. \cite{Anderson:1991zb,
Quiros:1999jp,Carrington:1991hz,Morrissey:2012db,Senaha:2020mop}.
In the context of the perturbative analysis, the sphaleron rate
criterion can be translated into an upper bound on the mass of the
Higgs boson, such as \(m_H \lesssim 42\) GeV \cite{Quiros:1999jp},
which showcases the lack of a strong phase transition in SM,
whereas this bound is \(m_H \lesssim 72 - 80\) GeV in lattice
calculations
\cite{Gurtler:1997hr,Laine:1998jb,Csikor:1998eu,Aoki:1999fi}.
While the height of the barrier in the effective potential and the
Higgs VEV in Fig. \ref{electroweak phase transition_in_SM} clearly change for different
renormalization scales, the critical temperature is slightly
dropped by less than \(1 \%\) for higher values of the
renormalization scale as it is presented in Table \ref{table:1}.

\section*{APPENDIX B: One-loop Beta Functions}

The RGEs for the parameters of the real singlet extension to the
SM are presented here:
\begin{equation}\label{beta_lamdbaH}
    16 \pi^2 \beta_{\lambda_H} = 24 \lambda^2_H - 3 \left(3 g^2 + g^{\prime 2} - 4 y^2_t \right)\lambda_H + \frac{1}{2} \lambda^2_{HS} + \frac{3}{8} \left(3 g^4 + 2 g^2 g^{\prime 2} + g^{\prime 2} \right) - 6 y^4_t,
\end{equation}
\begin{equation}\label{beta_lamdbaHS}
    16 \pi^2 \beta_{\lambda_{HS}} = 4 \lambda^2_{HS} + \left(12 \lambda_H + 6 \lambda_S + 6 y^2_t - \frac{9}{2} g^2 - \frac{3}{2} g^{\prime 2}  \right) \lambda_{HS},
\end{equation}
\begin{equation}\label{beta_lamdbaS}
    16 \pi^2 \beta_{\lambda_S} = 18 \lambda^2_S + 2 \lambda^2_{HS},
\end{equation}
\begin{equation}\label{beta_g}
    16 \pi^2 \beta_{g} = -\frac{19}{6} g^3,
\end{equation}
\begin{equation}\label{beta_g1}
    16 \pi^2 \beta_{g^{\prime}} = \frac{41}{6} g^{\prime 3},
\end{equation}
\begin{equation}
    16 \pi^2 \beta_{g_s} = - 7 g^3_s,
\end{equation}
\begin{equation}\label{beta_yt}
    16 \pi^2 \beta_{y_t} = \frac{9}{2} y^3_t - \left( \frac{9}{4} g^2 + \frac{17}{12} g^{\prime 2} + 8 g^2_s \right) y_t,
\end{equation}
where the one-loop beta function is defined as
\begin{equation}\label{beta}
    \beta_g = \mu \frac{dg}{d \mu}.
\end{equation}

\end{document}